\documentclass[12pt,hidelinks]{article}
\usepackage{amsmath,amsfonts}
\usepackage{enumerate}
\usepackage{natbib}
\usepackage{url} 
\usepackage{amsthm}
\usepackage[pdftex]{graphicx}
\usepackage{indentfirst}
\usepackage{verbatim}
\usepackage{setspace}
\usepackage{mathtools}
\usepackage{booktabs}
\usepackage{comment}
\usepackage{color}
\usepackage{bm}
\usepackage{subcaption}
\usepackage{changepage}
\usepackage{hyperref}
\usepackage{adjustbox}
\usepackage[toc,title,page]{appendix}

\newcommand{\zb}{\mathbf{z}}
\newcommand{\Zb}{\mathbf{Z}}
\newcommand{\xb}{\mathbf{x}}
\newcommand{\Xb}{\mathbf{X}}

\newcommand{\Var}{\mathrm{Var}}

\newcommand{\aipw}{\mathrm{aipw}}

\newcommand{\agg}{\mathrm{agg}}
\newcommand{\ind}{\mathrm{ind}}
\newcommand{\T}{\top}
\newcommand{\dist}{\mathrm{d}}
\newcommand{\MF}{\mathrm{mf}}
\newcommand{\FF}{\mathrm{ff}}
\newcommand{\M}{\mathrm{m}}
\newcommand{\F}{\mathrm{f}}

\def\+#1{\mathbb{#1}}
\def\*#1{\mathbf{#1}}

\newcommand{\gvec}{\mathbf{g}}
\newcommand{\Gvec}{\mathbf{G}}

\usepackage{mathrsfs}

\def\z(#1){\mathbf{z}_{(#1)}}

\def\c(#1){\mathbf{#1}_{c}}
\def\n(#1){{#1}_{c,(i)}}
\def\s(#1){{#1}_{c,i}}
\def\onenorm(#1){\lVert#1\rVert_1}

\newcommand{\norm}[1]{\left\lVert#1\right\rVert}

\definecolor{Blue}{rgb}{0,0,1}
\definecolor{Grey}{rgb}{.5,.5,.5}

\usepackage{multirow, rotating}
\usepackage{pst-node}
\usepackage{inconsolata}
\usepackage[T1]{fontenc}

\newtheorem{lemma}{Lemma}
\newtheorem{assumption}{Assumption}
\newtheorem{proposition}{Proposition}

\newtheorem{remark}{Remark}

\usepackage[margin=10pt,labelfont=bf]{caption}
\newcommand\tcaptab[1]{\captionsetup{position=top, font=normalsize, labelfont=bf, textfont=normalfont, justification=centering, margin=0mm, aboveskip=1mm, belowskip=0mm, labelsep=colon, singlelinecheck=false}\caption{#1}}
\newcommand\bnotetab[1]{\captionsetup{position=bottom, font=footnotesize,  textfont=normalfont, margin=1mm, skip=2mm, justification=justified, singlelinecheck=false}\caption*{#1}}

\newtheorem{theorem}{Theorem}
\newtheorem{corollary}{Corollary}
\usepackage{natbib}

\usepackage{amssymb}
\usepackage[final]{pdfpages}

\usepackage[inline]{enumitem}   
\makeatletter
\newcommand{\inlineitem}[1][]{%
\ifnum\enit@type=\tw@
    {\descriptionlabel{#1}}
  \hspace{\labelsep}%
\else
  \ifnum\enit@type=\z@
       \refstepcounter{\@listctr}\fi
    \quad\@itemlabel\hspace{\labelsep}%
\fi}
\makeatother

\usepackage{amsmath}
\usepackage{graphicx,psfrag,epsf}
\usepackage{enumerate}
\usepackage{natbib}
\usepackage{url} 

\begin{document}

\def\spacingset#1{\renewcommand{\baselinestretch}%
{#1}\small\normalsize} \spacingset{1}

	\title{Semiparametric Estimation of Treatment Effects in Observational Studies with Heterogeneous Partial Interference}
	\author{Zhaonan Qu$^\ast$\thanks{\scriptsize Stanford University, Department of Economics, \url{zhaonanq@stanford.edu}}
	\and Ruoxuan Xiong$^\ast$\thanks{\scriptsize Emory University, Department of Quantitative Theory and Methods, \url{ruoxuan.xiong@emory.edu}. Equal contribution.}
	\and Jizhou Liu\thanks{\scriptsize The University of Chicago Booth School of Business,  \url{jliu32@chicagobooth.edu}}
	\and Guido Imbens\thanks{\scriptsize Stanford University, Graduate School of Business, and Department of Economics,  \url{imbens@stanford.edu}}}
	
	\date{\vspace{-5ex}}
	\begin{doublespacing}
			\maketitle
			\begin{abstract}
				In many observational studies in social science and medicine, subjects or units are connected, and one unit's treatment and attributes may affect another's treatment and outcome, violating the stable unit treatment value assumption (SUTVA) and resulting in interference. To enable feasible estimation and inference, many previous works assume exchangeability of interfering units (neighbors). However, in many applications with distinctive units, interference is heterogeneous and needs to be modeled explicitly. In this paper, we focus on the partial interference setting, and only restrict units to be exchangeable conditional on observable characteristics. Under this framework, we propose generalized augmented inverse propensity weighted (AIPW) estimators for general causal estimands that include heterogeneous direct and spillover effects. We show that they are  semiparametric efficient and robust to heterogeneous interference as well as model misspecifications. We apply our methods to the Add Health dataset to study the direct effects of alcohol consumption on academic performance and the spillover effects of parental incarceration on adolescent well-being.
				
				\vspace{0.5cm}
				\noindent%
{\it Keywords:}  Exchangeability, SUTVA, Spillover Effects, Peer Effects, Augmented Inverse Propensity Weighting
\vfill
			\end{abstract}

  \newpage
		



        \section{Introduction}
        
        The classical treatment effect estimation literature typically relies on the stable unit treatment value assumption, or SUTVA \citep{rubin1974estimating,rubin1980randomization}, which assumes that a unit's potential outcomes do not depend on the treatment status of other units. However, SUTVA can be inappropriate for applications where individuals or units are connected or interact with other, for example directly through social networks \citep{banerjee2013diffusion,ogburn2017vaccines} and group memberships \citep{sacerdote2001peer,miguel2004worms,duflo2011peer,fadlon2019family}, or indirectly through equilibrium effects \citep{heckman1998general,johari2022experimental,munro2021treatment}. Motivated by these applications, there has been a growing literature on the identification and estimation of treatment and spillover effects under \emph{interference} \citep{cox1958planning,hudgens2008toward}, that allows a unit's potential outcomes to depend on the treatments of other units.\footnote{In observational studies, the same dependence structure that mediates interference can also lead to
correlations in treatment assignments, e.g., through the contagion of behaviors in a social network. This
implies that the classical unconfoundedness assumption (Rosenbaum and Rubin, 1983) is also likely to fail.}

\begin{figure}
			\centering
			\begin{subfigure}{0.5\textwidth}
				\centering
				\includegraphics[width=0.8\linewidth]{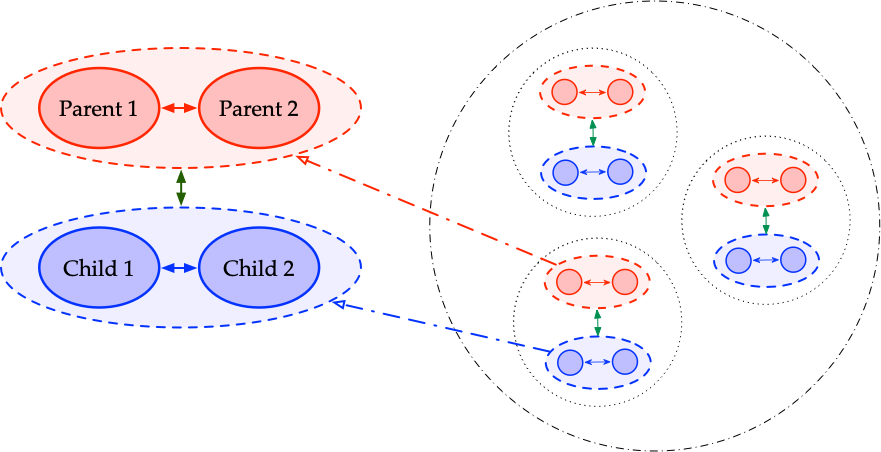}
			\caption{Household example}\label{fig:family}
			\end{subfigure}\hfill
			\begin{subfigure}{0.38\textwidth}
				\centering
				\includegraphics[width=0.8\linewidth]{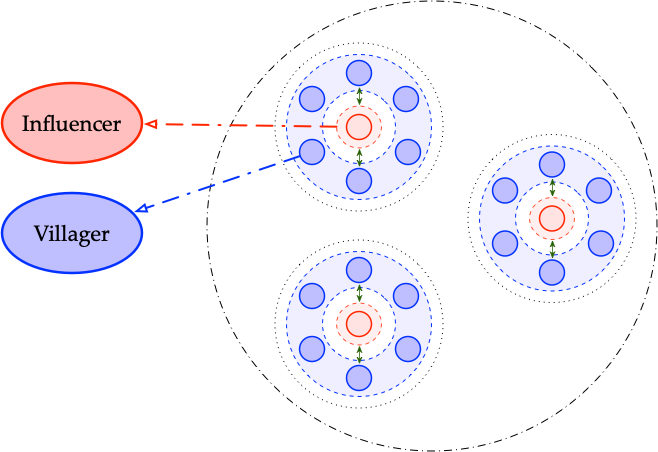}
			\caption{Village example}\label{fig:village}
			\end{subfigure}
			
    \caption{\footnotesize Illustration of our setting under heterogeneous partial interference. In Figure \ref{fig:family}, each household consists of two exchangeable subsets (see Section \ref{subsec:exchangeability}): parents and children, so that interference from units within each subset is homogeneous (denoted by links with the same color). In Figure \ref{fig:village}, each village consists of leaders/influencers and villagers, and interference depends on the  social status of a person in the village. Homogeneous interference among villagers is indicated by the blue-shaded ring. We allow for \emph{varying} cluster sizes in extensions.}
			\label{fig:basic-setting}
		\end{figure}

        Often in settings with interference, the mediating interactions among units are \emph{heterogeneous}. This heterogeneity can arise from different sources, depending on the distinct nature and strength of interactions. For example, interference among close friends can matter a lot more than that among acquaintances, and units within close geographical proximity tend to have stronger interference than distant units. Another  common scenario is where units have distinct \emph{types} based on observable information, and interactions across different types of units are expected to be heterogeneous. For example, for a child in a household in Figure \ref{fig:family}, spillover effects from a treated parent may be different from the spillover effect from a treated sibling, or may depend on the gender of the treated parent. Potential heterogeneities in interference abound in empirical applications, and failure to take them into account may lead to biased estimation and invalid inference, even if one is interested in aggregate or marginal treatment effects like the ATE \citep{forastiere2020identification}.
    
        In this paper, we focus on settings with \emph{partial interference} \citep{sobel2006randomized,hudgens2008toward}, where units can be naturally divided into disjoint clusters (e.g., households), and interference is restricted to occur only among units within the same cluster (see Figure \ref{fig:basic-setting}). 
        We further assume that within clusters, there is a partitioning of units into distinct types based on observable characteristics (e.g., parents vs. children or father vs. mother), and a unit's interactions with its neighbors of the same type are \emph{exchangeable}. However, the interactions can be arbitrarily distinct for units of different types. We refer to this setup as the \emph{conditional exchangeability} framework, which leads to an interference structure we call \emph{heterogeneous partial interference}. This structure can be expressed in the form of an \emph{exposure mapping} as established in previous works \citep{manski2013identification, aronow2017estimating,forastiere2020identification, vazquez2017identification}. Nevertheless, it enjoys the benefit of intuitive interpretation of various sources of heterogeneity, including that in treatment assignment probabilities and treatment effects.
        Our framework also includes the commonly used \emph{homogeneous partial interference} assumption as a special case, where all units in a cluster are exchangeable, not just those of the same type.

        Under the conditional exchangeability framework, we propose a class of generalized augmented inverse propensity weighted (AIPW) estimators for heterogeneous direct and spillover effects, which are shown to be doubly robust, asymptotically normal, and semiparametric efficient. To the best of our knowledge, there have not been any formal results on semiparametric efficiency and AIPW estimators in both homogeneous and heterogeneous partial interference settings for estimands defined under the corresponding exposure mapping. We further propose a data-driven framework to determine the appropriate interference structure by leveraging statistical tests using a matching-based variance estimator, which allows the practitioner to detect and account for heterogeneous partial interference in the absence of domain knowledge.
        
        Our estimators are natural and relevant in many applications, such as the optimal design of treatment allocation rules under resource constraints, where identifying heterogeneities in direct and spillover effects is important \citep{banerjee2013diffusion}.
        In addition, as shown in this work, even if our primary object of interest is an aggregated marginal treatment effect like the ATE, correctly accounting for the heterogeneity in interference is crucial to obtaining a \emph{consistent} and \emph{efficient} estimate of the aggregate effect. This is particularly the case when using \emph{observational data}, which requires at least one of the outcome and treatment assignment models to be consistently estimated. When the heterogeneity in interference is not appropriately accounted for,  both models may fail to be consistently estimated, resulting in biased estimates of the aggregate effect.

       We demonstrate the superior finite sample properties, robustness in estimation and inference, and overall robustness of our estimators through extensive simulation studies. We also apply our methods to the Add Health data \citep{harris2009national} to study two empirical questions. First, we investigate the effects of regular alcohol use on students' academic performance. Second, we examine the impact of parental incarceration on adolescent well-being. These applications showcase the practical relevance and effectiveness of our methods in real-world settings.

       In summary, our results provide practitioners with a data-driven toolkit to assess and account for the impact of heterogeneous interference in a wide range of applications with observational data.
			
\paragraph{Related Works} This paper contributes to the growing literature on treatment effect estimation under partial interference. Partial interference in randomized experiments has been studied by \cite{halloran1995causal, sobel2006randomized, hudgens2008toward, liu2014large, vazquez2017identification, basse2018, jagadeesan2020designs, liu2023inference}, often in the context of two-stage randomization. Other recent works have also studied the estimation and inference of causal quantities from \emph{observational} data \citep{tchetgen2012causal,perez2014assessing,liu2016inverse,barkley2020causal,park2020efficient,forastiere2020identification}. Both these works and our work account for endogenous treatment assignments in observational data under partial interference. Our work is distinct in the following aspects. First, in many previous works, the causal quantities of interest are typically an average treatment effect over the whole population when each neighbor is \emph{hypothetically} and \emph{independently} treated with probability $\alpha$, i.e., $\alpha$-allocation strategy. In contrast, our estimands are defined under the exposure mapping derived from the conditional exchangeability framework, and also include more granular estimands defined for distinct subpopulations to explicitly account for both heterogeneous interference and treatment effects. As a result, our work could help address more general empirical questions involving a wider class of causal quantities. Second, our generalized AIPW estimators are different from those proposed in
\citet{tchetgen2012causal,perez2014assessing,liu2016inverse,barkley2020causal}, which are generalized IPW estimators, and those in \citet{liu2019doubly} and \citet{park2020efficient} which are generalized AIPW estimators using different estimation strategies for the propensity and outcome models. Third, we provide the first semiparametric efficiency lower bound type results for estimands based on exposure mappings. Compared to the efficiency bounds based on the $\alpha$-allocation strategy in \cite{park2020efficient}, our bounds apply to a different class of estimands and explicitly quantify how various sources of heterogeneity affect estimation efficiency. Lastly, we also propose a data-driven approach to detect heterogeneous interference based on hypothesis testing using a matching-based variance estimator, which was not done in prior works. Our work is closely related to \cite{forastiere2020identification}, particularly with respect to the idea of defining estimands and constructing estimators through the use of a specific exposure mapping. While their work considers more general interference, our paper provides a specific interference structure under conditional exchangeability that is relevant and interpretable in many empirical settings. More importantly, our work complements this line of literature (e.g., \cite{forastiere2020identification, tortu2020modelling}) by offering efficient estimators and asymptotically valid inference methods that allow the practitioner to determine the appropriate interference structure.



		The rest of the paper is organized as follows. Section \ref{sec:model} presents model assumptions and the conditional exchageability framework. Section \ref{sec:estimator} defines the causal estimands of interest and proposes the estimation procedure based on generalized AIPW estimators. Section \ref{section:asymptotics} presents main asymptotic results. Sections \ref{section:simulations} and \ref{section:applications} present simulation results and applications to the Add Health dataset. Section \ref{section:conclusions} concludes the paper.

		\section{Model Setup}
		\label{sec:model}

		In this paper, we study the potential outcomes framework in the observational setting with partial interference, where we observe $N$ units that are divided into a large number $M=O(N)$ of clusters, and interference is restricted to among units within the same cluster. This clustering structure can arise from group memberships, geographical proximity, and sampling or randomization schemes. We assume that each cluster is drawn i.i.d. from a population distribution $\mathbb{P}$, but within each cluster, covariates and treatment assignments may have arbitrary dependence across units. In particular, we allow for phenomena such as \emph{homophily} of similar units \citep{mcpherson2001birds,bramoulle2012homophily} and \emph{contagion} of treatment assignments  \citep{centola2010spread,christakis2013social}. 
		We state the standard assumptions on the data generating process in Section \ref{subsec:assumption} and introduce our framework for heterogeneous interference in Section \ref{subsec:exchangeability}.
		
		\subsection{General Assumptions} \label{subsec:assumption}
		
		Let $c=1,\dots,M$ be the index of a cluster. For exposition, assume for now that each cluster has the same size $n = N/M$. We generalize our results to varying cluster sizes in Appendix \ref{sec:varying-cluster-size}. Let  $\mathbf{Y}_c \in \mathbb{R}^{n}$, $\mathbf{Z}_c \in \{0,1\}^n$, and $\mathbf{X}_c \in \mathbb{R}^{n\times d_x}$, be the observed outcomes, treatment assignments, and covariates of all the $n$ units in cluster $c$, where $d_x$ is the dimenionality of each unit's covariates.
		
		Let $Y_{c,i} \in \+R$, $Z_{c,i} \in \{0,1\}$, and $X_{c,i}\in \+R^{d_x}$ be the observed outcome, treatment assignment and covariates of unit $i$ in cluster $c$, where $Y_{c,i}$ and $Z_{c,i}$ are the $i$-th coordinate of $\mathbf{Y}$ and $\mathbf{Z}$, and $X_{c,i}$ is the $i$-th row of $\mathbf{X}_c$. We define unit $i$'s \textbf{neighbors} as all units in cluster $c$ other than unit $i$.\footnote{We can also extend to settings where there is a network structure within each cluster, so the definition of neighbors is unit-dependent.} Let $\*Y_{c,(i)} \in \+R^{n-1}$, $\*Z_{c,(i)} \in \{0,1\}^{n-1}$, and $\*X_{c,(i)} \in \+R^{(n-1) \times d_x}$ be the observed outcome, treatment assignment and covariates of $i$'s neighbors in cluster $c$.
		
		\begin{assumption}[i.i.d. Clusters]\label{ass:network}
		The $M$ clusters are i.i.d., i.e., tuples $(\c(X),\c(Y),\c(Z))$ for $c=1,\dots,M$ are drawn i.i.d. from some compactly-supported population distribution $\mathbb{P}$. 
		\end{assumption}

		Many applications exhibit a natural clustering structure. For example, clusters may be households \citep{vazquez2017identification} or dormitory rooms \citep{sacerdote2001peer}. We use households as a running example throughout the paper (see Figure \ref{fig:family}). The mechanism behind the clustering structure could be exogenous (e.g., assigned externally based on units' characteristics) or endogenous (e.g., through homophily). Even in general network settings where a clustering structure is not obvious, graph segmentation or sampling techniques can be used to partition the network into clusters \citep{ugander2013graph,saint2019using}. To address settings where clusters are potentially correlated with each other, we discuss the generalization to weakly connected clusters in Appendix \ref{subsec:network}.
  Note that our framework allows units' covariates and treatments to be arbitrarily correlated among units in the same cluster, thus allowing for contagion of treatments in addition to homophily.

        Next, we allow a unit's potential outcomes to depend on the treatment assignments of its \emph{neighbors} (in addition to its own treatment) within the same cluster, which relaxes the SUTVA assumption \citep{rubin1980randomization,rubin1986comment}. However, we impose the partial interference assumption \citep{sobel2006randomized}, where there is no interference between units in different clusters \citep{halloran1995causal}.  This assumption is reasonable for applications with disjoint or sufficiently separate clusters, e.g., in space or time.

		\begin{assumption}[Partial Interference]\label{ass:partial}
		For any unit $i$ in cluster $c$, unit $i$'s potential outcomes can only depend on the treatment assignments of units within the same cluster $c$.
		\end{assumption}

		Given Assumption \ref{ass:partial}, we can define the potential outcomes of unit $i$ in cluster $c$ as
		\begin{equation*}
		    Y_{c,i}(\s(z),\n(\zb))
		\end{equation*}
		for $c \in \{1, \cdots, M\}$, $i \in \{1, \cdots, n\}$, $\s(z) \in \{0,1\}$, and $\n(\zb) \in \{0,1\}^{n-1}$. Here $\s(z)$ and $\n(\zb)$ denote the deterministic values that $\s(Z)$ and $\n(\Zb)$ can take. 
		The observed outcome of unit $i$ in cluster $c$ is
		\[Y_{c,i} \equiv Y_{c,i}(Z_{c,i},\Zb_{c,(i)}). \]
		
		Now we consider the treatment assignment mechanism. In the classical observational setting, the unconfoundedness assumption is commonly imposed \citep{rosenbaum1983central} and states that treatment assignments are free from dependence on potential outcomes \emph{conditional} on units' own covariates. In the presence of interference, this may no longer be true: as units interact with each other, one unit's treatment assignment could also depend on its neighbors' characteristics and treatment assignments. Therefore, we impose a generalized unconfoundedness assumption, which is necessary for the identification of causal effects under interference in observational studies.
		
		\begin{assumption}[Generalized Unconfoundedness]\label{ass:unconfoundedness} 
			 For any cluster $c$, unit $i$, and treatment assignment values $\big(\s(z),\n(\zb)\big)$, 
			\begin{align}
			\s(Y)\big(\s(z), \n(\zb)\big) &\perp \big(\s(Z), \n(\Zb)\big) \mid \Xb_c. \label{eqn:unconfoundedness}
			\end{align}
		\end{assumption}
		
		Assumption \ref{ass:unconfoundedness} extends the classical unconfoundedness assumption and implies that the treatment assignment probability satisfies 
		\begin{align*}
		 &   P\big(\Zb_c \mid \s(Y)\big(\s(z), \n(\zb)\big), \Xb_c\big) = P(\Zb_c \mid \Xb_c) \qquad \forall i, \s(z),\n(\zb).
		\end{align*}
		We define $P(\Zb_c \mid \Xb_c)$ as the \textbf{propensity score} of cluster $c$. 
		Assumptions similar to Assumption \ref{ass:unconfoundedness} have been used in the literature \citep{liu2016inverse,forastiere2020identification,park2020efficient}.

		Lastly, for the purpose of identification, we impose the overlap assumption on cluster-level treatment probabilities (propensity scores). 
		
		\begin{assumption}[Overlap]\label{ass:overlap}
		There exist some $\underline{p}$ and $\overline{p}$ such that for any value of $\c(Z)$ and $\c(X)$,
			\begin{eqnarray}
			0 < \underline{p} \leq P(\c(Z) \mid \c(X)) \leq \overline{p} < 1.
			\end{eqnarray}
		\end{assumption}
Assumption \ref{ass:overlap} implies that $P(\s(Z) \mid \c(X))$ is bounded away from $0$ and $1$, i.e., the classical overlap assumption holds for every unit $i$ in cluster $c$ if we condition on cluster covariates $\c(X)$ instead of individual covariates only.\footnote{A unit's propensity scores are sometimes stated using the unit's own covariates only in some previous works. Note that in our framework, we can also define a unit's propensity score as $P(Z_{c,i} \mid \tilde{\*X}_{c,i})$ which uses the unit's ``own'' covariates $\tilde{\*X}_{c,i}$, and $\tilde{\*X}_{c,i}$ is defined as $\tilde{\*X}_{c,i} \coloneqq \left(X_{c,i}, \*X_{c,(i)} \right)$ that contains $i$'s neighbors' information.}

		\subsection{Conditional Exchangeability}
		\label{subsec:exchangeability}
		
        In many applications, units can have heterogeneous interactions and interference often depends on the particular \emph{type} of neighbors that are treated. 
      Building on \citet{vazquez2017identification} and \citet{forastiere2020identification}, we adopt a conditional exchangeability framework to formalize such heterogeneous interference. In our conditional exchangeability framework, we partition units within each cluster into $m \geq 1$ exchangeable and disjoint subsets (such as parents and children in a family), denoted by $\mathcal{I}_1, \cdots, \mathcal{I}_m$, that satisfy\footnote{Partitions can have varying sizes across clusters. $\mathcal{I}_j$ can be a singleton for any $j$. This partition implicitly imposes an ordering of subsets. In the household example with cluster size four in Figure \ref{fig:family}, units $i=1$ and $i=2$ are parents, and $i=3$ and $i=4$ are children for all clusters. Within each subset $\mathcal{I}_j$, the ordering of units can be arbitrary.} 
	     
	     \[\mathcal{I}_1 \cup \mathcal{I}_2 \cup \cdots \cup \mathcal{I}_m = \{1, 2, \cdots, n\}, \quad \text{and }\quad \mathcal{I}_j \cap \mathcal{I}_k = \emptyset \quad \text{ for $j \neq k$}. \]
	     We assume a unit's interference from treatments of neighboring units in the same subset are the same, but \emph{may} be different for neighboring units in distinct subsets, which is formalized in Assumption \ref{ass:cond-outcome-partial-exchangeable} below. For ease of exposition, we assume for now that the partition is the same for \emph{all} clusters and is unit-independent, but our results can be generalized to settings where the partition is unit-dependent or cluster-dependent in a straightforward manner, as discussed in Appendix \ref{subsection:general-cluster-structure}. 

        \begin{assumption}[Conditional Exchangeability of Potential Outcomes] 
		\label{ass:cond-outcome-partial-exchangeable} For each unit $i$, its potential outcomes are exchangeable with respect to arbitrary permutations of treatment assignments of other units in the same subset, i.e.,
			\begin{align}
			Y_{c,i}\big(\s(z), \zb_{c,(i),1}, \cdots, \zb_{c,(i),m}  \big) &= Y_{c,i}\big(\s(z), \pi_1(\zb_{c,(i),1}), \cdots, \pi_m(\zb_{c,(i),m})  \big) \, ,
			\end{align}
			where for $j \in \{1, \cdots, m\}$, $\zb_{c,(i),j}$ is the treatment realization that units in $\mathcal{I}_j\backslash i$ can take, and $\pi_j(\cdot) \in \mathbb{S}^{|\mathcal{I}_j\backslash i|}$ is an arbitrary permutation. $\mathcal{I}_j\backslash i$ is the subset of units in $\mathcal{I}_j$ but not in $\{i\}$. $|\mathcal{I}_j\backslash i|$ is the cardinality of $\mathcal{I}_j\backslash i$.\footnote{If $i \not\in \mathcal{I}_j$, then $\mathcal{I}_j\backslash i = \mathcal{I}_j$. If $i \in \mathcal{I}_j$ and $\mathcal{I}_j$ is singleton, then $\mathcal{I}_j\backslash i = \emptyset$ and therefore $\zb_{c,(i),j} = \emptyset$, $\Xb_{c,(i),j} = \emptyset$, $\pi_j( \zb_{c,(i),j}) = \emptyset$ and $\pi_j^{-1}(\Xb_{c,(i),j}) = \emptyset$ for any permutation $\pi_j(\cdot)$.} 
		\end{assumption}

	If a cluster only has two units or if $m  = n$, then Assumption \ref{ass:cond-outcome-partial-exchangeable} always holds. If $m = 1$, then Assumption \ref{ass:cond-outcome-partial-exchangeable} reduces to the \textbf{fully exchangeable} assumption (\cite{hudgens2008toward} calls this stratified interference). In this case, potential outcomes only depend on how many neighbors, but not which ones, are treated. This setting is commonly used in epidemiology, for example, to study the effect of vaccine coverage \citep{hudgens2008toward,tchetgen2012causal}. 
For general $m$, Assumption \ref{ass:cond-outcome-partial-exchangeable} implies  that potential outcomes depend on the \emph{numbers} of treated neighbors in each subset $\mathcal{I}_j$. Thus in the definition of potential outcomes, the ($n-$1)-dimensional vector of neighbors' treatment assignments $\n(\zb)$ can be summarized by the $m$-dimensional\footnote{When $m=n$, no units are exchangeable and we can identify $\gvec_{c,i}$ with the ($n-$1)-dimensional vector $\n(\zb)$ itself, instead of an $n$-dimensional vector with a trivial coordinate always equal to zero.} vector $\gvec \in \mathbb{Z}_{\geq 0}^m$ of the number of treated neighbors in each subset:
		\begin{equation*}
		\label{eq:potential-outcome-reduction}
		Y_{c,i}\big(\s(z), \zb_{c,(i),1}, \cdots, \zb_{c,(i),m}  \big) \equiv \s(Y)(\s(z), \underbrace{g_{c,1}, \cdots,g_{c,m}}_{\gvec_{c,i}} ),
		\end{equation*}
		for any $\zb_{c,(i),j}$ that satisfies $\norm{\zb_{c,(i),j}}_1 = g_{c,j}$ with $g_{c,j} \in \{0, \cdots,|\mathcal{I}_j \backslash i|\}$, where $\norm{\zb_{c,(i),j}}_1$ is the $\ell_1$ norm of $\zb_{c,(i),j}$.\footnote{In an extension in Section \ref{subsection:general-cluster-structure}, we consider the case where a unit's potential outcomes may depend on some, but not all, units' treatment assignments within the same cluster.}
		The number of potential outcomes can be significantly reduced by using $\s(Y)(\s(z),\gvec_{c,i})$. When $m = 1$, the number of potential outcomes is reduced from $2^n$ to $2(n-1)$.

		Our definition of potential outcomes $\s(Y)(\s(z),\gvec_{c,i})$ can be viewed as a form of exposure mapping \citep{aronow2017estimating} from $\zb_{c,(i)}$ to $\gvec_{c,i}$ defined through exchangeable subsets. Our conditional exchangeability framework thus complements \cite{bargagli2020heterogeneous}, \cite{tortu2020modelling} and \cite{forastiere2020identification} which use general treatment exposure mappings to capture heterogeneous interference. In contrast, we explicitly model heterogeneous interference using the exchangeable subsets $\mathcal{I}_j$ based on observable and interpretable characteristics, which aligns with many applications where such partitions arise naturally. These subsets are also used in Section \ref{sec:estimator} below for the definition and identification of \emph{heterogeneous} direct treatment and spillover effects for different subsets of units, whereas many previous works treat the heterogeneity of interference as a nuisance in the estimation of marginal treatment effects. A similar exchangeability condition is also discussed in \citet{vazquez2017identification} in the experimental setting. Our framework is designed for the observational setting and therefore imposes exchangeability on the propensity as well.
	
		 Given the conditional exchangeability of potential outcomes imposed in Assumption \ref{ass:cond-outcome-partial-exchangeable},  it follows immediately that the unconfoundedness assumption in Assumption \ref{ass:unconfoundedness} continues to hold when using  $\s(Y)(z, \gvec)$ as the definition of potential outcomes:
		\begin{equation*}
		\s(Y)(z, \gvec) \perp \left(\s(Z), \s(\Gvec)\right) \mid \Xb_c, \quad \forall c,i,z,\gvec,
		\end{equation*}
		where  $\Gvec_{c,i}  = \big(G_{c,i,1}, \cdots, G_{c,i,m}  \big)$, and $G_{c,i,j} = \sum_{k \in \mathcal{I}_j \backslash i} Z_{c,k}$ is the number of treated neighbors of unit $i$ in subset $j$ in cluster $c$.\footnote{If $i \in \mathcal{I}_j$ and $\mathcal{I}_j$ is singleton, then $\mathcal{I}_j \backslash i = \emptyset$ and $\sum_{k \in \mathcal{I}_j \backslash i} Z_{c,k} = 0$.}

    We define the \textbf{conditional outcome model} as 
    \begin{align*}
	    \mu_{i,(z,\gvec)}(\c(X)) \coloneqq&\mathbb{E}[Y_{c,i}(z,\gvec) \mid \s(X),\Xb_{c,(i),1}, \cdots, \Xb_{c,(i),m}].
	\end{align*}
	
	As shown in Lemma \ref{lemma:conditional-outcome-exchangeable} in Appendix \ref{subsec:conditional-exch}, the conditional outcome model is well-defined and satisfies the following permutation \emph{invariance} property over covariates:
	\begin{align*}
	    \mu_{i,(z,\gvec)}(\c(X))=&\mathbb{E}[Y_{c,i}(z,\gvec) \mid \s(X) ,\pi_1(\Xb_{c,(i),1}), \cdots, \pi_m(\Xb_{c,(i),m})].
	\end{align*}
    In other words, the effect of neighbors' covariates on unit $i$'s potential outcomes is invariant under permutations of all the neighbors in $\mathcal{I}_j\backslash i$. It is then possible to model the effect of $\n(\Xb)$ on $\mu_{i,(z,\gvec)}(X_{c,i},\n(\Xb))$ using the summary statistics of covariates $\Xb_{c,(i),j}$ in each subset $j$, such as the mean or second moment. 
	
Under our conditional exchangeability framework for observational studies, we also define the \textbf{joint propensity model} as 
	\begin{align*}
	    p_{i,(z,\gvec)}(\c(X)) \coloneqq& P\big(\s(Z) = z, \Gvec_{c,i}=\gvec \mid  X_{c,i}, \Xb_{c,(i),1}, \cdots, \Xb_{c,(i),m}\big).
	\end{align*}
Analogous to the invariance property of the conditional outcome model under conditional exchangeability, we state a similar condition for the propensity model.
	
	\begin{assumption}[Conditional Exchangeability of Propensity Models]\label{ass:partial-prop-exchangeable} For arbitrary permutation $\pi_j(\cdot)$, the joint propensity model satisfies
	\begin{align*} 
			p_{i,(z,\gvec)}(\c(X)) =& P\big(\s(Z) = z, \Gvec_{c,i}=\gvec \mid  X_{c,i}, \pi_1(\Xb_{c,(i),1}), \cdots, \pi_m(\Xb_{c,(i),m})\big),
		\end{align*}
	where for each $j$, $\Xb_{c,(i),j}$  is the covariates of units in $\mathcal{I}_j\backslash i$.
		\end{assumption}
		
		Under Assumption \ref{ass:partial-prop-exchangeable}, a unit's treatment assignment probability can be impacted differently by the treatments and covariates of neighbors in \emph{distinct} subsets, but is invariant under arbitrary permutations of units within the same subset.  
		
	    Lastly, the overlap assumption in Assumption \ref{ass:overlap} can also be simplified to $ 0 < \underline{p} \leq	p_{i,(z,\gvec)}(\c(X))  \leq \overline{p}<1$ for any $\c(X),i,z$ and $\gvec$ under Assumptions \ref{ass:cond-outcome-partial-exchangeable} and \ref{ass:partial-prop-exchangeable}.  In practice, this overlap assumption can be harder to assess compared to the classical setting, due to the high dimensionality of treatment levels induced by interference. This problem is of independent interest and left for future works.
		
		        \section{Semiparametric Treatment Effect Estimation under Conditional Exchangeability} \label{sec:estimator}
In this section, we first lay out the heterogeneous causal estimands under conditional exchangeability in Section \ref{subsec:estimand}, and then propose our estimators in Section \ref{subsec:estimator}. 

\subsection{Estimands}\label{subsec:estimand}
		We focus on two types of interference-based estimands. The first one is the \textbf{average direct effect (ADE)} for units in subset $\mathcal{I}_j$, denoted as $\beta_j(\gvec)$, that measures the average direct treatment effect for units in $\mathcal{I}_j$, given that the number of treated neighbors in all subsets is $\gvec$:
		\begin{equation}
		    \beta_j(\gvec) \coloneqq \frac{1}{|\mathcal{I}_j|} \sum_{i \in \mathcal{I}_j} \mathbb{E}[Y_{c,i}(1,\gvec )-Y_{c,i}(0,\gvec)].\label{eqn:beta-g-short-definition} 
		\end{equation}
		The second one is the \textbf{average spillover effect (ASE)} for units in subset $\mathcal{I}_j$, denoted as $\tau_j(z,\gvec,\gvec^\prime)$, that measures the average difference in expected outcomes for units in $\mathcal{I}_j$ when the number of treated neighbors is $\Gvec_{c,i} = \gvec$ versus when $\Gvec_{c,i} = \gvec^\prime$, and when unit $i$'s own treatment is $Z_{c,i} = z$:
		\begin{equation}
		    \tau_j(z,\gvec,\gvec^\prime)  \coloneqq \frac{1}{|\mathcal{I}_j|} \sum_{i \in \mathcal{I}_j} \mathbb{E}[Y_{c,i}(z,\gvec )-Y_{c,i}(z, \gvec^\prime )]  . \label{eqn:tau-g-short-definition}
		\end{equation}
        ADE $\beta_j(\gvec)$ relates to ASE $\tau_j(z,\gvec,\gvec^\prime)$ through the equality\footnote{Both ADE and ASE are special cases of the more general class of estimands 
	\begin{align}
	\label{eq:general-estimand}
	\psi_j(z,\gvec) - \psi_j(z^\prime,\gvec^\prime) = \frac{1}{|\mathcal{I}_j|}\sum_{i\in\mathcal{I}_j} \mathbb{E}[Y_{c,i}(z,\gvec)-Y_{c,i}(z^\prime,\gvec^\prime)],
	\end{align} 
		for which our proposed estimators and their asymptotic properties can be directly generalized. We provide the asymptotic distribution of these generalized estimands in Appendix \ref{subsec:additioinal-asymptotic-results}.}
        \[\beta_j(\gvec) + \tau_j(0,\gvec,\gvec^\prime)  = \beta_j(\gvec^\prime)+\tau_j(1,\gvec,\gvec^\prime),\]
        and when $\gvec^\prime=\mathbf{0}$ and $\|\gvec\|_1=n-1$, the sum on the left can be interpreted as a ``total treatment effect'' for units in subset $j$ \citep{chin2018central,savje2021average}.
        
        In the household example (Figure \ref{fig:family}) with $m = 2$, if $\mathcal{I}_1$ and $\mathcal{I}_2$ are the sets of parents and children, respectively, then $\beta_{2}((g_1,g_2))$ measures the ADE for children, given $g_1$ treated parents and $g_2$ treated siblings. $\tau_{2}(0, (g_1, g_2), (0,g_2))$ measures the ASE from $g_1$ treated parents to children, given the children are untreated  and have $g_2$ treated siblings.

        Our definitions of ADE and ASE are analogous to the direct and spillover effects defined in \cite{tchetgen2012causal,liu2016inverse,barkley2020causal,park2020efficient,forastiere2020identification,vazquez2017identification}. A key distinction is that we define ADE and ASE as averages over units in a specific subset, while prior works primarily focus on marginal effects over all units.
        When there is a natural and interpretable partition of clusters, our framework allows for the identification of \emph{heterogeneous} ADE and ASE across different subsets or types of units.\footnote{These subsets were defined in Section \ref{subsec:exchangeability} to model heterogeneities in interference. Although here they serve the dual function of 
       characterizing heterogeneities of treatment effects across units, in general, the subsets of units that are exchangeable in Assumption \ref{ass:cond-outcome-partial-exchangeable} and that are used to define treatment effects do not have to depend on the same partition. For example, we can define and estimate treatment effects for a strict subset of $\mathcal{I}_j$ for any $j$.} This is important in many applications, e.g., when we want to design better treatment targeting rules based on the estimated treatment effects for different types of units from observational data. 

       \begin{remark}[Aggregate Estimands]
       \normalfont
      In some settings, we may also be interested in treatment effects that aggregate over \emph{multiple} $\beta_j(\gvec)$ or $\tau_j(z,\gvec,\gvec^{\prime})$. One class of such effects aggregates over  multiple types of units (i.e., over $j$), such as (whenever $\gvec$ is feasible for all $j\in \mathcal{J}$),
		\[\beta_{\mathcal{J}}(\mathbf{g}) := \frac{1}{| \cup_{j \in \mathcal{J}} \mathcal{I}_j|}\sum_{i \in \cup_{j \in \mathcal{J}} \mathcal{I}_j} \mathbb{E}[Y_{c,i}(1,\gvec )-Y_{c,i}(0,\gvec)],  \qquad \mathcal{J} \subset \{1,\cdots, m\}.\]
			For example, if $\mathcal{J} = \{1,\cdots, m\}$, then $\beta_{\mathcal{J}}(\mathbf{g}) $ is the average direct effect of all units in a cluster. When $m=1$, $\beta_{\mathcal{J}}(\mathbf{g}) $ further reduces to the ADE in previous works that rely on full exchangeability assumptions.  
			Alternatively, we may be interested in aggregating over different $\gvec$. For example, for weights $\omega(\gvec)$ satisfying $\omega(\gvec) \geq 0$ and $\sum_{\gvec \in \mathcal{G}}\omega(\gvec)=1$ for some collection $\mathcal{G}$ of $\gvec$, we can define the aggregate quantity
		    \[ \beta_j(\mathcal{G}) \coloneqq
		    \sum_{\gvec \in \mathcal{G}} \omega(\gvec) \cdot \beta_j(\gvec).\]
			When $\mathcal{G}$ is the set of all possible $\gvec$ for units of type $j$, this estimand can be viewed as a natural analogue of the classical ATE under interference. Finally, we may aggregate over both $j \in \mathcal{J}$ and $\gvec \in \mathcal{G}$. An important example is the direct effect defined under the $\alpha$-allocation strategy \citep{tchetgen2012causal}. This is the special case with $m=n$, $\mathcal{J}=\{1,\cdots, m\}$, $\mathcal{G}=\{0,1\}^{n-1}$, and $\omega(\gvec)=\alpha^{\|\gvec\|_1}\cdot (1-\alpha)^{n-1-\|\gvec\|_1}$.
			
   In this paper, we primarily focus on the estimation of  $\beta_j(\mathbf{g})$ (and $\tau_j(z,\gvec,\gvec^\prime) $), which are not only of interest themselves, but also serve as important building blocks in the \emph{unbiased} and \emph{efficient} estimation of aggregate causal effects. In particular, the mis-specification of interference structures can lead to biased estimators of $\beta_{\mathcal{J}}(\mathbf{g})$. The intuition is that if the heterogeneity is overlooked, both the propensity and outcome models may be consistently estimated, resulting in biased estimates of treatment effects. On the other hand, specifying the heterogeneity in a more complicated way than needed could lead to less efficient estimates of treatment effects. We discuss this  bias-efficiency trade-off in detail in Appendix \ref{subsection:efficiency-loss} and illustrate with a numerical example in Table
   \ref{tab:tradeoff}. An important implication of our analysis is that previous estimation strategies proposed for estimands based on the $\alpha$-allocation strategy may not always be efficient. 
			\end{remark}
			\subsection{Estimators}\label{subsec:estimator}
		Motivated by the AIPW estimators in the classical observational setting without interference \citep{robins1994estimation,robins1995analysis}, we propose the generalized AIPW estimators for $\beta_j(\gvec)$ and  $\tau_j(z,\gvec,\gvec^\prime)$:
		\begin{align}
		    \hat{\beta}^\aipw_j(\gvec ) =& \hat \psi_j(1,\gvec) - \hat \psi_j(0,\gvec) \label{eqn:aipw-beta-estimator}  \\
		    \hat{\tau}^\aipw_j(z, \gvec,\gvec^\prime ) =& \hat \psi_j(z,\gvec) - \hat \psi_j(z,\gvec^\prime), \label{eqn:aipw-tau-estimator}
		\end{align}
		where
\begin{align*}
    \hat \psi_j(z,\gvec) =&  \frac{1}{M |\mathcal{I}_j|} \sum_{c=1}^{M} \sum_{i \in \mathcal{I}_j} \hat{\phi}_{c,i}(z, \gvec)
\end{align*}
and $\hat{\phi}_{c,i}(z, \gvec)$ is the estimated score of unit $i$ in cluster $c$ and is defined as
		\begin{align}
		    \hat{\phi}_{c,i}(z, \gvec) =& \underbrace{\frac{\boldsymbol{1}\{\s(Z)=z,\Gvec_{c,i}=\gvec \}\s(Y) }{\hat p_{i,(z,\gvec)}(\c(X)) } }_{\mathrm{IPW} }  + \underbrace{ \bigg( 1 - \frac{\boldsymbol{1}\{\s(Z)=z,\Gvec_{c,i}=\gvec \}  }{\hat p_{i,(z,\gvec)}(\c(X)) } \bigg) \cdot \hat{\mu}_{i,(z,\gvec)}(\c(X)) }_{\mathrm{augmentation} } ,  \label{eqn:aipw-decompose}
		\end{align}
        $\hat p_{i,(z,\gvec)}(\c(X))$ and $\hat{\mu}_{i,(z,\gvec)}(\c(X))$ are unit $i$'s estimated joint propensity and conditional outcome models.
        
        Analogous to the AIPW in the classical setting \citep{robins1994estimation,robins1995analysis}, 
        both $\hat{\beta}^\aipw_j(\gvec)$  and $\hat{\tau}^\aipw_j(z, \gvec,\gvec^\prime )$ can be decomposed into two parts:
		the first part is the generalized IPW estimator, and the second part is an augmentation term that is a weighted average of conditional outcomes. As a result, the doubly robust property of the classical AIPW estimator carries over to $\hat{\beta}^\aipw_j(\gvec)$  and $\hat{\tau}^\aipw_j(z, \gvec,\gvec^\prime )$, as will be shown in Theorem \ref{thm:consistency} in Section \ref{section:asymptotics}. Double robustness means that the estimated average treatment effect is consistent if either the outcome model  or the propensity model can be consistently estimated (e.g.,  \cite{robins1994estimation,scharfstein1999adjusting,kang2007demystifying,tsiatis2007comment}).
		
		
		We first discuss the estimation of the conditional outcome models $\mu_{i,(z,\gvec)}(\*x)$ and joint propensity models $p_{i,(z,\gvec)}(\*x)$, for generic $\*x \in \mathbb{R}^{n\times d_x}$ and $i\in \{1,\cdots, n\}$, using nonparametric series estimators (a.k.a. sieve estimators) \citep{newey1997convergence,chen2007large,hirano2003efficient,cattaneo2010efficient}, on which our asymptotic results in Section \ref{section:asymptotics} are built. Sieve estimators are a sequence of estimators that progressively use more basis functions and more complex models to approximate $\mu_{i,(z,\gvec)}(\*x)$ and $p_{i,(z,\gvec)}(\*x)$. Let $\{r_k(\*x)\}_{k = 1}^\infty$ be such a sequence of known functions (e.g., polynomials).  In the sequence of estimators for $\mu_{i,(z,\gvec)}(\*x)$, let $\hat{\mu}_{i,K,(z,\gvec)}(\*x)$ be the estimator that uses the first $K$ approximation functions $R_K(\*x) = \big(r_1(\*x) ~ \cdots ~ r_K(\*x) \big)^\T$ and takes the form of 
		\[ \hat{\mu}_{i,K,(z,\gvec)}(\*x) = R_K(\*x)^\T \hat{\bm{\theta}}_{i,K,(z,\gvec)}, \]
		where 
		$\hat{\bm{\theta}}_{i,K,(z,\gvec)}$ is estimated from the ordinary least squares estimator, using the outcomes of the $i$-th units across all clusters $c$ that satisfy $Z_{c,i} = z$ and $\Gvec_{c,i} = \gvec$. See Internet Appendix IA.A  for the formula to obtain $\hat{ \bm{\theta}}_{i,K,(z,\gvec)}$.
		Intuitively, $\hat{\mu}_{i,K,(z,\gvec)}(\*x)$ better approximates $\mu_{i,(z,\gvec)}(\*x)$ as $K$ increases.

		Similarly, in the sequence of estimators for $p_{i,(z,\gvec)}(\*x)$, let $\hat{p}_{i,K,(z,\gvec)}(\*x)$ be the estimator that uses $K$ approximation functions, i.e., (a possibly different) $R_K(\*x)$, and satisfies
		\[\ln \frac{\hat{p}_{i,K,(z,\gvec)}(\*x)  }{\hat{p}_{i,K,(0,\bm{0})}(\*x) } = R_K(\*x)^\T \hat{\bm{\gamma}}_{i,K,(z,\gvec)},\]
		where $p_{i,(0,\bm{0})}(\*x)$ is chosen as the ``pivot'' for identification purposes, and $\hat{\bm{\gamma}}_{i,K,(z,\gvec)} $ maximizes the log-likelihood function using the treatment assignments of the $i$-th units across all clusters $c$ that satisfy $Z_{c,i} \in  \{z,0\}$ and $\Gvec_{c,i} \in \{\gvec,\bm{0}\}$. See Internet Appendix IA.A for the objective function for $\hat{\bm{\gamma}}_{i,K,(z,\gvec)}$.

		\begin{remark}[Alternative Estimators]
		\normalfont
		  We focus on nonparametric series estimators which require few functional form assumptions on the propensity and outcome models, and enjoy estimation consistency properties.
		    In practice, one could consider alternative parametric or nonparametric estimators, such as matching, kernel regression, and random forests, for the propensity and outcome models. It is possible to generalize our results in Section \ref{section:asymptotics} to some of these alterantive estimators, as long as the estimated conditional outcome and propensity models satisfy certain rate conditions. In Internet Appendix IA.A, we discuss some parametric simplifications of the estimation problem. In particular, when $n$ or $m$ are large so that there is a large number of pairs of $(z,\gvec)$, estimating a separate model $\hat p_{i,(z,\gvec)}(\c(X))$ and $\hat{\mu}_{i,(z,\gvec)}(\c(X))$  for each $(z,\gvec)$ can be infeasible. In this case, one may consider a universal propensity model $p(\c(X), z, \gvec)$ and conditional outcome model $\mu(\c(X), z, \gvec)$ for all $i,z,\gvec$.  
		\end{remark}

				\section{Main Asymptotic Results}\label{section:asymptotics}
		
		In this section, we show that our generalized AIPW estimators are doubly robust, asymptotically normal, and semiparametric efficient. For exposition, we present our results for ADE $\beta_j(\gvec)$. The results for ASE ${\tau}_j(z,\gvec, \gvec^\prime)$ and general causal estimands $\psi_j(z,\gvec) - \psi_j(z^\prime,\gvec^\prime)$ are conceptually the same, and are provided in Corollary \ref{thm:tauzg} and Theorem \ref{thm:normality-general} in Appendix \ref{subsec:additioinal-asymptotic-results}. We first show that, if either the propensity or the outcome model is estimated from the sieve estimator in Section \ref{subsec:estimator} and standard regularity conditions (Assumption \ref{ass:continuity-boundedness} in Appendix \ref{subsec:additioinal-asymptotic-results}) hold, then our AIPW estimators are consistent.
		
		\begin{theorem}[Consistency, ADE]\label{thm:consistency}
			Suppose Assumptions \ref{ass:network}-\ref{ass:partial-prop-exchangeable} hold. 
			As $M \rightarrow \infty$, for any $z$ and $\gvec$, 
			if either the estimated joint propensity $\hat{p}_{i,(z,\gvec)}(\c(X))$ or the estimated outcome $\hat \mu_{i,(z,\gvec)}(\c(X))$ is uniformly consistent in $\c(X)$, then the AIPW estimators are consistent, i.e.,  
			\begin{align}
			\hat{\beta}^\aipw_j(\gvec) \xrightarrow{P} \beta_j(\gvec).\label{eqn:direct-effect-consistency}
			\end{align}
			
		In particular, if at least one of  $\hat p_{i,(z,\gvec)}(\c(X))$ and  $\hat \mu_{i,(z,\gvec)}(\c(X))$ is estimated from the sieve estimators in Section \ref{subsec:estimator} and the regularity conditions in Assumption \ref{ass:continuity-boundedness} in Appendix \ref{subsec:additioinal-asymptotic-results} hold, then Equation \eqref{eqn:direct-effect-consistency} holds.
			
		\end{theorem}
		
		The key challenge in showing Theorem \ref{thm:consistency} is that for any two units $i$ and $i^\prime$ in cluster $c$, $(X_{c,i}, Y_{c,i}, Z_{c,i})$ and $(X_{c,i^\prime}, Y_{c,i^\prime}, Z_{c,i^\prime})$ are correlated, and consequently the estimated scores of these two units, $\hat{\phi}_{c,i}(z, \gvec)$ and $\hat{\phi}_{c,i^\prime}(z, \gvec)$, used in the AIPW estimators are correlated. Therefore,  the independence assumption of units, which is commonly used to show the doubly robust property \citep{robins1994estimation}, is violated. However, from Assumption \ref{ass:network}, the correlation is limited to the units within a cluster, and for units $i$ and $i^\prime$ that are in two distinct clusters $c$ and $c^\prime$, $(X_{c,i}, Y_{c,i}, Z_{c,i})$ and $(X_{c^\prime,i^\prime}, Y_{c^\prime,i^\prime}, Z_{c^\prime,i^\prime})$ are independent. Using this property, we can show that the AIPW estimators are consistent even with correlated observations within a cluster, as long as the number of clusters $M$ grows to infinity.\footnote{Alternatively, we can show that AIPW estimators constructed from matching-based estimators $\hat{\mu}_{i,(z,\gvec)}(\c(X))$ of ${\mu}_{i,(z,\gvec)}(\c(X))$ and kernel regression estimators $\hat p_{i,(z,\gvec)}(\c(X))$ of $p_{i,(z,\gvec)}(\c(X))$ (see Section \ref{sec:variance-estimators} for details) are consistent, even though $\hat{\mu}_{i,(z,\gvec)}(\c(X))$ and $\hat p_{i,(z,\gvec)}(\c(X))$ are only (uniformly) asymptotically unbiased instead of consistent.}

        Next we derive the asymptotic distribution of $\hat \beta^\aipw_j(\gvec)$. Even though the correlation among units within a cluster does not affect consistency, it affects the asymptotic variance of $\hat \beta^\aipw_j(\gvec)$. In Theorem \ref{thm:normality}, we provide the semiparametric efficiency bound for $\beta_j(\gvec)$ in the case of correlated observations, and we show that $\hat \beta^\aipw_j(\gvec)$ is asymptotically normal and attains this efficiency bound.

		
        		\begin{theorem}[Asymptotic Normality and Semiparametric Efficiency, ADE]\label{thm:normality}
		Suppose Assumptions \ref{ass:network}-\ref{ass:continuity-boundedness} hold, 
		 then $\hat \beta^\aipw_j(\gvec)$ is asymptotically normal. If, in addition, for any $c$,
	\begin{equation}
		Y_{c,i}(\s(Z), \n(\Zb)) \perp Y_{c,i^\prime}(Z_{c,i^\prime}, \mathbf{Z}_{c,(i^\prime)}) \mid \c(Z), \c(X) \qquad  \forall i\neq i^\prime, \label{eqn:indep-error}
			\end{equation}
			then as $M \rightarrow \infty$, for any subset $j$ and neighbors' treatment $\gvec$, we have
			\begin{equation*}
			\sqrt{M}\big(\hat{\beta}^\aipw_j(\gvec) - {\beta}_j(\gvec) \big)  \overset{d}{\rightarrow}\mathcal{N}\big(0,V_{j,\gvec}\big),
			\end{equation*}
			where $V_{j,\gvec}$ is the semiparametric efficiency bound for $\beta_{j}(\gvec)$, and can be decomposed into
			\[V_{j,\gvec} =  V_{j,\gvec,\mathrm{var}} +  V_{j,\gvec,\mathrm{cov}}.\]
			The first term $V_{j,\gvec,\mathrm{var}}$ is analogous to the classical efficiency bound and is defined as
			\begin{align} \label{var_bound}
			V_{j,\gvec,\mathrm{var}} =& \frac{1}{|\mathcal{I}_j|^2}\sum_{i \in \mathcal{I}_j}\mathbb{E}\left[\frac{\sigma_{i,(1,\gvec)}^{2}(\c(X))}{p_{i,(1,\gvec)}(\c(X))}+\frac{\sigma_{i,(0,\gvec)}^{2}(\c(X))}{p_{i,(0,\gvec)}(\c(X))}+(\beta_{i,\gvec}(\c(X))-\beta_{i,\gvec})^{2}\right],	\end{align}
		and the second term $V_{j,\gvec,\mathrm{cov}}$ is unique to our problem that quantifies the effect of interference on estimation efficiency, and is defined as	\begin{align}
		V_{j,\gvec,\mathrm{cov}}=&\frac{1}{|\mathcal{I}_j|^2}\sum_{i, i^\prime \in \mathcal{I}_j, i \neq i^\prime }\mathbb{E}\left[(\beta_{i,\gvec}(\c(X))-\beta_{i,\gvec})(\beta_{i',\gvec}(\c(X))-\beta_{i',\gvec})\right], \label{eqn:var_bound_second_term}
			\end{align}
            where
			$\sigma_{i,(z,\gvec)}^2(\c(X)) = \Var\left[\s(Y)(z,\gvec)\mid \c(X)\right]$, $\mu_{i,(z ,\gvec)} = \+E[\s(Y)(z,\gvec)]$, $\beta_{i,\gvec}(\c(X)) = \mu_{i,(1 ,\gvec)}(\c(X)) - \mu_{i,(0 ,\gvec)}(\c(X)) $ and $\beta_{i,\gvec} = \mu_{i,(1,\gvec)} - \mu_{i,(0,\gvec)}$.
		\end{theorem}
		
		
	The convergence rate of $\beta^\aipw_j(\gvec)$ is $\sqrt{M}$ because there are $M$ independent clusters. As units within a cluster are dependent, we will show that a valid influence function should be defined at the \emph{cluster} level, rather than at the individual level as in the classical setting without interference \citep{hahn1998role}. Theorem \ref{thm:normality} states that $V_{j,\gvec}$ can be decomposed into two terms, $V_{j,\gvec,\mathrm{var}}$ and $V_{j,\gvec,\mathrm{cov}}$. The term $V_{j,\gvec,\mathrm{var}}$ scales with $1/|\mathcal{I}_j|$ and is equal to the average of
    \[ \+E\left[\left(\phi_{c,i}(1,\gvec) - \phi_{c,i}(0,\gvec) - \beta_{i, \gvec}\right)^2\right] \]
    over units $i \in\mathcal{I}_j$, and is analogous to the efficiency bound derived in \cite{hahn1998role} and \cite{hirano2003efficient} under SUTVA. Here $\phi_{c,i}(z,\gvec) $ is the score of unit $i$ in cluster $c$ and is the population version of $\hat{\phi}_{c,i}(z,\gvec) $ defined in Equation \eqref{eqn:aipw-decompose}.\footnote{$\phi_{c,i}(z,\gvec) $ equals to $\hat{\phi}_{c,i}(z,\gvec) $ with $\hat{p}_{i,(z,\gvec)}(\c(X))$ and $\hat{\mu}_{i,(z ,\gvec)}(\c(X))$ replaced by $p_{i,(z,\gvec)}(\c(X))$ and $\mu_{i,(z ,\gvec)}(\c(X))$.}

In contrast, the term $V_{j,\gvec,\mathrm{cov}}$ is unique to our problem and comes from the the interference between units. $V_{j,\gvec,\mathrm{cov}}$ scales with the average of 
		\[\sigma_{i,i^\prime}  \coloneqq \+E\left[\left(\phi_{c,i}(1,\gvec) - \phi_{c,i}(0,\gvec) - \beta_{i, \gvec}\right) \left(\phi_{c,i^\prime}(1,\gvec) - \phi_{c,i^\prime}(0,\gvec) - \beta_{i^\prime, \gvec}\right)\right]  \]
		over any two distinct units $i$ and $i^\prime$ in $\mathcal{I}_j$, where the above term has the interpretation of the ``covariance'' between $i$ and $i^\prime$. From the expression of $V_{j,\gvec,\mathrm{cov}}$ in Theorem \ref{thm:normality}, $\sigma_{i,i^\prime}$ is equal to the covariance between the direct effects of $i$ and $i^\prime$ conditional on $\c(X)$, i.e., $\sigma_{i,i^\prime} = \+E\left[(\beta_{i,\gvec}(\c(X))-\beta_{i,\gvec})(\beta_{i',\gvec}(\c(X))-\beta_{i',\gvec})\right].$

		If $\sigma_{i,i^\prime} = 0$ for all distinct $i$ and $i^\prime$, then $V_{j,\gvec,\mathrm{cov}} = 0$ and $V_{j,\gvec} = V_{j,\gvec,\mathrm{var}}$. Furthermore, if units within a cluster are i.i.d. 
		and we set $m = 1$, then $V_{j,\gvec}$ equals to 
		\[V_{j,\gvec} = V_{j,\gvec,\mathrm{var}} =  \frac{1}{n} \cdot \mathbb{E}\left[\frac{\sigma_{i,(1,\gvec)}^{2}(\c(X))}{p_{i,(1,\gvec)}(\c(X))}+\frac{\sigma_{i,(0,\gvec)}^{2}(\c(X))}{p_{i,(0,\gvec)}(\c(X))}+(\beta_{i,\gvec}(\c(X))-\beta_{i,\gvec})^{2}\right], \]
		which is identical to the efficiency bound in \cite{hahn1998role} divided by $n$. The factor $n$ adjusts the rate $\sqrt{N} = \sqrt{Mn}$ in  \cite{hahn1998role} to the rate $\sqrt{M}$ in Theorem \ref{thm:normality}. At the other extreme, $V_{j,\gvec,\mathrm{cov}}$ is maximized when the conditional direct effects for units in a cluster are perfectly correlated, i.e., $\beta_{i,\gvec}(\c(X)) = \beta_{i^\prime,\gvec}(\c(X))$, for all distinct $i$ and $ i^\prime$, but are not constant. In this case, the effective sample size is minimized at $M$, which is the least efficient case.
		 
		 \begin{remark}[Conditional Independence]
		 \label{remark:independent-error}
		 \normalfont In Theorem \ref{thm:normality}, the condition in \eqref{eqn:indep-error}  states that outcomes of any two units in a cluster are independent \emph{conditional} on $\c(Z)$ and $\c(X)$. This condition is similar to the assumption of independent error terms in linear models. Essentially, it requires that the available covariates capture enough information about units so that no unobserved variables can cause correlations in outcomes between different units. In some applications, we may be concerned that there are unobserved variables that invalidate Equation \eqref{eqn:indep-error}. In this case, Theorem \ref{thm:normality} can still hold, but with a more complicated form of $V_{j,\gvec,\mathrm{cov}}$ containing the covariance between the residuals $\s(Y) - \mu_{i,(z,\gvec)}(\*X_c)$ of two units. The propensity scores will then play a role in  $V_{j,\gvec,\mathrm{cov}}$.
		 \end{remark}

\begin{remark}[Heterogeneous Interference]
\normalfont
A main motivation of our work is the heterogeneity of interference, which is captured by the conditional exchangeability framework. In practice, an important consideration is how to specify the exchangeable subsets. Not surprisingly, a trade-off arises. A more granular partition of each cluster can capture more complicated heterogeneities and reduce bias, but could result in less efficient estimators:
  \begin{center}
  \small
            \begin{tabular}{ll}
        \multirow{5}*{\rotatebox[origin=c]{270}{{\large  $\xRightarrow[\text{Efficiency Loss}]{\text{Reduced Bias}}$} }} & No Interference \\
        & Partial Interference with Full Exchangeability  \\ & Partial Interference with Conditional Exchangeability \\ & Network Interference with Conditional Exchangeability \\ & General Interference
        \end{tabular}
        \end{center}
    We formalize this trade-off in Section \ref{subsection:efficiency-loss} and illustrate with a numerical example in Table \ref{tab:tradeoff}. Our characterization of the asymptotic variance of proposed estimators paves the way for data-driven selection of the appropriate interference structure by leveraging statistical tests on the heterogeneity of interference. 
   We discuss these ideas in detail in Appendix \ref{sec:testing}. As  statistical tests require the use of feasible and valid variance estimators, we also propose a matching-based variance estimator in Appendix \ref{sec:variance-estimators} that is consistent and performs well in simulation studies. Together, these results allow practitioners to assess the impacts of heterogeneous interference in a wide array of applications, from identifying effective targets of candidate policies to constructing interference-robust treatment effect estimators.
\end{remark}

  So far, we have assumed that all clusters have the same size $n$. However, in many applications, such as those with family or classroom as a cluster, clusters may have different sizes. In Appendix \ref{sec:varying-cluster-size}, we extend our framework and results to the setting with varying cluster sizes under a mixture model.

        		\section{Simulation Studies} \label{section:simulations}
		In this section, we demonstrate the finite sample properties and practical relevance of hypothesis testing based on our asymptotic results, and show that our AIPW estimators are robust to model mis-specifications.\footnote{Code for implementations of our estimators is available at https://github.com/freshtaste/CausalModel as part of an actively maintained package.} 
		
		We start by introducing the data generating process for the simulated data used in this section. We generate $M = 5,000$ clusters of size $n = 4$, i.e., $N=20,000$ units in total. Each cluster has two exchangeable subsets, $\mathcal{I}_1$ and $\mathcal{I}_2$, and each subset consists of 2 units. We generate covariates from $\s(X) \stackrel{\mathrm{i.i.d.}}{\sim} \mathcal{N}(0, 0.2)$ for all $c$ and $i$. The treatment variable $Z_{c,i}$ is randomly and independently sampled from a Bernoulli distribution:
		\[ P(Z_{c,i}=1|\c(X)) = \frac{1}{1 + \exp(-0.5 \s(X) - 0.5/m \cdot \sum_{j=1}^m \bar{X}_{c,j}+1) } \qquad \text{ for all $c$ and $i \in \mathcal{I}_j$,}  \]
		where $\bar{X}_{c,j} = \frac{1}{|\mathcal{I}_j|}  \sum_{i \in \mathcal{I}_j} X_{c,i}$ is the average covariate of units in subset $\mathcal{I}_j$ in cluster $c$. 

		We generate the outcomes from the following model:
		\begin{align}\label{eqn:sim-outcome}
		    Y_{c,i}=\omega \cdot Z_{c,i}+ \left(f(\Gvec_{c,i}) + \s(X) + \bar{X}_{c,1} \right) \cdot Z_{c,i}+ \s(X) + \bar{X}_{c,1} +\varepsilon_{c,i} \qquad \text{ for all $c$ and $i$,} 
		\end{align}
		where 
		$\Gvec_{c,i} = \left(\sum_{i^\prime\in \mathcal{I}_1,i^\prime \neq i} Z_{c,i^\prime}, \sum_{i^\prime\in \mathcal{I}_2,i^\prime \neq i} Z_{c,i^\prime} \right)$ is the number of treated neighbors in $\mathcal{I}_1$ and $\mathcal{I}_2$, and $\varepsilon_{c,i}  \stackrel{\mathrm{i.i.d.}}{\sim} \mathcal{N}(0, 1)$ for all $c$ and $i$. In addition, $\omega \in \+R$ is a parameter that governs the level of direct treatment effect, and $f(\cdot,\cdot): \+Z^{2} \rightarrow \+R$ is an interference function that specifies how a unit's outcome is affected by its treated neighbors. We will consider various functional forms of $f(\cdot,\cdot)$. In this model, the direct and spillover effects of two subsets $\mathcal{I}_1$  and $\mathcal{I}_2$ are the same. Since $X_{c,i}$ has mean zero, the average direct effects equal to $\beta_1((g_1, g_2)) = \beta_2((g_1, g_2))  = \omega + f(g_1, g_2)$, and average spillover effects equal to $\tau_1(1, (g_1, g_2)) = \tau_2(1, (g_1, g_2))  = f(g_1, g_2)$ and $\tau_1(0, (g_1, g_2)) = \tau_2(0, (g_1, g_2))  = 0$.

		\subsection{Inference and Hypothesis Tests of Treatment Effects}\label{subsec:simulation-inference-test}
		
		We examine the finite sample properties of our treatment effect estimators, and the size and power of hypothesis tests for treatment effects. We present the results for $\beta_1\left((g_1, g_2)\right)$ to conserve space. The results for other estimands, e.g., $\beta_2\left((g_1, g_2)\right)$, $\tau_1\left(z,(g_1, g_2)\right)$ and $\tau_2\left(z,(g_1, g_2)\right)$, are similar. We consider three different interference mapping functions $f(\cdot, \cdot)$ in (\ref{eqn:sim-outcome}) for parameter $\gamma \in \+R$:
		\begin{enumerate}
		    \item (\textbf{HO}) $f(\mathbf{g}) = \gamma \cdot (g_1 + g_2)$.
		    \item (\textbf{HE1})
		    $f(\mathbf{g}) = \gamma \cdot (g_1 + g_2 + g_1 g_2)$.
		    \item (\textbf{HE2}) $f(\mathbf{g}) = \gamma \cdot (g_1 + 2 \cdot g_2)$.
		\end{enumerate}
		
		For \textbf{HO} (homogeneous interference), the direct and spillover effects do not vary with which neighbors are treated, as long as the total number of treated neighbors $g_1 + g_2$ is the same. In this case, the specification of interference structure with two exchangeable subsets $\mathcal{I}_1$ and $\mathcal{I}_2$ is in fact more granular than needed, since $m=1$ satisfies Assumption \ref{ass:cond-outcome-partial-exchangeable} as well. For both \textbf{HE1} and \textbf{HE2} (heterogeneous interference), the direct and spillover effects vary with which neighbors are treated. Specifically, the direct and spillover effects are different for $(g_1, g_2) = (0,2)$ and $(g_1, g_2) = (1,1)$ under \textbf{HE1} and \textbf{HE2}. In addition, the direct and spillover effects are also different for $(g_1, g_2) = (0,1)$ and $(g_1, g_2) = (1,0)$ under \textbf{HE2}. Under both \textbf{HE1} and \textbf{HE2}, the specification with the partition $\mathcal{I}_1$ and $\mathcal{I}_2$ is the most parsimonious one that satisfies Assumption \ref{ass:cond-outcome-partial-exchangeable}.
		
		We will consider tests with the following null hypotheses:

  \begin{enumerate*}
       \item $\mathcal{H}_0: \beta_1((0,0)) = 0$. 
		    \item $\mathcal{H}_0: \beta_1((0,1)) = 0$.
		    \item $\mathcal{H}_0: \beta_1((0,0))=\beta_1((0,1))$.
		    \item $\mathcal{H}_0: \beta_1((0,0))=\beta_1((0,2))$.
		    \item $\mathcal{H}_0: \beta_1((0,1))=\beta_1((1,0))$.
		    \item $\mathcal{H}_0: \beta_1((0,2)) = \beta_1((1,1))$.
		    \item $\mathcal{H}_0: \beta_1((0,1)) = \beta_1((1,0))$ \& $\beta_1((0,2)) = \beta_1((1,1))$.
\end{enumerate*}
  
		
	   For each hypothesis test, we first use our generalized AIPW estimators to estimate all the $\beta_1((g_1, g_2))$ parameters involved in the hypothesis test. For example, in the third hypothesis, we need to estimate both $\beta_1((0, 0))$ and $\beta_1((0, 1))$. In the generalized AIPW estimator, we estimate the individual propensity by logistic regression, neighborhood propensity by multinomial logistic regression, and the outcome by linear regression. We use the correct propensity and outcome model specifications. Using variance estimators based on Theorems \ref{thm:normality} and \ref{thm:consistent-variance}, we conduct the hypothesis test and report the rejection probability in Table \ref{tab:reject-prob} for various values of treatment effect parameters $\omega$ and $\gamma$ under $K = 2,000$ Monte Carlo trials.\footnote{Details on variance estimators and hypothesis testings are provided in Section \ref{sec:variance-estimators} and \ref{sec:testing}.}

        In Table \ref{tab:reject-prob}, we find that under the null hypotheses, the rejection probability is close to the nominal level of $\alpha = 0.05$. This finding is consistent across all cases where the null hypotheses hold. Specifically, the null hypotheses are true in the following cases: (1) If $\omega = \gamma = 0$, then $\beta_1((g_1, g_2)) = 0$ for all $g_1$ and $g_2$ in \textbf{HO}, \textbf{HE1}, and \textbf{HE2}. All the seven null hypotheses are true; (2) If $\omega = 1$ and $\gamma = 0$, then $\beta_1((g_1, g_2)) = 1$ for all $g_1$ and $g_2$ in \textbf{HO}, \textbf{HE1}, and \textbf{HE2}. The third to seventh null hypotheses are true; (3) If $\omega = \gamma = 1$, then the fifth to seventh null hypotheses are true for \textbf{HO}, and the fifth null hypothesis is true for \textbf{HE1}. Additionally, we verify the asymptotic normality in Theorem \ref{thm:normality} with the histograms provided in Internet Appendix IA.D.
		
		\begin{table}[t]
			\centering
			\tcaptab{Rejection probabilities of hypothesis tests}
			\label{tab:reject-prob}
			\begin{adjustbox}{max width=\linewidth,center}
			\begin{tabular}{c ccc c ccc c ccc}
				\toprule
				\centering
				& \multicolumn{3}{c}{$\omega = \gamma = 0$} & & \multicolumn{3}{c}{$\omega=1, \gamma=0$} & & \multicolumn{3}{c}{$\omega=\gamma=1$}\\ \cmidrule{2-4} \cmidrule{6-8} \cmidrule{10-12}
				$\mathcal{H}_0$ & \textbf{HO} &  \textbf{HE1} & \textbf{HE2} & &\textbf{HO} &  \textbf{HE1} & \textbf{HE2} & & \textbf{HO} &  \textbf{HE1} & \textbf{HE2} \tabularnewline
				\midrule
				1 & 0.0445 &  0.0365 & 0.0360 & & 0.9740 &  0.9690 & 0.9715 & & 0.9660 &  0.9550 & 0.9655 \\
				2 & 0.0470 &  0.0535 & 0.0500 & & 0.9620 &  0.9705 & 0.9705 & & 1.0000 &  1.0000 & 1.0000 \\ 
				3 & 0.0435 &  0.0440 & 0.0420 & & 0.0525 &  0.0490 & 0.0560 & & 0.7860 &  0.7210 & 1.0000 \\ 
				4 & 0.0455 &  0.0530 & 0.0530 & & 0.0520 &  0.0420 & 0.0555 & & 0.9995 &  0.9995 & 1.0000 \\ 
				5 & 0.0510 &  0.0415 & 0.0440 & & 0.0445 &  0.0450 & 0.0450 & & 0.0395 & 0.0330 & 0.7845 \\
				6 & 0.0530 &  0.0410 & 0.0435 & & 0.0460 &  0.0415 & 0.0425 & & 0.0520 & 0.7085 & 0.7570 \\
				7 & 0.0440 &  0.0455 & 0.0500 & & 0.0575 &  0.0395 & 0.0530 & & 0.0390 & 0.7460 & 0.9670 \\
				\bottomrule
				\addlinespace
			\end{tabular}
			\end{adjustbox}
			\bnotetab{Rejection probabilities are calculated using 5,000 clusters of size four and $K = 2,000$ Monte Carlo simulations. Significance level is set to be 0.05. }
		\end{table}

		\subsection{Robustness Properties of Our Estimators}
		
		In this section, we compare the performance of our AIPW estimators with  two common alternative estimators in estimating treatment effects:
		\begin{enumerate}
		    \item Ordinary least squares (OLS): Run the following linear regression
		    \[Y_{c,i} = \alpha + \theta_z \cdot Z_{c,i} + \bm{\theta}^\T_x  \cdot \*X_c +  \left( \theta_{zg1} \cdot G_{c,i,1} + \theta_{zg2} \cdot G_{c,i,2} + \bm{\theta}_{zx}^\T \cdot \*X_c \right)\cdot Z_{c,i} + \varepsilon_{c,i}, \]
		    where $\*X_c=\left(X_{c,i}, \bar{X}_{c,1}\right)$. Then estimate $\beta_1((g_1, g_2))$ by $\hat{\theta}_{zg1} \cdot g_1 + \hat{\theta}_{zg2} \cdot g_2$, where $\hat{\theta}_{zg1} $ and $\hat{\theta}_{zg2}$ are the estimated coefficients from the above regression.
		    \item Orthogonal Random Forest (ORF) for CATE: Suppose SUTVA holds and let $Y_{c,i}(z)$ be the potential outcomes. Treat $\Gvec_{c,i}$ as additional covariates and apply off-the-shelf forest based methods.\footnote{The CATE estimator is implemented in the python package \textsc{econml} by \cite{econml}. Among all the available CATE estimators, we adopt DROrthoForest, which is the AIPW-based CATE estimator \citep{chernozhukov2018double}.} Then use the fitted forest to estimate $\+E[Y_{c,i}(1) -  Y_{c,i}(0)\mid G_{c,i,1} = g_1, G_{c,i,2} = g_2]$, which can be viewed as an approximation of $\beta_1((g_1, g_2))$.
		\end{enumerate}
		
		We draw simulated data $K = 1,000$ times and estimate $\beta_1((g_1, g_2))$ on each simulated data set using three different estimators. We evaluate the performance of these estimators using two metrics: mean-squared error (MSE) and coverage rate. As shown in Table \ref{tab:coverage}, our AIPW estimators consistently exhibit much smaller MSEs compared to both OLS and ORF, suggesting that our AIPW estimators can most accurately estimate $\beta_1((g_1, g_2))$ for various interference functions $f(\gvec)$, even under misspecifications of the outcome model. Moreover, ORF has a much smaller MSE than OLS, indicating that in the presence of interference, tree-based methods, with their greater flexibility, tend to perform better than the more restrictive regression-based methods.
		
		Furthermore, as shown in Table \ref{tab:coverage}, our AIPW estimators achieve the correct coverage rate (i.e., close to 95\%) for various specifications of $f(\gvec)$, while OLS and ORF do not. OLS tends to have a lower coverage rate when $f(\gvec)$ is nonlinear (e.g., quadratic, reciprocal). This low coverage rate is likely due to its failure to obtain an accurate point estimate of the direct effects. On the other hand, ORF tends to have a higher coverage rate than the nominal rate. In fact, the coverage rate of ORF is very close to 100\%, implying that the estimated confidence interval from tree-based methods can be too wide, making hypothesis tests using tree-based methods overly conservative.

		Importantly, the outcome model in our AIPW estimators is misspecified when $f((g_1, g_2))$ is not linear in $g_1$ and $g_2$. However, even in such cases, our AIPW estimators can outperform OLS and ORF, thanks to the double robustness property of AIPW estimators. Therefore, we suggest that explicitly modeling the interference structure and using a doubly robust estimator can be crucial for accurate estimation and valid inference of treatment effects.

		\begin{table}[t]
			\centering
			\tcaptab{Coverage rate and MSE of various estimators for $\beta_1((g_1,g_2))$}
			\label{tab:coverage}
			{\footnotesize
			\begin{tabular}{c ccc c ccc}
				\toprule
				 & \multicolumn{3}{c}{Coverage Rate} & &  \multicolumn{3}{c}{MSE} \\ \cline{2-4} \cline{6-8}
				\centering
				$f(\gvec)$ & OLS & ORF & AIPW & & OLS & ORF & AIPW \tabularnewline
				\midrule
				$g_1 + 2 g_2$ & 94.52\% & 98.93\% & 95.60\% & & 0.0003 & 0.0027 & 0.0038\tabularnewline
				$\sqrt{g_1 + 2 g_2}$ & 14.37\% & 99.07\% & 95.08\% & & 0.0482 & 0.0027 & 0.0039 \tabularnewline
				$1/(g_1 + 2 g_2+1)$ & 15.35\% & 98.97\% &  94.88\% & & 0.0191 & 0.0027 & 0.0039 \tabularnewline
				$0.1 (g_1 + 2 g_2)^2 +  (g_1 + 2 g_2)$& 0.47\% & 99.18\% & 95.53\% & & 0.0627 & 0.0027 & 0.0038 \tabularnewline
				\bottomrule
				\addlinespace
			\end{tabular}
			}
			\bnotetab{We report the average coverage rate and MSE of $\hat{\beta}((g_1,g_2))$ over $(g_1, g_2) = (0,0), (0,1), (0,2), (1,0), (1,1)$ and $(1,2)$. We run $K=1,000$ Monte Carlo simulations for OLS, ORF, and AIPW.  }
		\end{table}

				\section{Two Applications to the Add Health Dataset}\label{section:applications}
		
		In this section, we demonstrate our methods through two empirical applications using the National Longitudinal Study of Adolescent to Adult Health (Add Health) dataset \citep{harris2009national}. The Add Health data has been frequently used in methodological and empirical studies on peer effects and interference because of its rich information on respondents' social and familial connections (e.g., \cite{bramoulle2009identification,goldsmith2013social,swisher2015paternal,forastiere2020identification}).
        
        In the first application, we investigate the effect of alcohol consumption on academic performance. We construct direct effect estimators under various specifications of the interference structure and find negative effects of regular alcohol consumption on students' academic performance. This finding is robust across different specifications of interference structures. However, the confidence intervals are wider under more complex interference structures due to finite sample efficiency loss. In the second application, we use our spillover effect estimators to study the impact of parental incarceration on adolescent well-being and find heterogeneous effects in the gender of the incarcerated parent.

        \subsection{Alcohol Consumption and Academic Performance}

        Prior studies have found negative associations between alcohol use and academic performance \citep{mcgrath1999academic,jeynes2002relationship,diego2003academic}. Other works have also studied peer effects of alcohol use among friends \citep{clark2007wasn,eisenberg2014peer}. In this paper, we adjust for the peer effects of alcohol use through the joint propensity model of alcohol use and estimate the effect of alcohol use on academic performance.

We use the Add Health data to construct a cohort of clusters, each with a small number of adolescents. Then we apply our average direct effect (ADE) estimators under various interference specifications to this cohort. We construct the cohort using adolescents' nominations of close friends. For each adolescent of our interest (which we refer to as the \textbf{centroid} of a cluster), we construct a cluster that consists of this adolescent, his/her best female friend, and his/her best male friend. Therefore each cluster has a size of three. We also perform sub-sampling to ensure minimal overlaps among clusters and obtain 7905 clusters in total. We define regular alcohol consumption as drinking at least once or twice a week. We measure academic performance by achieving a grade of B or better in mathematics, although the estimates are similar for other subjects. 
		
		We consider three interference specifications corresponding to the top three levels of interference structures in Section \ref{subsection:efficiency-loss}. The first specification assumes that there is no interference, and thus an individual's alcohol use and academic performance are not affected by his/her friends' alcohol use. The second specification assumes that there is homogeneous interference, so that an adolescent's alcohol use and academic performance are allowed to be affected interchangeably by their two best friends, regardless of the friends' genders. The third specification assumes interference is heterogeneous, so that a best friend's influence on an individual's alcohol use and academic performance is potentially gender-dependent.
		
		For each specification, we estimate the average direct effect of alcohol consumption on academic performance for centroids, using the corresponding AIPW estimator, where the propensity and outcome models are estimated under the corresponding interference structures. We adjust for covariates of both the centroids and their friends, including age, gender, frequency of skipping classes or missing school, and parents' educational backgrounds.

			\begin{table}[t!]
			\tcaptab{AIPW estimators of direct effects under different interference specifications}
			\centering
			\setlength{\tabcolsep}{6pt} 
			\renewcommand{\arraystretch}{1} 
			\begin{adjustbox}{max width=\linewidth,center}
				\begin{tabular}{l|c|c|c|c|c|c|c|c}
					\toprule
				 specification& \multicolumn{8}{c}{$\hat\beta$}   \tabularnewline
					\midrule
					no interference &  \multicolumn{8}{c}{-0.049***}
 					\\
 					 & \multicolumn{8}{c}{(0.010)}
 					\\
				\midrule
	& \multicolumn{4}{c|}{$\hat\beta_\M$} &  \multicolumn{4}{c}{$\hat\beta_\F$}  \tabularnewline
	\midrule
					no interference 
					& \multicolumn{4}{c|}{-0.054***} &  \multicolumn{4}{c}{-0.044***} 
 					\\
 					& \multicolumn{4}{c|}{(0.013)} &  \multicolumn{4}{c}{(0.015)}  
 					\\
 					\midrule \midrule
				& \multicolumn{2}{c|}{$\hat\beta(0)$}  & \multicolumn{4}{c|}{$\hat\beta(1)$} & \multicolumn{2}{c}{$\hat\beta(2)$} \tabularnewline
					\midrule
					hom. interference & \multicolumn{2}{c|}{-0.073***}  & \multicolumn{4}{c|}{-0.065***} & \multicolumn{2}{c}{-0.006}
 					\\
 					 & \multicolumn{2}{c|}{(0.013)}  & \multicolumn{4}{c|}{(0.020)} & \multicolumn{2}{c}{(0.068)}
 					\\
 					\midrule
				& $\hat\beta_\M(0)$ & $\hat\beta_\F(0)$ & \multicolumn{2}{c|}{$\hat\beta_\M(1)$} & \multicolumn{2}{c|}{$\hat\beta_\F(1)$} &  $\hat\beta_\M(2)$ & $\hat\beta_\F(2)$
                     \tabularnewline 
                     \midrule 
					hom. interference & -0.062*** & 	-0.084*** & \multicolumn{2}{c|}{-0.106***} & \multicolumn{2}{c|}{	-0.025} &  -0.002 & -0.010
 					\\
 					 & (0.018) & (0.020) & \multicolumn{2}{c|}{(0.027)} & \multicolumn{2}{c|}{(0.031)} &  (0.091) & (0.102)
 					\\
 					\midrule \midrule
				& $\hat\beta_\M(0,0)$ & $\hat\beta_\F(0,0)$ &
	            $\hat\beta_\M(1,0)$ &
	            $\hat\beta_\M(0,1)$ &
                     $\hat\beta_\F(1,0)$ &   $\hat\beta_\F(0,1)$  &
                     $\hat\beta_\M(1,1)$ &
                     $\hat\beta_\F(1,1)$
                     \tabularnewline 
                     \midrule
			het. interference & -0.062***& -0.084*** & -0.053& -0.144***& -0.102**&  0.105 & -0.002 & -0.005
 					\\
 					&(0.018) & (0.019) & (0.033) & (0.042)& (0.037)& (0.053) & (0.098) & (0.117) \tabularnewline
					\bottomrule
				\end{tabular}
			\end{adjustbox}
			\bnotetab{Under the specification of no interference, $\beta$ is the ATE for all centroids, and $\beta_\M$ is the ATE for male centroids, and similarly for $\beta_\F$. Under the specification of homogeneous interference, the direct effect is defined as $\beta(g)$, where $g \in \{0,1,2\}$ denotes the number of treated friends. 
			Under the specification of heterogeneous interference, the direct effects are defined as $\beta_\M(z_{\MF},z_{\FF})$ for male centroids and $\beta_\F(z_{\MF},z_{\FF})$ for female centroids, where $z_{\MF}\in \{0,1\}$ denotes the treatment status of male friend and $ z_{\FF} \in \{0,1\}$ denotes the treatment status of female friend. For each specification, covariates are adjusted in the propensity and outcome models.
			}
			\label{tab:empirical}
		\end{table}

		In Table \ref{tab:empirical}, we report the estimated direct effects under different interference structures. There are three main findings. First, all estimated treatment effects are negative, implying that regular alcohol use could have a negative effect on academic performance, which is robust to the particular specification of interference structures. Second, the magnitude of direct effects varies with the gender of centroids and their numbers of treated friends. The impact is diminished with an increase in the number of friends who drink regularly. Furthermore, direct effects are also heterogeneous in the \emph{gender} of treated friends, suggesting that interference could be heterogeneous. Third, standard errors increase with the complexity of the interference structure. There are two potential contributing factors. First, our bias-variance tradeoff analysis in Section \ref{subsection:efficiency-loss} suggests that estimators based on more complex specifications of the interference structure tend to have larger \emph{asymptotic} variances. Second, fewer samples are available when estimating each heterogeneous effect, highlighting the cost of a complex specification of interference in practice.
			
		\subsection{Paternal Incarceration and Adolescent Well-Being}

  The impact of parental incarceration on children's health, education, and economic outcomes is an important topic that has generated much attention in empirical works \citep{lee2013impact,miller2015association,wildeman2018parental,austin2022parental,jones2022mental}. Following the empirical study of \cite{swisher2015paternal}, we apply the spillover effect estimators from  \eqref{eqn:aipw-tau-estimator} to examine the impact of paternal incarceration on children's well-being, specifically delinquency and depression (both binary outcomes). In this context, families naturally form independent clusters, with heterogeneous groups within each family comprising of the father, mother, and children. We consider three spillover effects of parental incarceration prior to measuring the outcome: only the mother incarcerated, only the father incarcerated, and both parents incarcerated ($\hat\tau(1,0)$, $\hat\tau(0,1)$, and $\hat\tau(1,1)$). The set of control variables includes age, gender, ethnicity, physical abuse, and sexual abuse.\footnote{Outcomes and covariates are obtained from Wave I while treatments are obtained from Wave IV questionnaires.} The sample size is 4692. We compare our estimators to an OLS regression similar to the original study by regressing outcomes on the three incarceration indicators and covariates. In Table \ref{tab:empirical2}, we present the OLS and AIPW estimators for the spillover effects. Both estimators yield similar point estimates for delinquency across various spillover exposures. However, the standard errors associated with our AIPW estimators are consistently smaller than those of the OLS estimators, demonstrating the efficiency gain achieved with our method. In addition, our estimators reveal a larger effect on depression when the father is incarcerated compared to the OLS estimators.
        
        \begin{table}[h]
        \centering
        \tcaptab{OLS and AIPW estimators of spillover effects}
        \label{tab:empirical2}
        
        \begin{tabular}{lccccccc}
        \toprule
				\centering
                          & \multicolumn{3}{c}{delinquency}                         &               & \multicolumn{3}{c}{depression}                      \\
                         \cline{2-4} \cline{6-8}
                          & $\hat\tau(1,0)$ & $\hat\tau(0,1)$ & \multicolumn{2}{l}{$\hat\tau(1,1)$} & $\hat\tau(1,0)$ & $\hat\tau(0,1)$ & $\hat\tau(1,1)$ \\ \midrule
        OLS & 0.1892***          & 0.1099***          & 0.2142***              &               & -0.0013         & 0.0363          & 0.1008          \\
                          & (0.059)           & (0.021)           & (0.062)               &               & (0.056)           & (0.021)           & (0.060)            \\ \\
        AIPW              & 0.1854***          & 0.1177***          & 0.1993***              &               & 0.0077          & 0.0424***          & 0.1728***          \\
                          & (0.037)           & (0.013)           & (0.040)                &               & (0.037)           & (0.012)           & (0.039)        \\
        \bottomrule
        \end{tabular}
        \end{table}

        		\section{Concluding Remarks}\label{section:conclusions}
    
        In this paper, we propose to explicitly model heterogeneous interference in the observational setting relevant in many empirical works through a conditional exchangeability framework. While this framework is an instance of the more general exposure mapping framework, it is applicable to many applications where heterogeneities in interference and treatment effects are determined by observable characteristics.  We construct doubly robust and semiparametric efficient AIPW estimators for granular average direct and spillover effects based on the conditional exchangeability framework. Our asymptotic results provide off-the-shelf estimation and inference methods for various types of causal estimands relevant in practice, such as optimal policy targeting. We also propose a data-driven method based on hypothesis testing that allows practitioners to detect and account for heterogeneities in interference without relying heavily on domain knowledge. We demonstrate the validity and practical appeal of our estimators through extensive simulation studies and two relevant applications to the Add Health dataset. Lastly, our work is also relevant for researchers interested in estimating aggregate treatment effects, such as analogs of classical ATEs in the presence of potential interference. By aggregating our AIPW estimators for granular effects, one can construct interference-robust estimators at the cost of some efficiency loss.

	\end{doublespacing}

  \onehalfspacing
    {\small 
    \section*{Acknowledgements}
		We are very grateful for many helpful comments and suggestions from Eric Auerbach, Jianfei Cao, Park Chan, Juan Estrada, Laura Forastiere, Hong Han, Yuchen Hu, Hyungseok Kang, Yongchan Kwon, Michael Leung,  Haodong Li, Fabrizia Mealli, Markus Pelger, and Michael Pollmann, Evan Rose, Brad Ross, Martin Rotemberg, Rose Tan, Merrill Warnick, Jason Weitze, Keli Xu, Johan Ugander, and participants of the Online Causal Inference Seminar, NASMES 2022, California Econometrics Conference, the Stanford econometrics lunch seminar, and the LinkedIn Tech Talk series. We would like to thank Joshua Quan and Christopher Fraga at the Stanford Institute for Research in the Social Sciences for their assistance in accessing the Add Health dataset. 
		
		This research uses data from Add Health, a program project directed by Kathleen Mullan Harris and designed by J. Richard Udry, Peter S. Bearman, and Kathleen Mullan Harris at the University of North Carolina at Chapel Hill, and funded by grant P01-HD31921 from the Eunice Kennedy Shriver National Institute of Child Health and Human Development, with cooperative funding from 23 other federal agencies and foundations. Special acknowledgment is due Ronald R. Rindfuss and Barbara Entwisle for assistance in the original design. Information on how to obtain the Add Health data files is available on the Add Health website. No direct support was received from grant P01-HD31921 for this analysis.
    }
    \newpage
	\bibliographystyle{apalike}
 {\small
\bibliography{interference_reference}
 }
	
	\newpage	
 \appendix
\renewcommand{\thesubsection}{\Alph{section}.\arabic{subsection}}
\setcounter{table}{0}
\setcounter{figure}{0}
\renewcommand{\thetable}{A.\arabic{table}}
\renewcommand{\thefigure}{A.\arabic{figure}}
 
    {\small
    \onehalfspacing
    \bigskip
\begin{center}
{\large\bf SUPPLEMENTARY MATERIAL}
\end{center}  
    \section{Additional Results}\label{sec:additional-result}
    	\subsection{Lemma for Conditional Exchangeability}\label{subsec:conditional-exch}
	
		\begin{lemma}\label{lemma:conditional-outcome-exchangeable}
			Suppose Assumption \ref{ass:cond-outcome-partial-exchangeable} holds. If 
			for any $c$, $i$, and any permutation $\pi_j(\cdot) \in \mathbb{S}^{|\mathcal{I}_j\backslash i|}$ for all $j$, $\zb\in \{0,1\}^n$ and $\xb\in\mathbb{R}^{n\times d_x}$
		    \begin{equation*}
		\begin{aligned}
		&    \mathbb{E}[Y_{c,i}(z,\zb_{c,(i)}) \mid \c(X)] \\
      =&\mathbb{E}[Y_{c,i}(z_{c,i}, \pi_1(\zb_{c,(i),1}), \cdots, \pi_m(\zb_{c,(i),m})) \mid \s(X) ,\pi_1(\Xb_{c,(i),1}), \cdots,\pi_m(\Xb_{c,(i),m})],
		\end{aligned}
			\end{equation*}
			then the following holds:
			\begin{equation*}
			\label{eq:cond-exchangeable}
			\begin{aligned}
		&	    \mu_{i,(z,\gvec)}(X_{c,i},\n(\Xb)):=\mathbb{E}[Y_{c,i}(z,g) \mid \s(X),\n(\Xb)] \\
       =&\mathbb{E}[Y_{c,i}(z,\gvec) \mid \s(X) ,\pi_1(\Xb_{c,(i),1}), \cdots,\pi_m(\Xb_{c,(i),m}))].
			\end{aligned}
			\end{equation*}
		\end{lemma}
 
 \subsection{Generalized AIPW Estimator}\label{subsec:additioinal-asymptotic-results}
	We state the omitted result on oracle AIPW estimators in Proposition \ref{prop:oracle-consistency} below, as well as additional asymptotic results for spillover effect in Corollary \ref{thm:tauzg} and more general causal estimands in Theorem \ref{thm:normality-general} below. 
	Corollary \ref{thm:tauzg} is analogous to Theorem \ref{thm:normality} and states that
		$\hat{\tau}^\aipw(z,\gvec)$ is asymptotically normal and semiparametrically efficient, and provides the expression for the asymptotic variance. Theorem \ref{thm:normality-general} is the most general result that implies Theorem \ref{thm:normality} and Corollary \ref{thm:tauzg}, and we will provide its proof in the Internet Appendix.
		
		\begin{proposition}[Oracle AIPW Estimator]
			\label{prop:oracle-consistency}
			Suppose Assumptions \ref{ass:network}-\ref{ass:overlap} hold. For all $j \in \{1,\cdots,m\}$, the oracle AIPW estimator ${\beta}^\aipw_j(\gvec)$ is unbiased and consistent for ${\beta}_j(\gvec)$, where ${\beta}^\aipw_j(\gvec)$ replaces $\hat{p}_{i,(z,\gvec)}(\c(X))$ with ${p}_{i,(z,\gvec)}(\c(X))$ and $\hat{\mu}_{i,(z,\gvec)}(\c(X))$ with ${\mu}_{i,(z,\gvec)}(\c(X))$ in $\hat {\beta}^\aipw_j(\gvec)$.
		\end{proposition}

  We impose the following continuity assumption on propensity and conditional outcome functions and boundedness assumption on covariates and outcome. Under this assumption,  we can show $\hat {\bm{\theta}}_{i,K}$  $\hat{\bm{\gamma}}_{i,K}$ is consistent ollowing similar arguments as \cite{cattaneo2010efficient}, which is an important intermediate step to show the asymptotic distribution of our estimators.
		
		\begin{assumption}[Continuity and Boundedness]\label{ass:continuity-boundedness}
\texttt{} 
			\begin{enumerate}
				\item For all $(z,\gvec)$, $p_{i,(z,\gvec)}(\cdot) $ and $\mu_{i,(z,\gvec)}(\cdot)$ are $s$ times differentiable with $s/d_x > 5 \eta/2 + 1/2$ where where $\eta = 1$ or $\eta = 1/2$ depending on whether power series or splines are used as basis functions.
				\item $\c(X)$ is continuously distributed with density bounded  and bounded away from zero on its compact support $\mathcal{X}$, and $|\s(Y)(z,\gvec)| < \infty$.
			\end{enumerate}
		\end{assumption}

	\begin{theorem}[Asymptotic Normality, General Estimand]\label{thm:normality-general}
			Suppose the assumptions in Theorem \ref{thm:normality} hold.
			As $M \rightarrow \infty$, for any subset $j$ and treatment assignments $(z,\gvec)$ and $(z^\prime,\gvec^\prime)$,  we have 
			\begin{align}
			\sqrt{M}\big((\hat{\psi}^\aipw_j(z,\gvec) - \hat{\psi}^\aipw_j(z^\prime,\gvec^\prime))-({\psi}_j(z,\gvec) - {\psi}_j(z^\prime,\gvec^\prime)) \big) & \overset{d}{\rightarrow}\mathcal{N}\big(0,V_{j,z,z^\prime,\gvec,\gvec^\prime}\big) \label{eqn:aipw-betag-asy-normal-general}
			\end{align}
			where $V_{j,z,z^\prime,\gvec,\gvec^\prime}$ equals
			\begin{align} 
			V_{j,z,z^\prime,\gvec,\gvec^\prime} =& \frac{1}{|\mathcal{I}_j|^2}\sum_{i \in \mathcal{I}_j}\mathbb{E}\left[\frac{\sigma_{i,(z,\gvec)}^{2}(\c(X))}{p_{i,(z,\gvec)}(\c(X))}+\frac{\sigma_{i,(z^\prime,\gvec^\prime)}^{2}(\c(X))}{p_{i,(z^\prime,\gvec^\prime)}(\c(X))}+\big(\mu_{i,z,z^\prime,\gvec,\gvec^\prime}(\c(X))-\mu_{i,z,z^\prime,\gvec,\gvec^\prime} \big)^{2}\right]\\
			&\quad +\frac{1}{|\mathcal{I}_j|^2}\sum_{i, i^\prime \in \mathcal{I}_j, i \neq i^\prime } \mathbb{E}\big[\big(\mu_{i,z,z^\prime,\gvec,\gvec^\prime}(\c(X))-\mu_{i,z,z^\prime,\gvec,\gvec^\prime} \big) \big(\mu_{i^\prime,z,z^\prime,\gvec,\gvec^\prime}(\c(X))-\mu_{i^\prime,z,z^\prime,\gvec,\gvec^\prime} \big)\big] 
			\label{eqn:general-var-bound}
			\end{align}
			where $\sigma_{i,(z,\gvec)}^2(\c(X)) = \Var\left[\s(Y)(z,\gvec)|\c(X)\right]$, $\mu_{i,(z ,\gvec)} = \+E[\s(Y)(z,\gvec)]$, $\mu_{i,z,z^\prime,\gvec,\gvec^\prime}(\c(X)) = \mu_{i,(z ,\gvec)}(\c(X)) - \mu_{i,(z^\prime ,\gvec^\prime)}(\c(X)) $ and $\mu_{i,z,z^\prime,\gvec,\gvec^\prime} = \mu_{i,(z ,\gvec)} - \mu_{i,(z^\prime ,\gvec^\prime)}$.
		\end{theorem}
	When $z = 1, z^\prime = 0$ and $ \gvec^\prime = \gvec$, Theorem \ref{thm:normality-general} reduces to Theorem \ref{thm:normality}. When $z^\prime = z$ and $\gvec^\prime = 0$,  Theorem \ref{thm:normality-general} reduces to the following result for spillover effects $\tau(z,\gvec)$. 
	
		\begin{corollary}[Asymptotic Normality, ASE]\label{thm:tauzg}
		Suppose the assumptions in Theorem \ref{thm:consistency} hold.
		As $M \rightarrow \infty$, for any $\gvec$,  we have  
		$\hat{\tau}^\aipw(z,\gvec) \xrightarrow{P} \tau(z,\gvec)$ if either $\hat{p}_{i,(z,\gvec)}(\c(X))$ or $\hat{\mu}_{i,(z,\gvec)}(\c(X))$ is uniformly consistent, and
		\begin{align}
		\sqrt{M}(\hat{\tau}^\aipw(z,\gvec)-\tau(z,\gvec)) & \overset{d}{\rightarrow}\mathcal{N}(0,V_{j,z,\gvec}),
		\end{align}
		where $V_{j,z,\gvec}$ is the asymptotic variance bound for $\beta(\gvec)$ and equals
		\begin{align} 
		V_{j,z,\gvec} =& \frac{1}{|\mathcal{I}_j|^2}\sum_{i \in \mathcal{I}_j}\mathbb{E}\left[\frac{\sigma_{i,z,\gvec}^{2}(\c(X))}{p_{i,(z,\gvec)}(\c(X))}+\frac{\sigma_{i,z,0}^{2}(\c(X))}{p_{i,(z,0)}(\c(X))}+(\tau_{i,z,\gvec}(\c(X))-\tau_{i,z,\gvec})^{2}\right] \label{tauzg_var_bound}\\
		&+\frac{1}{n^{2}}\sum_{i\neq j}\mathbb{E}\left[(\tau_{i,z,\gvec}(\c(X))-\tau_{i,z,\gvec})(\tau_{j,z,\gvec}(\c(X))-\tau_{j,z,\gvec})\right] \label{eqn:tauzg_var_bound_second_term}
		\end{align}
		where $p_{i,\gvec}(\c(X)) = P(\s(\mathbf{G}) = \gvec|\c(X)) $, $q_{i,z}(\c(X)) = P(\s(Z) = z|\c(X)) $, $\sigma_{i,z,\gvec}^2(\c(X)) = Var\left[\s(Y)(z,\gvec)|\c(X)\right]$, $\tau_{i,z,\gvec}(\c(X)) = \+E\left[\s(Y)(z,\gvec) - \s(Y)(z,0)|\c(X)\right]$ and $\tau_{i,z,\gvec} = \+E\left[\s(Y)(z,\gvec) - \s(Y)(z,0) \right]$. 
	\end{corollary}

 		\subsection{Varying Cluster Sizes}
		\label{sec:varying-cluster-size}
  In the main text, we primarily focus on the case where all clusters have the same size $n$. This is primarily for the exposition purpose. In this section, we show how our estimands and results can be generalized to allow for varying cluster sizes.  

		We start with the sampling scheme with varying cluster sizes. For each cluster $c$, its size $n_c$ is drawn from a fixed finite set\footnote{It is possible to have $\mathcal{S} = \mathbb{Z}_{+}$, as long as $p_n$ decays sufficiently fast with $n$.} $\mathcal{S} \subset  \mathbb{Z}_{+}$ with $|\mathcal{S}|=\Bar{n}$ following the distribution $p_n \coloneqq P(n_c = n) \in (0,1)$ for all $n \in\mathcal{S}$. Conditional on the cluster size $n_c = n$, $\left(\c(Y), \c(X), \c(Z)\right)$ are sampled from a joint population $\mathbb{P}_{n}$. Suppose the \emph{types} of exchangeable units are consistent across clusters.\footnote{Formally speaking, there is a ($\mathbb{P}_n$-dependent) partition $\mathcal{I}_{n,1}, \mathcal{I}_{n,2}, \cdots, \mathcal{I}_{n,m}$ of $\{1, 2, \cdots, n\}$ into $m$ disjoint and exchangeable subsets for clusters drawn from $\mathbb{P}_n$, where $\sum_{j = 1}^m |\mathcal{I}_{n,j}| = n$. Note that we allow the partition to depend on $\mathbb{P}_n$, but assume $m$ is universal across all $\mathbb{P}_n$.} For example, the types could be parents and children in families with different sizes, or male and female students in classrooms with different sizes.\footnote{ As there are clusters that may not have a particular type of unit (e.g., families with no children), we allow some $\mathcal{I}_{n,j}$ to be empty, and as before $\mathcal{I}_{n,j}$ can also be singleton. 
		}

  \begin{figure}[t]
			\centering
			\begin{subfigure}{.5\textwidth}
			\centering
				\includegraphics[width=0.6\textwidth]{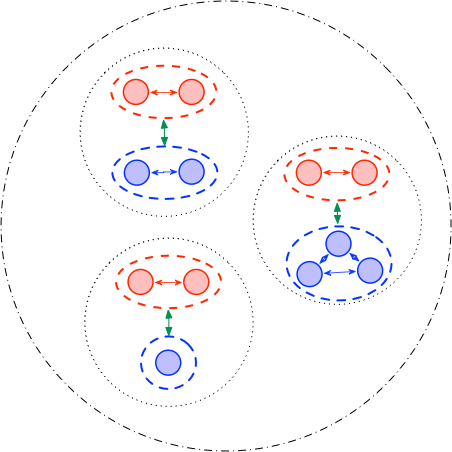} 
				\subcaption{Varying cluster sizes}
				\label{subfig:varying-cluster-size}
			\end{subfigure}\hfill
			\begin{subfigure}{.5\textwidth}
			\centering
				\includegraphics[width=0.6\textwidth]{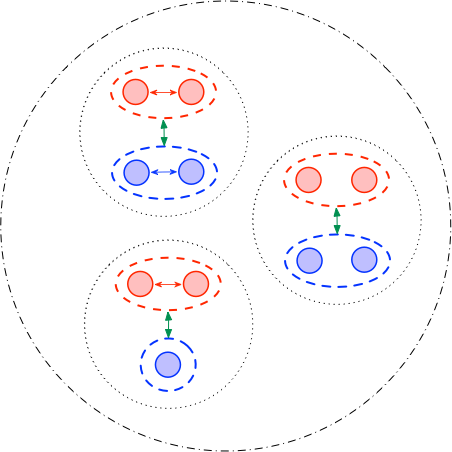} 
				\subcaption{Arbitrary cluster structures}
				\label{subfig:arbitrary}
			\end{subfigure}
			\caption{Illustration of the extension with varying cluster sizes (Figure \ref{subfig:varying-cluster-size}, see Section \ref{sec:varying-cluster-size}) and the extension with arbitrary cluster structures (Figure \ref{subfig:arbitrary}, see Section \ref{subsection:general-cluster-structure}). Now each family consists of heterosexual parents and a variable number of children. For Figure \ref{subfig:arbitrary}, family members may not be fully connected.}
			\label{varying-size}
		\end{figure}

		When the cluster size varies, the direct effect can be defined as 
        \begin{equation}
		\beta_j(\gvec)  = \sum_{n \in \mathcal{S}} \omega_{n,j}  \cdot \beta_{n,j}(\gvec), \label{eqn:beta-varying-n}
		\end{equation}
		where $\beta_{n,j}(\gvec) $ is the direct effect for units in subset $\mathcal{I}_j$ and in clusters with size $n$, and
		the weight $\omega_{n,j}$ is proportional to the fraction of clusters with size $n$ and is defined as 
		\[\omega_{n,j} = \frac{p_{n} \boldsymbol{1}\{\gvec \leq \gvec_{n,j,\max}\}}{\sum_{n^\prime \in \mathcal{S}} p_{n^\prime} \boldsymbol{1}\{\gvec \leq \gvec_{n^\prime,j,\max}\}}.\]
		Here $\gvec_{n,j,\max} \in \mathbb{R}^m$ denotes the the maximum treated neighbors a unit in $\mathcal{I}_j$ could have in clusters with size $n$.\footnote{ $\gvec_{n,j,\max} \in \mathbb{R}^m$ has its $k$-th entry $g_{n,j,\max,k} = |\mathcal{I}_{n,j}|$ if $k \neq j$ and otherwise $g_{n,j,\max,k} = |\mathcal{I}_{n,j}| - 1$, and $|\mathcal{I}_{n,j}|$ is the cardinality of $j$-th subset with cluster size $n$.} The definition of $\omega_{n,j}$ accounts for the cases where $\gvec$ is larger than $\gvec_{n^\prime,j,\max}$ in some coordinate(s) for some $n^\prime$. In such cases, $\beta_{n,j}(\gvec)$ cannot be identified and $\omega_{n,j} = 0$. Besides direct effects, we can also consider the following more general estimands
        \begin{equation*}
		\psi_j(z,\gvec) - \psi_j(z^\prime,\gvec^\prime) = \sum_{n \in \mathcal{S}} \omega_{n,j}   \big(\psi_{n,j}(z,\gvec) - \psi_{n,j}(z^\prime,\gvec^\prime) \big),
		\end{equation*}
		where 
		\[\omega_{n,j} = \frac{p_{n} \boldsymbol{1}\{\gvec,\gvec' \leq \gvec_{n,j,\max}\}}{\sum_{n^\prime \in \mathcal{S}} p_{n^\prime} \boldsymbol{1}\{\gvec,\gvec' \leq \gvec_{n^\prime,j,\max}\}} \]
		and $\gvec_{n,j,\max} \in \mathbb{R}^m$ denotes the the maximum treated neighbors a unit in $\mathcal{I}_j$ could have in clusters with size $n$.
  Our use of a mixture model and our definition of direct effects and general estimands are conceptually similar to \cite{park2020efficient} that study the treatment effect estimation with varying cluster sizes under the $\alpha$-allocation strategy. 

  To estimate the treatment effects with varying cluster size, we can use a three-step approach that is a generalization of the estimator in Section \ref{subsec:estimator}. Take the direct effect as an example. In the first step, 
		 we estimate $\beta_{n,j}(\gvec) $ using the generalized AIPW estimators from Equation \eqref{eqn:aipw-beta-estimator} for every valid $n\in\mathcal{S}$ (denote the estimator as $\hat{\beta}^\aipw_{n,j}(\gvec) $). In the second step, we estimate $p_n$ by taking the ratio of the number of clusters with size $n$ to $M$ (denote the estimator as $\hat{p}_n$). In the third step, we estimate $\beta_j(\gvec)$ by plugging $\hat{\beta}^\aipw_{n,j}(\gvec) $ and $\hat{p}_n$ into Equation \eqref{eqn:beta-varying-n} (denote the estimator as $\hat{\beta}_j^\aipw(\gvec)$).
   
The following theorem shows that the treatment effects estimated from this three-step approach are consistent and  asymptotically normal. 

	\begin{theorem}[Varying Cluster Sizes, General Estimand]\label{theorem:vary-cluster-size-general}
			Suppose the assumptions in Theorem \ref{thm:consistency} hold,
			$|\mathcal{S}|$ is finite and $p_n$ is bounded away from 0 for all $n \in \mathcal{S}$. As $M \rightarrow \infty$, for any subset $j$, treatment assignments $(z,\gvec)$ and $(z^\prime,\gvec^\prime)$, 
			$\hat\psi^\aipw_j(z,\gvec) - \hat\psi^\aipw_j(z^\prime,\gvec^\prime)$ are consistent, and 
			\begin{align*}
		\sqrt{M}\big( (\hat\psi^\aipw_j(z,\gvec) - \hat\psi^\aipw_j(z^\prime,\gvec^\prime) )- (\psi_j(z,\gvec) - \psi_j(z^\prime,\gvec^\prime)) \big) \overset{d}{\rightarrow}\mathcal{N}\left(0,V^{(1)}_{j,z,z^\prime,\gvec,\gvec^\prime} + V^{(2)}_{j,z,z^\prime,\gvec,\gvec^\prime}\right)
		\end{align*}
			where
			\begin{align*}
			V^{(1)}_{j,z,z^\prime,\gvec,\gvec^\prime} =& \sum_{n \in \mathcal{S}} \left(\frac{p_{n} \boldsymbol{1}\{\gvec,\gvec' \leq \gvec_n-e_j\}}{\sum_{n^\prime \in \mathcal{S}} p_{n^\prime} \boldsymbol{1}\{\gvec,\gvec' \leq \gvec_{n'}-e_j\}} \right)^2  \frac{1}{p_n} V_{n,j,z,z^\prime,\gvec,\gvec^\prime} \\
			V^{(2)}_{j,z,z^\prime,\gvec,\gvec^\prime} =& \frac{\sum_{n \in \mathcal{S}} c_{n,j,z,z^\prime,\gvec,\gvec^\prime}^2 (1-p_n)p_n - \sum_{n \neq n^\prime} c_{n,j,z,z^\prime,\gvec,\gvec^\prime} c_{n^\prime,j,z,z^\prime,\gvec,\gvec^\prime} p_n p_{n^\prime}}{\big(\sum_{n^\prime \in \mathcal{S}} p_{n^\prime} \boldsymbol{1}\{\gvec,\gvec' \leq \gvec_{n'}-e_j\} \big)^4 } 
			\end{align*}
			with $V_{n,j,z,z^\prime,\gvec,\gvec^\prime}$ to be the semiparametric bound in \eqref{eqn:general-var-bound} for cluster size $n$ and 
			\begin{align*}
			    c_{n,j,z,z^\prime,\gvec,\gvec^\prime} =&  \boldsymbol{1}_{\onenorm(\gvec)\leq n} \Big[ (\psi_{n,j}(z,\gvec) - \psi_{n,j}(z^\prime,\gvec^\prime)) \Big(\sum_{n^\prime \in \mathcal{S}} p_{n^\prime} \boldsymbol{1}_{\onenorm(\gvec)\leq n^\prime} \Big) \\ & \quad  - \Big(\sum_{n^\prime \in \mathcal{S}} p_{n^\prime} \boldsymbol{1}_{\onenorm(\gvec)\leq n^\prime} (\psi_{n^\prime,j}(z,\gvec) - \psi_{n^\prime,j}(z^\prime,\gvec^\prime)) \Big) \Big].
			\end{align*}
		\end{theorem}

  The proof and the finite sample properties of Theorem \ref{theorem:vary-cluster-size-general} are provided in the Internet Appendix.
  As the general estimand in Theorem \ref{theorem:vary-cluster-size-general} covers the direct effect as a special case, Theorem \ref{theorem:vary-cluster-size-general} implies that the estimated direct effect from the three-step approach is consistent and asymptotically normal. The following corollary formally states this result and therefore generalizes Theorems \ref{thm:consistency} and \ref{thm:normality} in Section \ref{section:asymptotics} to the case with varying cluster sizes.
  
		\begin{corollary}
		    [Varying Cluster Sizes, Direct Effect]\label{theorem:vary-cluster-size}
			Suppose the assumptions in Theorem \ref{thm:normality} hold,
			$|\mathcal{S}|$ is finite and $p_n$ is bounded away from 0 for all $n \in \mathcal{S}$. As $M \rightarrow \infty$, for any subset $j$ and neighbors' treatment assignments $\gvec$,  $\hat{\beta}_j^\aipw(\gvec)$ are consistent, and 
			\begin{align*}
		\sqrt{M}\big( \hat{\beta}_j^\aipw(\gvec)- \beta_j^\aipw(\gvec) \big) \overset{d}{\rightarrow}\mathcal{N}\left(0,V_{j,\gvec} + V_{j,\gvec,p}\right) \, ,
		\end{align*}
			where
			\begin{align*}
			V_{j,\gvec} =& \sum_{n \in \mathcal{S}}  \frac{\omega_{n,j}^2}{p_n} V_{n,j,\gvec} \\
			V_{j,\gvec,p} =& \sum_{n \in \mathcal{S}} \frac{1-p_n}{p_n}  \omega_{n,j}^2 \Big( \beta_{n,j}(\gvec) - S_{j,\gvec} \Big)^2- \sum_{n \neq n^\prime} \omega_{n,j} \omega_{n^\prime,j} \Big( \beta_{n,j}(\gvec) - S_{j,\gvec} \Big) \Big( \beta_{n^\prime,j}(\gvec) - S_{j,\gvec} \Big),
			\end{align*}
			$S_{j,\gvec}  = \sum_{n^\prime} \omega_{n^\prime,j}\beta_{n^\prime,j}(\gvec)$,
			and $V_{n,j,\gvec}$ is the asymptotic variance of $\hat{\beta}^\aipw_{n,j}(\gvec)$ with size $n$ (see \eqref{var_bound}).
		\end{corollary}

		The asymptotic variance of $\hat{\beta}_j^\aipw(\gvec)$ consists of two terms. The first term $V_{j,\gvec}$  is analogous to the asymptotic variance of $\hat{\beta}^\aipw_{n,j}(\gvec)$ in Theorem \ref{thm:normality}. 
		If $\omega_{n,j} = p_n$ (i.e., $\boldsymbol{1}\{\gvec \leq \gvec_{n,j,\max}\} = 1$ for all $n$), then the first term is simplified to $V_{j,\gvec} = \sum_{n \in \mathcal{S}}  p_n V_{n,j,\gvec} $, which is the weighted average of $V_{n,j,\gvec}$ by the fraction of clusters with size $n$.

		The second term $V_{j,\gvec,p}$ is unique to the setting with varying cluster sizes, and this term comes from the estimation error of $\hat{p}_n$.\footnote{Note that the estimation error of $\hat{p}_n$ also appears in the efficient influence function in \cite{park2020efficient}. Compared to \cite{park2020efficient}, we explicitly quantify how the estimation error of $\hat{p}_n$ for different $n$ affects the efficiency bound.} Note that $V_{j,\gvec,p}$ consists of two sums. The first sum comes from the variance of the estimation error of $\hat{p}_n$, and the second sum comes from the covariance between estimation error of $\hat{p}_n$ and $\hat{p}_{n^\prime}$ for $n \neq n^\prime$. To see this point clearer, if $\omega_{n,j} = p_n$, then in the first sum $\frac{1-p_n}{p_n}  \omega_{n,j}^2 = p_n(1-p_n)$, which is the asymptotic variance of $\hat{p}_{n^\prime}$, and in the second sum $- \omega_{n,j} \omega_{n^\prime,j} = -p_n p_{n^\prime}$, which is the asymptotic covariance between $\hat{p}_n$ and $\hat{p}_{n^\prime}$.




   \subsection{Robustness to Heterogeneous Interference vs. Estimation Efficiency: a Bias-Variance Tradeoff}
		 \label{subsection:efficiency-loss}

  In this section, we formally demonstrate that the specification of interference structures discussed in Section \ref{subsec:exchangeability} induces a bias-variance tradeoff in treatment effect estimation. We start with a simple example to illustrate the intuition. Suppose we are interested in estimating the average treatment effect from observational data for a population with a clustering structure, and we suspect that there \emph{may} be (homogeneous) interference within clusters. In this case, we can set $m=1$ in our conditional exchangeability framework and define the overall average treatment effect as\footnote{When all units are i.i.d. and also independent of the clustering structure, this effect $\beta$ has a simpler representation $\mathbb{E}[Y_{c,i}(1,G_{c,i})-Y_{c,i}(0,G_{c,i})]$, which is a direct generalization of the classical average treatment effect.} 
 \begin{align}
 \label{eq:tradeoff-simple}
     \beta \coloneqq \sum_{g_1=0}^{n-1} P(g_1) \cdot \beta_1(g_1),
 \end{align}
  where $P(g_1):=P(G_{c,i} = g_1)$ is the marginal probability of unit $i$ having $g_1$ treated neighbors that is assumed to be the same for all $i$ here. We can consistently estimate $\beta$ by estimating the direct effect $\beta_1(g_1)$ with AIPW estimators and the marginal probability $P(g_1)$ with empirical probability for each $g_1$, and plugging them into the sum. We refer to this estimator as the plug-in estimator.
  
  If, however, there is in fact \emph{no} interference among units and hence SUTVA holds, then $\beta_1(g_1) = \beta$ for all $g_1$, where $\beta$ can be interpreted as the classical average treatment effect (e.g., \cite{imbens2015causal}).\footnote{Under SUTVA, the potential outcomes can be written as $Y_{c,i}(z)$ for $z \in \{0,1\}$ and $\beta = \frac{1}{n} \sum_{i=1}^n \+E\left[Y_{c,i}(1) - Y_{c,i}(0) \right]$. If $(X_{c,i}, Y_{c,i}, Z_{c,i})$ is randomly sampled from the same superpopulation for all $i$, then $\beta = \+E\left[Y_{c,i}(1) - Y_{c,i}(0) \right]$, which is identical to the classical definition of average treatment effects (e.g., \cite{imbens2015causal}).  }
  In this case, the conventional AIPW estimator (e.g., \cite{robins1994estimation}) for $\beta$ is also consistent. Additionally, it is semiparametric efficient, and is therefore more efficient than the plug-in estimator in general.
  On the other hand, if there is indeed homogeneous interference, then in general only the plug-in estimator is consistent for $\beta$.\footnote{When SUTVA fails, the conventional AIPW estimator is a consistent estimator of $\beta$ only under some special settings, e.g., completely random treatment assignments \citep{savje2021average}, which are generally not satisfied in observational studies, the primary settings studied in this paper. } 
  
    \begin{table}
		\caption{Bias-variance tradeoff in a simulation study}
\begin{centering}
\begin{adjustbox}{max width=\linewidth,center}
\begin{tabular}{cccccccccc}
\toprule 
& \multicolumn{3}{c}{Bias} &  \multicolumn{3}{c}{Variance} &  \multicolumn{3}{c}{MSE}
\tabularnewline
\cmidrule(l){2-4} \cmidrule(l){5-7} \cmidrule(l){8-10}
 DGP & $\hat{\beta}_{\text{no}}$ & $\hat{\beta}_{\text{homo}}$ & $\hat{\beta}_{\text{heter}}$ & $\hat{\beta}_{\text{no}}$ & $\hat{\beta}_{\text{homo}}$ & $\hat{\beta}_{\text{heter}}$ & $\hat{\beta}_{\text{no}}$ & $\hat{\beta}_{\text{homo}}$ &  $\hat{\beta}_{\text{heter}}$ \tabularnewline
\midrule 
no interference & \textbf{0.001} & $0.0009$ & $0.0003$ & \textbf{0.002} & $0.003$ & $0.006$ &\textbf{0.002} &$0.003$ &$0.005$\tabularnewline
\midrule 
homo. interference & $-0.076$ & $0.001$ & \textbf{0.0001} & --- & \textbf{0.0008} &$0.0017$ &$0.084$ & \textbf{0.001}& $0.002$ \tabularnewline
\midrule 
heter. interference & $-0.123$ & $-0.049$ & \textbf{0.001} & --- & --- &\textbf{0.001} &$0.146$ &$0.018$ & \textbf{0.002} \tabularnewline
\bottomrule
\end{tabular}
\end{adjustbox}

\par\end{centering}
\bnotetab{Average bias, variance, and MSE of estimators based on three different specifications of the interference
structure. In this table, ``homo'' stands for homogeneous and ``heter'' stands for heterogeneous. The simulation setup is the same as that in Section \ref{section:simulations}. 
Variance is calculated based on the formula in Theorem \ref{thm:normality}. It is invalid under misspecified interference structures (below the diagonal), and is therefore omitted. $\hat{\beta}_{\text{no}}$ and $\hat{\beta}_{\text{homo}}$ are defined in the text, while $\hat{\beta}_{\text{heter}}$ is the plug-in estimator of $\sum_{g_1,g_2} P(g_1,g_2)\cdot\beta_1(g_1,g_2)$.}
\label{tab:tradeoff}
\end{table}
  
  To illustrate the tradeoff in this example more intuitively, we provide results from a simulation study in Table \ref{tab:tradeoff}, where $\hat{\beta}_{\text{homo}}$ is the plug-in estimator and $\hat{\beta}_{\text{no}}$ is the conventional AIPW estimator. We additionally provide an estimator $\hat{\beta}_{\text{heter}}$ based on heterogeneous interference with $m=2$, using the plug-in estimator of $\sum_{g_1,g_2} P(g_1,g_2)\cdot\beta_1(g_1,g_2)$. Accordingly, we consider three data generating processes, with no, homogeneous, and heterogeneous interference among units, respectively. When the estimators do not capture the interference structure, e.g., $\hat{\beta}_{\text{no}}$ when the data generating process (DGP) has homogeneous interference, they are biased; on the other hand, if the specification is more complex than necessary, e.g., $\hat{\beta}_{\text{homo}}$ when the DGP has no interfernece, estimators have larger variances compared to the estimators based on the correct specification.


  Based on the discussion above, we propose that this tradeoff between estimation bias and efficiency holds for a more general \emph{hierarchy} of interference structures on which estimators are built:
		 
        \begin{center}
            \begin{tabular}{ll}
        \multirow{5}*{\rotatebox[origin=c]{270}{{  $\xRightarrow[\text{Efficiency Loss}]{\text{Reduced Bias}}$} }} & No Interference \\
        & Partial Interference with Full Exchangeability  \\ & Partial Interference with Conditional Exchangeability \\ & Network Interference with Conditional Exchangeability \\ & General Interference
        \end{tabular}
        \end{center}
        
    Specifically, let us consider two estimation approaches: one is a sophisticated estimation approach (like the plug-in estimator) that accounts for possible complex interference structures, and the other one is a naive approach (like the conventional AIPW estimator) that neglects such potential interference. Intuitively, the sophisticated approach generally has less bias than the naive approach in the presence of complex interference, but is generally less efficient when the true interference structure is simple. 
    
        
In this section, we formally show the tradeoff among the top three levels in the hierarchy above, for which we have fully understood the asymptotic theory. To start, let us consider two \emph{nested} candidate partitions of a cluster. The first partition is $\mathcal{I}_1, \cdots, \mathcal{I}_m$, which is a granular partition. The second partition is $\mathcal{I}^\prime_1, \cdots,\mathcal{I}^\prime_\ell$, for $0 \leq \ell < m$, which combines some of the subsets in the first partition together and is a coarse partition. $\ell = 1$ denotes homogeneous interference (i.e., the second level in the hierarchy) for the second partition. 
$\ell = 0$ is used to denote no interference (i.e., the top level in the hierarchy). This setup thus includes all of the top three levels of interference structure.
        
        Given a vector of neighbors' treatments $\mathbf{z}_{c,(i)}\in \{0,1\}^{n-1}$ for any $c,i$, let $\gvec \in \+Z_{\geq 0}^m$ be the number of treated neighbors (calculated using $\mathbf{z}_{c,(i)}$) in each of the $m$ subsets in $\mathcal{I}_1, \cdots, \mathcal{I}_m$, and $\*h \in \+Z_{\geq 0}^\ell$ be the number of treated neighbors in each of the $\ell$ subsets in $\mathcal{I}^\prime_1, \cdots,\mathcal{I}^\prime_\ell$, for $0 \leq \ell < m$. 
        Note that since the two partitions are nested, i.e., the second partition combines some of the subsets in the first partition together, $\*g$ and $\*h$ are related through the following binary matrix $A = [A_{kj}] \in \{0,1\}^{\ell \times m}$:
        \[\sum_{k=1}^\ell \sum_{j=1}^m A_{kj} = m  \qquad \text{and} \qquad \*h = A \cdot \gvec. \]
        For example, if $\ell = 1$, then $A = \bm{1}_{1 \times m}$ is a $1\times m$ vector of ones, and $\*h$ is the scalar that equals to the total number of treated neighbors, i.e., 
        $\*h = \bm{1}_{1 \times m} \cdot \gvec = \sum_{j = 1}^m g_j$. If $\ell = m$, then $A = \*I_{m \times m}$ is the $m \times m$ identity matrix, and $\*h$ equals to $\*g$, i.e., $\*h = \*I_{m \times m} \cdot \gvec$.  
        
        Note that the mapping from $\gvec$ to $\*h$ is a many-to-one mapping, so we use $\mathcal{G}_A(\*h)  = \{\gvec: \*h = A \gvec\}$ to denote the set of all $\gvec$ that map to $\*h$. If $\ell = 1$, then $\*h$ is a scalar and $\mathcal{G}_A(\*h) $ denotes all possible $\*g$ that satisfy $ \|\gvec\|_1 = \*h$. If $\ell = 0$, we let $\mathcal{G}_A(\*h)$ denote all possible values of $\*g$.\footnote{Specifically, if $\ell = 0$, then $\mathcal{G}_A(\*h) = \{\*g: \sum_{j=1}^m g_j \leq n, 0\leq g_j \leq |\mathcal{I}_j \backslash i|\}$. } 
        
        %
        We will formally show the bias-variance tradeoff in the estimation of the following estimand, which is related to  both candidate partitions:
		 \begin{align}\label{eqn:weighted-estimand}
		    \tilde{\beta}_j(\*h) \coloneqq
		    \sum_{\gvec \in \mathcal{G}_A(\*h)} \omega_{A}(\gvec) \cdot \beta_j(\gvec),
		\end{align}
		where $\omega_A(\gvec) \geq 0$ is a generic weight of $\gvec$ that satisfies $\sum_{\gvec \in \mathcal{G}_A(\*h)} \omega_{A}(\gvec) = 1$, and $j \in [m]$. For example, $\omega_A(\gvec)$ can be the same for every possible $\gvec$, or $\omega_A(\gvec)$ can be proportional to $P(\Gvec_{c,i} = \gvec)$.\footnote{If $\omega_A(\gvec)$ is the same for every $\gvec$, then  $\omega_A(\gvec) = \frac{1}{|\mathcal{G}_A(\*h)|}$. If $\omega_A(\gvec)$ is proportional to $P(\Gvec_{c,i} = \gvec)$, then $\omega_A(\gvec) = \frac{P(\Gvec_{c,i} = \gvec)}{P(\Gvec_{c,i} \in \mathcal{G}_A(\*h))}$ with the assumption that $P(\Gvec_{c,i} = \gvec)$ is the same for every $i \in \mathcal{I}_j$. }
		
		\begin{remark}
		\label{remark:tradeoff-alpha-allocation}
		\normalfont
		   If $\ell = 0$, then $\tilde{\beta}_j(\*h)$ does not depend on $\*h$, and a natural choice of $\omega_A(\gvec)$ is $P(\Gvec_{c,i} = \gvec)$ for any $i$, which generalizes the example in \eqref{eq:tradeoff-simple} with $m=1$. 
		%
		%
		Thus our framework includes the case of no interference vs. homogeneous partial interference as a special case. On the other hand, if $\ell=0, m=n$, and $\omega_A(\gvec)=\alpha^{\|\gvec\|_1}\cdot(1-\alpha)^{n-1-\|\gvec\|_1}$, $\tilde{\beta}_j(\*h)$ reduces to the direct effect under the $\alpha$-allocation strategy \citep{hudgens2008toward,park2020efficient}. Thus our framework includes this setting as a special case, and in particular, our bias-variance tradeoff analysis applies.
		\end{remark}
		 
		We will consider two estimation approaches for $\tilde{\beta}_j(\*h)$, each based on one of the two candidate partitions. The first approach is a sophisticated approach that is based on the granular partition $\mathcal{I}_1, \cdots, \mathcal{I}_m$.
		 This approach first estimates $\beta_j(\gvec)$ and (if necessary) $\omega_{A}(\gvec)$ for $\gvec \in \mathcal{G}_A(\*h)$, and then plugs them into \eqref{eqn:weighted-estimand} to estimate $\tilde \beta_j(\*h)$. Denote this estimator as $\hat\beta^\ind_j(\*h)$. 
		 
		 The second approach is a simplified approach that is based on the coarse partition $\mathcal{I}^\prime_1, \cdots, \mathcal{I}^\prime_\ell$. If this coarse partition is sufficient to capture the interference structure, i.e., it satisfies Assumption \ref{ass:cond-outcome-partial-exchangeable}, then units in $\mathcal{I}^\prime_k$ are exchangeable for $k \in \{1,\cdots, \ell\}$, and the potential outcomes satisfy
		 \[Y_{c,i}(z,\gvec) = Y_{c,i}(z,\gvec^\prime) \qquad \forall  \gvec, \gvec^\prime \in \mathcal{G}_A(\*h).\] 
	 Consequently, $\beta_j(\gvec)$ is the same for all $\gvec \in \mathcal{G}_A(\*h)$, and $\tilde{\beta}_j(\*h) $ is invariant to any choice of $\omega_A(\gvec)$, and is in fact equal to the ADE $\beta_j(\*h)$ based on the coarse partition.\footnote{The direct effect for the subset of units corresponding to $\mathcal{I}_j$ in the granular partition is still well defined under the coarse partition based on \eqref{eqn:beta-g-short-definition}, even if they are included in a larger $\mathcal{I}^{\prime}_k$ in the coarse partition, i.e., $\mathcal{I}_j \subsetneq \mathcal{I}_k^\prime$.} The second approach then uses our generalized AIPW estimator to estimate $\tilde{\beta}_j(\*h)$. Denote this estimator as $\hat\beta^\agg_j(\*h)$.

		 
		 If the coarse partition satisfies Assumption 	\ref{ass:cond-outcome-partial-exchangeable} (hence so will the granular partition), we will show in Theorem \ref{thm:flexible-model-efficiency} that both $\hat\beta^\ind_j(\*h)$ and  $\hat\beta^\agg_j(\*h)$ are consistent estimators of $\tilde\beta_j(\*h)$, but $\hat{\beta}_j^\ind(\*h)$ is weakly less efficient than $\hat{\beta}_j^\agg(\*h)$. The key insight is when both partitions satisfy Assumption \ref{ass:cond-outcome-partial-exchangeable}, 
		 \begin{align*}
		  \Var\left(\hat{\beta}_j^\ind(\*h) \right) \propto\+E\Bigg[ \underbrace{ \sum_{\gvec \in \mathcal{G}_A(\*h)}  \frac{\omega_{A}(\gvec)^2 }{p_{i,(z,\gvec)}(\c(X))}}_{\textbf{I}}    \Bigg] \qquad \Var\left(\hat{\beta}_j^\agg(\*h) \right) \propto \+E\Bigg[\underbrace{ \frac{1 }{p_{i,(z,\*h)}(\c(X))} }_{\textbf{II}} \Bigg],
		 \end{align*}
		 where the expectation is taken with respect to $\c(X)$. For any fixed $\c(X)$, we have 
		 \begin{align*}
		 \textbf{I}= \sum_{\gvec \in \mathcal{G}_A(\*h)} \omega_{A}(\gvec) \cdot \frac{1 }{p_{i,(z,\gvec)}(\c(X))/\omega_{A}(\gvec)}  \geq \frac{1 }{\sum_{\gvec \in \mathcal{G}_A(\*h)} \omega_{A}(\gvec) \cdot p_{i,(z,\gvec)}(\c(X))/\omega_{A}(\gvec)} =\textbf{II},
		 \end{align*}
		 where we have used the \emph{convexity} of the function $x: \rightarrow 1/x$ for $x > 0$.
		 
		 On the other hand, if the coarse partition $\mathcal{I}^\prime_1, \cdots, \mathcal{I}^\prime_\ell$ does not satisfy Assumption	\ref{ass:cond-outcome-partial-exchangeable} but the granular partition $\mathcal{I}_1, \cdots, \mathcal{I}_m$ does, then $\hat{\beta}_j^\ind(\*h)$ is consistent, but $\hat{\beta}_j^\agg(\*h)$ is generally inconsistent. A similar estimation bias result when $\ell=0$ and $m=1$ was observed in \cite{forastiere2020identification}, but in Theorem \ref{thm:flexible-model-efficiency} below we provide the complete bias-variance tradeoff between robustness to interference and estimation efficiency in the general conditional exchangeability framework.\footnote{For exposition, we will assume that if a partition does not satisfy Assumption \ref{ass:cond-outcome-partial-exchangeable}, it also does not satisfy Assumption \ref{ass:partial-prop-exchangeable}. If the coarse partition satisfies Assumption \ref{ass:partial-prop-exchangeable} but not Assumption \ref{ass:cond-outcome-partial-exchangeable}, a similar tradeoff result holds if $\omega(\gvec)=\frac{1}{| \mathcal{G}_A(\*h)|}$ for all $\gvec \in \mathcal{G}_A(\*h)$, due to the double robustness of AIPW estimators.}

		 		\begin{theorem}[Bias-Variance Tradeoff between Robustness vs. Efficiency]\label{thm:flexible-model-efficiency}
		      Suppose Assumptions \ref{ass:network}-\ref{ass:overlap} hold, and that the granular partition $\mathcal{I}_1, \cdots, \mathcal{I}_m$ satisfies Assumptions \ref{ass:cond-outcome-partial-exchangeable}-\ref{ass:continuity-boundedness}.
		      
		  If the coarse partition $\mathcal{I}^\prime_1, \cdots, \mathcal{I}^\prime_\ell$ also satisfies Assumptions \ref{ass:cond-outcome-partial-exchangeable}-\ref{ass:continuity-boundedness}, then both $\hat\beta^{\agg}_j(\*h)$ and $\hat\beta^{\ind}_j(\*h)$ based on sieve estimators are consistent estimators for $\tilde{\beta}_j(\*h)$; otherwise, 
		  only $\hat\beta^{\ind}_j(\*h)$ is consistent for $\tilde{\beta}_j(\*h)$, unless for any $\gvec$,  $\frac{p_{i,(z,\gvec)}(\c(X))}{\sum_{\gvec \in \mathcal{G}_A(\*h)}p_{i,(z,\gvec)}(\c(X)) } \equiv \omega_A(\gvec)$ for all $\c(X)$ and $z$.
		       
		       On the other hand, if $\mu_{i,(z,\gvec)}(\c(X))$ and $\sigma^2_{i,(z,\gvec)}(\c(X))$ are the same for $\gvec \in \mathcal{G}_A(\*h) $, then $\hat \beta^{\ind}_j(\*h) $ is weakly less efficient than $\hat \beta_j^{\agg}(\*h)$, i.e., 
		       the asymptotic variance of $\hat \beta^{\ind}_j(\*h)$ is bounded below by the asymptotic variance of $\hat \beta^{\agg}_j(\*h)$. 
		       The inequality is strict unless the following two conditions hold: (1)
		       $\big( \hat \omega_A(\gvec) - \omega_A(\gvec) \big)\cdot \beta_j(\gvec) =  o_p \big(M^{-1/2} \big)$; (2)
		       for any $\gvec$, $\frac{ \omega_A(\gvec)}{p_{i,(z,\gvec)}(\c(X))}$ is the same for all $\c(X)$ and $z$.
		\end{theorem}

		For the variance comparison in Theorem \ref{thm:flexible-model-efficiency}, if the coarse partition is correctly specified, i.e., it satisfies Assumption \ref{ass:cond-outcome-partial-exchangeable}, then the condition for $\hat \beta_j^{\ind}(\*h) $ to be weakly less efficient is satisfied. In this case, $\hat \beta_j^{\ind}(\*h) $ is generally \emph{strictly} less efficient than $\hat \beta_j^{\agg}(\*h) $ except for some special cases (see Remark \ref{remark:efficiency}). Theorem \ref{thm:flexible-model-efficiency} highlights an important aspect of interference that, although intuitive, has not been formalized in the literature before. It implies that there is no free lunch when modeling interference: if we specify a general structure that allows complex interference patterns, the resulting estimator is robust to bias, but at the cost of efficiency loss. A less sophisticated specification potentially increases estimation efficiency, but at the cost of increased risk of bias. As an example, recall the causal estimands defined based on the $\alpha$-allocation strategy discussed in Remark \ref{remark:tradeoff-alpha-allocation}. 
		Theorem \ref{thm:flexible-model-efficiency} implies that estimation strategies that assume \emph{no} units are exchangeable, i.e. $m=n$, are potentially inefficient. 
		
		Therefore, we suggest using the most parsimonious, but correct, conditional exchangeability structure, whenever possible.
		In Section \ref{sec:testing}, we develop hypothesis tests that can be used to test for the heterogeneity of interference and treatment effects, that can help practitioners determine the appropriate specification of the interference structure.
		
		\begin{remark}[Bias]
		            \normalfont 
		  In general, estimators of $\tilde \beta_j(\*h)$ are consistent only if they are based on a partition that satisfies Assumption  \ref{ass:cond-outcome-partial-exchangeable}. One exception is that when treatment assignments are fully \emph{randomized}, $\hat\beta^{\agg}_j(\*h)$ is a consistent estimator of $\tilde \beta_j(\*h), $ 
		  even if the partition $\mathcal{I}^\prime_1, \cdots,\mathcal{I}^\prime_\ell$ is \emph{misspecified}. This type of robustness result has been observed for example in \cite{savje2021average}. Our results highlight the fundamental difference in the observational setting. Moreover, even if treatments within a cluster are independent conditional on $\c(X)$, naive estimators are still biased in general. 
		\end{remark}
		
		\begin{remark}[Efficiency]\label{remark:efficiency}
		   \normalfont 
		   The two conditions for efficiency equality are usually violated. Specifically, if $\hat{\omega}_A(\gvec)$ is estimated from a sample with $O(M)$ observations, then the convergence rate of most estimators is no more than $\sqrt{M}$, violating the first condition for efficiency equality. Furthermore, if the propensity score $p_{i,(z,\gvec)}(\c(X))$ vary with either covariates $\c(X)$ or $z$ (which is commonly the case), then the second condition for efficiency equality is violated.  
		   Only under some special cases can the two conditions hold. For example, if for each $\gvec \in \mathcal{G}_A(\*h)$, we either know the true value of $\omega_A(\gvec)$ or $\gvec \in$ $\beta_j(\gvec) = 0$, then the first condition holds. Furthermore, if the treatment assignments are completely random, then the second condition holds. 
		\end{remark}


    		\subsection{Feasible Variance Estimators}
		\label{sec:variance-estimators}
		
		In this section, we discuss feasible estimators for the asymptotic variance $V_{j,\gvec}$ in Theorem \ref{thm:normality}.\footnote{The feasible estimators for the asymptotic variance of the estimated spillover effects (and other general estimands) can be constructed analogously.} We first review two standard variance estimators and emphasize that the validity of these two estimators rely on stronger assumptions. We then propose an alternative variance estimator that is consistent under much weaker assumptions.
		
		The first standard variance estimator is the sample variance. Specifically, we can calculate the sample variance of the cluster average score $\frac{1}{|\mathcal{I}_j|} \sum_{i \in \mathcal{I}_j} \left(\hat{\phi}_{c,i}(1, \gvec)-\hat{\phi}_{c,i}(0, \gvec) \right)$ over all clusters $c$.\footnote{The formula to calculate the sample variance is \begin{equation}\label{bound:empirical}
        \begin{aligned}
            \hat{V}^{\mathrm{smp}}_{\beta_j}(\gvec) \coloneqq&  \frac{1}{M } \sum_{c=1}^{M}  \left( \frac{1}{|\mathcal{I}_j|} \sum_{i \in \mathcal{I}_j}\left(  \hat{\phi}_{c,i}(1, \gvec) - \hat{\phi}_{c,i}(0, \gvec) \right)- \hat\beta^\aipw_j(\gvec)\right)^2
        \end{aligned}
    \end{equation}} 
    The resulting estimator is consistent for $V_{j,\gvec}$ if both $\hat \beta_{i,\gvec}(\c(X))$ and $\hat p_{i,(z,\gvec)}(\c(X))$ are uniformly consistent, which can be restrictive in practice.\footnote{The double robustness property of AIPW estimators does not carry over to their second moments.} 
    Furthermore, we find that test statistics constructed from sample variances do not typically have good finite sample performances in Section \ref{section:simulations}. 
    
    The second standard variance estimator is the plug-in estimator. We first estimate $\beta_{i,\gvec}(\c(X))$, $\sigma_{i,(z,\gvec)}^{2}(\c(X))$, and $p_{i,(z,\gvec)}(\c(X))$, and then plug their estimators into the formula of $V_{j,\gvec}$ in Theorem \ref{thm:normality}. Similar to the sample variance estimator, this standard plug-in estimator is consistent if $\hat\beta_{i,\gvec}(\c(X))$, $\hat\sigma_{i,(z,\gvec)}^{2}(\c(X))$, and $\hat p_{i,(z,\gvec)}(\c(X))$ are uniformly consistent.\footnote{Under the weaker condition of asymptotic unbiasedness, our proposed bias correction procedure in Step 3 below can also be adapted to produce a consistent plug-in estimator.}

    In this section, we propose a consistent estimation strategy for $V_{j,\gvec}$ that is based on uniformly \emph{asymptotically unbiased} estimates of $\beta_{i,\gvec}(\c(X))$, $\sigma_{i,(z,\gvec)}^{2}(\c(X))$, and $p_{i,(z,\gvec)}(\c(X))$, which significantly relaxes the uniform consistency requirements. Our proposed estimation strategy has three main steps: first estimate $1/p_{i,(z,\gvec)}(\c(X))$, then estimate $\beta_{i,\gvec}(\c(X))$ and $\sigma_{i,(z,\gvec)}^{2}(\c(X))$, and lastly estimate $V_{j,\gvec}$. We elaborate on these three steps below.
    
    \paragraph{Step 1: Estimate $1/p_{i,(z,\gvec)}(\c(X))$.} A natural idea is to first estimate $p_{i,(z,\gvec)}(\c(X))$ and take its reciprocal. However, the asymptotic unbiasedness of $\hat{p}_{i,(z,\gvec)}(\c(X))$ does not guarantee the asymptotic unbiasedness of $1/\hat{p}_{i,(z,\gvec)}(\c(X))$. To address this challenge, we leverage the idea from \cite{blanchet2015unbiased,moka2019unbiased} to obtain an unbiased estimator of the reciprocal mean. The idea is as follows. Suppose we seek to estimate $\gamma = 1/\+E U$ for a random variable $U \in (0,1)$. From Taylor expansion, $\gamma = 1/\+E U = \sum_{k = 0}^\infty \left( 1 - \+E U \right)^k$. Let $\{U_i: i \geq 0\}$ be a sequence of i.i.d. copies of $U$ and let $K$ be a non-negative integer-valued random variable with $q_k \coloneqq P(K = k) > 0$ for all $k \geq 0$. The key observation is the following identity:
    \[\frac{1}{\+EU} = \sum_{k = 0}^\infty \left( 1 - \+E U \right)^k = \sum_{k = 0}^\infty q_k \frac{\+E\prod_{i = 1}^k (1 - U_i)}{q_k} = \+E \left[ q_K \prod_{i = 1}^K (1 - U_i) \right]. \]
    The following estimator $\hat{\gamma}$ suggested by \cite{blanchet2015unbiased,moka2019unbiased} is clearly unbiased: 
    \[\hat{\gamma} \coloneqq \frac{1}{q_K} \prod_{i = 1}^K (1 - U_i). \]
    We use this idea to construct asymptotically unbiased estimators of $1/p_{i,(z,\gvec)}(\c(X))$ for any fixed $\c(X)$. First, we generate the integer $K$ and let $\tilde{K}:=\max(K,M/h_M)$, where $h_M$ is a slowly increasing sequence in $M$. 
    Next, we split the clusters into $\tilde K$ folds, and estimate $ p_{i,(z,\gvec)}(\c(X))$ in each fold using an asymptotically unbiased estimator (e.g., kernel regression). Let $\hat p^\ell_{i,(z,\gvec)}(\c(X))$ be the estimator based on fold $\ell$. Finally, we estimate $1/p_{i,(z,\gvec)}(\c(X))$ by 
    \[\widehat{{p}^{-1}_{i,(z,\gvec)}}(\c(X)) = \frac{1}{q_{\tilde K}} \prod_{\ell = 1}^{\tilde K} \left(1 - \hat p^\ell_{i,(z,\gvec)}(\c(X)) \right). \]
     We can show this estimator that is asymptotically unbiased as $M\rightarrow \infty$, following similar arguments as \cite{blanchet2015unbiased,moka2019unbiased}.\footnote{Incidentally, AIPW estimators for $\beta_{j}(\gvec)$ based on this estimation approach for $1/p_{i,(z,\gvec)}(\c(X))$ and matching for $\beta_{i,\gvec}(\c(X))$ detailed in Step 2 are guaranteed to be consistent. Thus we have also provided an alternative nonparametric approach to sieve estimators that yields a consistent estimator for $\beta_{j}(\gvec)$ and is easier to implement in practice.}
     
     \paragraph{Step 2: Estimate $\beta_{i,\gvec}(\c(X))$ and $\sigma_{i,(z,\gvec)}^{2}(\c(X))$.} To achieve asymptotic unbiasedness, we propose to use matching. Conceptually, we first match unit $i$ in cluster $c$ with a few units of the same type as itself (i.e., in the same subset $\mathcal{I}_j$) whose neighbors' treatments are $\gvec$ and whose covariates are ``close'' to $\left(X_{c,i}, \*X_{c,(i)}\right)$. Then we use $i$ and its matched units to estimate $\beta_{i,\gvec}(\c(X))$ and $\sigma_{i,(z,\gvec)}^{2}(\c(X))$.

     To measure the proximity between
     $\left(X_{c,i}, \*X_{c,(i)}\right)$ and $\left(X_{c^\prime,i^\prime}, \*X_{c^\prime,(i^\prime)}\right)$, we need a distance metric, denoted by $\dist(\cdot,\cdot)$. For example, $\dist(\cdot,\cdot)$ can be 
     \begin{eqnarray*}
	    \dist\left((X_{c,i},\n(\Xb)),(X_{c',i'},\Xb_{c',(i')})\right)\coloneqq \sqrt{\norm{X_{c^\prime,i^\prime} -\s(X)}_2^2+\dist^{\mathrm{neigh}}\left(\*X_{c^\prime,(i^\prime)}, \n(\Xb)\right)^2},
	\end{eqnarray*}
	where $\norm{x}_2 = (x^\T x)^{1/2}$ is the standard Euclidean vector norm of a generic vector $x$.\footnote{We can use other metrics as well. For example, in the household example, we can choose to only match a parent with parents in other households of the same gender, and similarly for children.} $\dist^{\mathrm{neigh}}(\cdot, \cdot)$ is a distance metric for neighbors' covariates that is \emph{invariant} with respect to permutations of units within each $\mathcal{I}_j$. For example, if $m = 1$, then $\dist^{\mathrm{neigh}}(\cdot, \cdot)$ can be the distance between two units' averaged neighbors' covariate values, i.e.,
	\begin{equation*}
	    \dist^{\mathrm{neigh}}\left(\Xb_{c^\prime,(i^\prime)}, \n(\Xb)\right) = \norm{\frac{1}{n-1} \sum_{k : k \neq i^\prime} X_{c^\prime,k} - \frac{1}{n-1} \sum_{k : k \neq i} X_{c,k}  }_2.
	\end{equation*}
	
	Suppose we match unit $i$ in cluster $c$ with $l \geq 2$ units in the same $\mathcal{I}_j$ as itself and whose own treatment is $z$ and neighbors' treatments are $\gvec$, and whose covariates are the closest to $\left(X_{c,i}, \*X_{c,(i)}\right)$ measured by $\dist(\cdot,\cdot)$.\footnote{The implicit assumption here is that $\beta_{i,\gvec}(\c(X))$ and $\sigma_{i,(z,\gvec)}^{2}(\c(X))$ do not vary with $i\in \mathcal{I}_j$, so that we may match with any unit $i' \in \mathcal{I}_j$ from any cluster. If we are concerned about this assumption, 
	we can use only a subset of units in $\mathcal{I}_j$ in each cluster as possible matching candidates, e.g., only match with parents with the same gender.} Let $\mathcal{J}_{l,(z,\gvec)}(c,i)$ be the set of indices $(c^\prime,i^\prime)$ of these $l$ units. As a special case, if $\s(Z) = z$ and $\s(\mathbf{G}) = \gvec$, then unit $(c,i) \in \mathcal{J}_{l,(z,\gvec)}(c,i)$. We focus on matching with replacement  \citep{abadie2006large}, allowing each unit to be matched to multiple $(c,i)$.
	
	We estimate $\beta_{i,\gvec}(\c(X))$ by the difference-in-means estimator 
	\[\hat \beta_{i,\gvec}(\c(X)) = \bar{Y}_{c,i}(1, \gvec) - \bar{Y}_{c,i}(0, \gvec),\qquad \text{ where } \qquad \bar{Y}_{c,i}(z, \gvec) = \frac{1}{l} \sum_{(c^\prime, i^\prime) \in \mathcal{J}_{l,(z,\gvec)}(c,i) } Y_{c^\prime, i^\prime}.\]
	We estimate $\sigma_{i,(z,\gvec)}^{2}(\c(X))$ by the sample variance in the matched group $\mathcal{J}_{l,(z,\gvec)}(c,i)$
	\[\hat \sigma_{i,(z,\gvec)}^{2}(\c(X)) = \frac{1}{l-1} \sum_{(c^\prime, i^\prime) \in \mathcal{J}_{l,(z,\gvec)}(c,i) } \big( Y_{c^\prime, i^\prime} - \bar{Y}_{c,i}(z, \gvec) \big)^2, \qquad \forall z \in \{0,1\}.\]
	
	\paragraph{Step 3: Estimate $V_{j,\gvec}$.} A natural idea is to plug the estimators from steps 1 and 2 into the formula of $V_{j,\gvec}$. Denote this plug-in estimator by $\hat{V}_{j,\gvec}$. Note that $\hat{V}_{j,\gvec}$ is generally inconsistent, because $(\hat  \beta_{i,\gvec}(\c(X))-\hat  \beta_{i,\gvec})^{2}$ is \emph{not} asymptotically unbiased (even though $\hat  \beta_{i,\gvec}(\c(X))$ is asymptotically unbiased), where $\hat  \beta_{i,\gvec}$ is the average of $\hat  \beta_{i,\gvec}(\c(X))$ over $c$ and $i$. More specifically, the estimation error of $\hat  \beta_{i,\gvec}(\c(X))-\hat  \beta_{i,\gvec}$ and its squared are at the order of $O(l)$. Since the error squared is always positive, 
	it cannot be averaged out over $c$ and $i$ in the plug-in estimator $\hat{V}_{j,\gvec}$. Fortunately, we can explicitly calculate the asymptotic bias of $\hat{V}_{j,\gvec}$ and we propose the following bias-corrected estimator for $V_{j,\gvec}$:
		\begin{align}\label{eqn:bias-correct-variance}	
		\hat{V}^{\mathrm{bc}}_{j,\gvec} \coloneqq \hat{V}_{j,\gvec}-\frac{1}{l}
	\frac{1}{M} \sum_{c = 1}^M	\frac{1}{|\mathcal{I}_j|^2}\sum_{i \in \mathcal{I}_j} \big(\hat  \sigma_{i,(1,\gvec)}^{2}(\c(X))+\hat  \sigma_{i,(0,\gvec)}^{2}(\c(X))\big).
			\end{align}
			Note that $\hat{V}^{\mathrm{bc}}_{j,\gvec}$ is strictly smaller than $\hat{V}_{j,\gvec}$, and the difference shrinks with the number of matches $l$. In Theorem \ref{thm:consistent-variance} below, we show that $\hat{V}^{\mathrm{bc}}_{j,\gvec}$ is consistent. Therefore, we can use $\hat{V}^{\mathrm{bc}}_{j,\gvec}$ to construct valid test statistics and confidence intervals, while those based on $\hat{V}_{j,\gvec}$ are conservative.

			\begin{theorem}[Consistency of Bias-Corrected Variance Estimator]\label{thm:consistent-variance}
			Suppose for all $i$, $z$ and $\gvec$, ${\beta}_{i,\gvec}(\c(X))$ and ${\sigma}_{i,(z,\gvec)}^{2}(\c(X))$ are bounded and $L$-Lipschitz in $\c(X)$ with respect to the metric $\dist(\cdot,\cdot)$, and that $p_{i,(z,\gvec)}(\c(X))$ is estimated with uniformly asymptotically unbiased methods, such as kernel regression. Moreover, suppose $\c(X)$ does not contain identical rows almost surely. 
			Then our estimators for $1/p_{i,(z,\gvec)}(\c(X))$, ${\beta}_{i,\gvec}(\c(X))$ and ${\sigma}_{i,(z,\gvec)}^{2}(\c(X))$ are uniformly asymptotically unbiased, and our proposed variance estimator $\hat{V}^{\mathrm{bc}}_{j,\gvec}$ is a consistent estimator of $V_{j,\gvec}$.  
			
		\end{theorem}
		


		\begin{remark}[Other Estimands]
		\normalfont
		We can also design similar matching-based bias-corrected estimators for the asymptotic variance of ASE $\tau_{j}(z,\gvec,\gvec^\prime)$ and general causal effects $\hat{\psi}^\aipw_j(z,\gvec) - \hat{\psi}^\aipw_j(z^\prime,\gvec^\prime)$. The consistency of the corresponding variance estimators can be derived similarly.
		\end{remark}

     \subsection{Hypothesis Testing for Treatment Effects and Interference}
 \label{sec:testing}
 
 With consistent variance estimators, we can construct asymptotically valid confidence intervals and run one-sided or two-sided hypothesis tests for direct and spillover effects. For example, the two-sided hypothesis test for $\beta_j(\gvec)$ for a specific $\gvec$ can take the form 
    \[\mathcal{H}_0:  \beta_j(\gvec) = 0, \qquad \mathcal{H}_1: \beta_j(\gvec) \neq 0. \]
    See Internet Appendix IA.B for additional hypothesis tests.

 Importantly, running hypothesis tests can help practitioners decide on a parsimonious interference structure that satisfies Assumption \ref{ass:cond-outcome-partial-exchangeable}, which is an important consideration given our bias-variance tradeoff analysis. In practice, one could start with a coarse partition, and run hypothesis tests to decide whether to split some subset(s), which is conceptually similar to tree-building in tree-based methods. Alternatively, one could start with a fine partition, and decide whether to merge some subsets, which is conceptually similar to tree-pruning in tree-based methods. 
 Below we discuss some useful hypothesis tests that one may consider when choosing between candidate interference structures in the two procedures.
 
 First, we can test for the heterogeneity of treatment effects for units in two subsets $\mathcal{I}_j$ and $\mathcal{I}_k$. For example, the two-sided joint hypothesis test for the direct effects of units in $\mathcal{I}_j$ and $\mathcal{I}_k$ (i.e., $\beta_j(\gvec)$ and $\beta_k(\gvec)$) can take the form\footnote{Such tests complement those in \cite{athey2019generalized} that use random forests to test for heterogeneity of conditional average treatment effects (CATE).} 
 \[\mathcal{H}_0:\beta_j(\gvec) = \beta_k(\gvec)  \quad \forall \gvec \in \mathcal{G}, \qquad \mathcal{H}_1:\beta_j(\gvec) \neq \beta_k(\gvec)  \quad \exists \gvec \in \mathcal{G},\]
 where $\mathcal{G}$ is a set of $\gvec$ valid for both $j$ and $k$.
 
 Second, we can focus on the treatment effects of units in one subset $\mathcal{I}_j$ and test if the treatment effects given different $\gvec$ are the same. For example, the two-sided joint hypothesis test for the direct effects of units in $\mathcal{I}_j$ can take the form 
 \[\mathcal{H}_0:\beta_j(\gvec) = \beta_j(\gvec^\prime)  \quad \forall \gvec, \gvec^\prime \in \mathcal{G}, \qquad \mathcal{H}_1:\beta_j(\gvec) \neq \beta_j(\gvec^\prime)  \quad \exists \gvec, \gvec^\prime \in \mathcal{G}.\]
 This type of test is particularly useful when we try to determine whether some subsets in a fine partition $\mathcal{I}_1, \cdots, \mathcal{I}_m$ can be merged to form a coarser partition $\mathcal{I}^\prime_1, \cdots,\mathcal{I}^\prime_\ell$ (or vice versa), by setting  $\mathcal{G} = \mathcal{G}_A(\*h)$ defined in Section \ref{subsection:efficiency-loss}, for some $\*h \in \+Z_{\geq 0}^\ell$. 

 There are two issues requiring our attention when constructing the test statistics and $p$-values for the two types of tests discussed above. First, since both types of tests are global tests involving multiple nulls, the issue of multiple testing occurs. One could use the Bonferroni-corrected $p$-values from the local hypothesis tests to address this issue. 
 
 Second, for each local null in the joint hypothesis, estimators for the two quantities involved could be correlated.\footnote{This is always the case for the second type of tests, when estimates of $\beta_j(\gvec)$ and $\beta_j(\gvec^\prime)$ are based on the same observations. For the first type of tests, this can also happen if, for example, we assume some models that we need to estimate do not depend on $i$.} There are two possible solutions to address this issue. First, we can derive the asymptotic covariance between these two estimates. Second, as a conceptually simpler but less efficient solution, we can use the sample splitting idea. For example, we can randomly choose half of the clusters to estimate $\beta_j(\gvec)$ and its asymptotic variance, and the other half to estimate $\beta_j(\gvec^\prime)$ and its asymptotic variance. Since clusters are independent, these two estimates are independent. 
 
 To conclude, we suggest that hypothesis testing can be useful when choosing the interference structure, as our proposed approach is essentially testing for the correct specification of exposure models. For empirical applications, as a complementary solution, we can report treatment effect estimates and confidence intervals based on various specifications of the interference structure. This can be helpful for understanding the robustness of effect size to model specification.

		\section{Extensions}
		\label{section:extensions}
		
		In this section, we discuss the extensions of our framework to more general cluster and interference structures, and to alternative estimands. 
		

		\subsection{General Interference Structures within Clusters}
		\label{subsection:general-cluster-structure}
		
		We have assumed that units within a cluster are fully connected, and the partition of neighbors into heterogeneous subsets is universal for every unit. These assumptions are expositional, and our results can be easily generalized to the case without these assumptions.
		
		If units within a cluster are not fully connected, e.g., they interact through some network, then we can generalize the definition of $\Gvec_{c,i}$, which only counts the treated units connected to unit $i$ in the cluster $c$ (as opposed to all the other treated units in the same cluster). Then we can define direct and spillover effects with this generalized $\Gvec_{c,i}$, and use our proposed approach in Section \ref{sec:estimator} to estimate treatment effects. 
		
		Furthermore, if the adjacency matrix varies with clusters, then we can consider the mixture model for the sampling of clusters with varying adjacency matrices similar to Section \ref{sec:varying-cluster-size}. We can use a similar approach as that in Section \ref{sec:varying-cluster-size} to define and estimate treatment effects. 
		
		If the partition of neighbors varies with units, then we can factor the unit-specific partition into the definition of treatment effects. We may consider unit-specific treatment effects rather than subset-specific treatment effects, i.e., index treatment effects by unit $i$ rather than by subset $\mathcal{I}_j$. Our estimation approach can be adapted accordingly to estimate these treatment effects.

		
		\subsection{Weakly Connected Clusters}
		\label{subsec:network}
		
		Our methods and results are shown under the assumption that clusters are disjoint, and interference is restricted to units within a cluster. We emphasize that our methods and results can still be valid with some minor modifications under certain relaxations of this assumption. For example, if units in a cluster can only interfere with a small and finite number of units outside this cluster, then we can generalize the definition of $\*X_{c,(i)}$ and $\*Z_{c,(i)}$ to account for all the adjacent units of $i$, including those in a different cluster as $i$. Suppose all the assumptions hold with the new definition of $\*X_{c,(i)}$ and $\*Z_{c,(i)}$. Then we can follow conceptually the same approach to define and estimate treatment effects, and derive similar asymptotic results as those in Section \ref{section:asymptotics} in this generalized setting, with possibly a different asymptotic variance to account for inter-cluster correlations. The technical details are left for future work.

		\subsection{Alternative Estimands}
	     \label{subsec:att-hte}
	     
	   Our estimands in Section \ref{sec:estimator} are defined as the average effects across units of the same type (e.g., in $\mathcal{I}_j$). For some applications, we may be interested in the average effects of some subpopulations. 
	   
	   For example, one may be interested in the direct effects for the directly treated group defined as
	     \[\tilde{\beta}_j(\gvec) = \frac{1}{|\mathcal{I}_j|} \sum_{i \in \mathcal{I}_j} \mathbb{E}[Y_{c,i}(1,\gvec )-Y_{c,i}(0,\gvec) \mid Z_{c,i}=1].\]
	     One may also be interested in the direct effects for the subpopulations with cluster-level covariates equal to $\*x$, such as
	     \[\tilde{\beta}_j(\gvec, \mathbf{x}) =  \frac{1}{|\mathcal{I}_j|} \sum_{i \in \mathcal{I}_j} \mathbb{E}[Y_{c,i}(1,\gvec )-Y_{c,i}(0,\gvec) \mid \c(X)=\mathbf{x}].\]
	     We can use the generalized AIPW estimators for $\tilde{\beta}_j(\gvec) $, but with a different propensity score weighting scheme similar to the ATT weighting in conventional IPW/AIPW estimators. For $\tilde{\beta}_j(\gvec, \mathbf{x})$, it is possible to use tree-based methods similar to \cite{athey2019generalized,bargagli2020heterogeneous,yuan2021causal}, and the technical details are left for future work.

	     
	     \subsection{Indexing Potential Outcomes with Treated Fractions}
	     
	      Under our conditional exchangeability framework, we can index potential outcomes by $(z,\gvec)$, where $\gvec$ is the vector of numbers of treated neighbors in each exchangeable subset $\mathcal{I}_j$, and we have
	      defined treatment effects based on $\gvec$. It is also possible to define treatment effects using the fraction of treated neighbors. When the number of neighbors is fixed for all units, these two definitions are equivalent. When the number of neighbors varies, there are subtle differences between these two definitions: using the number of treated neighbors implicitly assumes that the treatment effects are the same given the same \emph{number} of treated neighbors, even though the total number of neighbors varies; using the fraction of treated neighbors implicitly assumes that the treatment effects are the same given the same \emph{saturation} of neighbors' treatments, even though the number of treated neighbors varies.
	      Which definition is more appropriate may depend on the particular application.
	      
	      When the treated fraction is used, it is possible to use estimators that are conceptually the same as those in Section \ref{sec:estimator}. A potential benefit of this approach is that if we are willing to impose additional (smoothness) assumptions on how the treatment effects vary with the fraction of treated neighbors, then it is possible to use samples with different treated fractions in the estimation of treatment effects, by weighting each sample inversely proportional to the distance between its fraction of treated neighbors and the target treated fractions in the estimand.



    }

    

\section*{Supplementary material}

# Appendix B  Estimation Considerations

Our theoretical results are primarily developed under non-parametric sieve estimators for the propensity and outcome models. We therefore provide additional details for the sieve estimators in Appendix B.1.

Note that our theoretical results also work on parametric estimators under suitable assumptions. When data is limited, parametric estimators may be preferable. For this case, we provide a few simplifications for the estimation of propensity and outcome models, that can be used when efficiency is a primary concern.

## B.1  Non-parametric Estimation Approach

In cases where we are concerned with the model misspecification when using parametric estimators for the propensity and outcome models, we could consider nonparametric series estimators (Newey, 1997; Chen, 2007; Hirano et al., 2003; Cattaneo, 2010). Let $\{r_k(\mathbf{x})\}_{k=1}^{\infty}$ be a sequence of known approximation functions. For the conditional outcome $\mu_{i,(z,\mathbf{g})}(\mathbf{X}_c)$, we only use the units whose index in the cluster is $i$, own treatment is $z$, number of treated neighbors is $g$, and approximate $\mu_{i,(z,g)}(\mathbf{X}_c)$ by $\hat{\mu}_{i,(z,\mathbf{g})}(\mathbf{X}_c) = R_K(\mathbf{X}_c)^\top \hat{\boldsymbol{\theta}}_{i,K,(z,\mathbf{g})}$, where $R_K(\mathbf{x}) = \begin{pmatrix} r_1(\mathbf{x}) & \cdots & r_K(\mathbf{x}) \end{pmatrix}^\top$ and $\hat{\boldsymbol{\theta}}_{i,K,(z,\mathbf{g})}$ is estimated from the ordinary least squares estimator

$$\hat{\boldsymbol{\theta}}_{i,K,(z,\mathbf{g})} = \left( \sum_{c=1}^{M} \mathbf{1}\{Z_{c,i} = z, \mathbf{G}_{c,i} = \mathbf{g}\} R_K(\mathbf{X}_c) R_K(\mathbf{X}_c)^\top \right)^{-1}$$
$$\times \sum_{c=1}^{M} \mathbf{1}\{Z_{c,i} = z, \mathbf{G}_{c,i} = \mathbf{g}\} R_K(\mathbf{X}_c)^\top Y_{c,i}. \tag{26}$$

For the propensity score $p_{i,(z,\mathbf{g})}(\mathbf{X}_c)$, we use $h(\mathbf{X}_c; \boldsymbol{\gamma}_{i,K,(z,\mathbf{g})}) = R_K(\mathbf{X}_c)^\top \boldsymbol{\gamma}_{i,K,(z,\mathbf{g})}$ to approximate the log of the odds $p_{i,(z,\mathbf{g})}(\mathbf{X}_c)/p_{i,(0,\mathbf{0})}(\mathbf{X}_c)$ for all $(z,\mathbf{g}) \neq (0,\mathbf{0})$ and for all unit $i$, where $(0,\mathbf{0})$ denotes all units in the cluster are under control. Since there are many possible realizations for $(z,\mathbf{g})$, we use the multinomial logistic sieve estimator (MLSE) similar to Cattaneo (2010) to estimate $p_{i,(z,\mathbf{g})}(\mathbf{X}_c)$ that maximizes the following log-likelihood function

$$\hat{\boldsymbol{\gamma}}_{i,K} = \arg\max_{\boldsymbol{\gamma}_{i,K}: \boldsymbol{\gamma}_{i,K,(0,\mathbf{0})} = \mathbf{0}} \sum_{c=1}^{M} \sum_{z,\mathbf{g}} \mathbf{1}\{Z_{c,i} = z, \mathbf{G}_{c,i} = \mathbf{g}\} \log \left( \frac{\exp\left(h(\mathbf{X}_c; \boldsymbol{\gamma}_{i,K,(z,\mathbf{g})})\right)}{\sum_{z',\mathbf{g}'} \exp\left(h(\mathbf{X}_c; \boldsymbol{\gamma}_{i,K,(z',\mathbf{g}')})\right)} \right) \tag{27}$$

where we restrict $\boldsymbol{\gamma}_{i,K,(0,\mathbf{0})} = \mathbf{0}$ for the identification purpose. The objective function (27) is reduced to the classical multinomial logistic regression when $R_K(\mathbf{x}) = \mathbf{x}$.

The major difference compared to conventional (sieve) maximum likelihood estimation of propensity models under SUTVA is that a unit's log-likelihood depends on covariates of all units in the same cluster. Treatments of units in different clusters are still independent, so (sieve) maximum likelihood estimation can be applied to the joint propensity model. Let $\hat{\boldsymbol{\gamma}}_i$ be the solution that maximizes the log-likelihood function $\ell_{M,i}(\boldsymbol{\gamma}_i)$ defined in Eq. (27).

Following similar arguments as Cattaneo (2010), we can show $\hat{\boldsymbol{\theta}}_{i,K}$ $\hat{\boldsymbol{\gamma}}_{i,K}$ is consistent under the additional continuity assumption on propensity and conditional outcome functions and boundedness assumption on covariates and outcome.

**Assumption 7** (Continuity and Boundedness).

1. For all $(z,\mathbf{g})$, $p_{i,(z,\mathbf{g})}(\cdot)$ and $\mu_{i,(z,\mathbf{g})}(\cdot)$ are $s$ times differentiable with $s/k_x > 5\eta/2 + 1/2$ where where $\eta = 1$ or $\eta = 1/2$ depending on whether power series or splines are used as basis functions.

2. $\mathbf{X}_c$ is continuously distributed with density bounded and bounded away from zero on its compact support $\mathcal{X}$, and $|Y_{c,i}(z, \mathbf{g})| < \infty$.

## B.2 Parametric Simplifications of Estimation Methods

In practice, estimating a propensity model $p_{i,(z,\mathbf{g})}(\mathbf{X}_c)$ and an outcome model $\mu_{i,(z,\mathbf{g})}(\mathbf{X}_c)$ for every $i$ can be challenging, especially when we do not have a sufficiently large sample size to precisely estimate the model parameters for every $i$. One may consider various simplifications of the propensity and conditional outcome models and reduce the number of model parameters.

First, one could leverage the permutation invariance property of the covariates of neighbors in the same subset $\mathcal{I}_j$, as is assumed of $p_{i,(z,\mathbf{g})}(\mathbf{X}_c)$ in Assumption 6 and shown in Lemma 1 in Appendix A for $\mu_{i,(z,\mathbf{g})}(\mathbf{X}_c)$. Then one could consider models that only use permutation-invariant statistics of $\mathbf{X}_c$, such as the mean (over each $\mathcal{I}_j$), instead of $\mathbf{X}_c$ itself.

Second, one could assume that the outcome and joint propensity model parameters $\boldsymbol{\theta}_{i,K,(z,\mathbf{g})}$ and $\boldsymbol{\gamma}_{i,K,(z,\mathbf{g})}$ in the sieve estmiators do not depend on $i$, which is reasonable when units' heterogeneity can be completely captured by covariates $\mathbf{x}$. Then the number of parameters is reduced and one can use all units in a cluster to estimate $\boldsymbol{\theta}_{K,(z,\mathbf{g})}$ and $\boldsymbol{\gamma}_{K,(z,\mathbf{g})}$. Note that this simplification is not restricted to sieve estimators, since we can assume model is the same for any $i$ for any parametric estimation approach.

Third, based on the factorization of the joint propensity model

$$p_{i,(z,\mathbf{g})}(\mathbf{X}_c) = P(Z_{c,i} = z \mid \mathbf{X}_c) \cdot \prod_{j=1}^{m} P(\mathbf{G}_{c,i,j} = \mathbf{g}_j \mid \mathbf{X}_c, Z_{c,i} = z, \cdots, \mathbf{G}_{c,i,j-1} = \mathbf{g}_{j-1}),$$

one could estimate the first term with logistic regression and estimate each term in the product with a multinomial logistic regression. In practice, this simplification proves to be a sensible approximation and can significantly reduce the number of parameters in the joint propensity model.

Last but not least, in some applications it is reasonable to assume that the treatment assignments of all units in a cluster are conditionally independent given $\mathbf{X}_c$,

$$P(\mathbf{Z}_c \mid \mathbf{X}_c) = \prod_i P(Z_{c,i} = z \mid \mathbf{X}_c). \tag{28}$$

If we further assume $P(Z_{c,i} = z \mid \mathbf{X}_c)$ is the same for all $i$, then the estimation of $P(\mathbf{Z}_c \mid \mathbf{X}_c)$ is simplified to estimating $P(Z_{c,i} = 1 \mid \mathbf{X}_c)$ (e.g. from logistic regression), which is essentially the same problem as that in the classical setting without interference. In this case, the neighborhood propensity $P(\mathbf{G}_{c,i} = \mathbf{g} \mid \mathbf{X}_c)$ can be written as an analytical function of $P(Z_{c,j} = z_j \mid \mathbf{X}_c)$. For example, when $m = 1$ and all units in a cluster are exchangeable, then both $\mathbf{G}_{c,i}$ and $\mathbf{g}$ are scalars (i.e., $\mathbf{G}_{c,i} = G_{c,i,1}$ and $\mathbf{g} = g_1$, where $g_1$ is the first coordinate in $\mathbf{g}$), and $p_{i,\mathbf{g}}(\mathbf{X}_c)$ takes the form of

$$P(\mathbf{G}_{c,i} = \mathbf{g} \mid \mathbf{X}_c) = P(G_{c,i,1} = g_1 \mid \mathbf{X}_c) = \sum_{\mathbf{z} \in \{0,1\}^{n-1}} \mathbf{1}\{\|\mathbf{z}\|_1 = g_1\} \prod_{j \neq i} P(Z_{c,j} = z_j \mid \mathbf{X}_c),$$

where $\mathbf{z}$ denotes the treatment realization of the $n-1$ neighbors of unit $i$, and $z_j$ denotes the treatment realization of unit $j$ defined by $\mathbf{z}$.

A further simplification involves aggregating models of $p_{i,(z,\mathbf{g})}(\mathbf{X}_c)$ and $\mu_{i,(z,\mathbf{g})}(\mathbf{X}_c)$ over different $(z, \mathbf{g})$ pairs. For example, assuming that there is a universal propensity model $p(\mathbf{X}_c, z, \mathbf{g})$ and outcome model $\mu(\mathbf{X}_c, z, \mathbf{g})$ for every $i, z, \mathbf{g}$. This simplification can be particularly useful when the cluster size $n$ or number

of subsets $m$ in the conditional exchangeability assumption is large, so that there are a large number of pairs of $(z, \mathbf{g})$. In this case, estimating a separate model for each $(z, \mathbf{g})$ can be infeasible.

## Appendix C  Additional Hypothesis Tests

In the following, we list two groups of null hypotheses that may be of practical interest. The first are about average direct treatment effects $\beta_j(\mathbf{g})$, and the second are about average spillover effects $\tau_j(z, \mathbf{g}, \mathbf{g}')$. In particular, they include tests for the *presence* of interference in observational data.

For $\beta_j(\mathbf{g})$, the first null hypothesis is to test if the direct treatment effect is zero given a specific vector of number of treated neighbors $\mathbf{g}$, i.e.

$$\mathcal{H}_0 : \beta_j(\mathbf{g}) = 0 \text{ for a fixed } \mathbf{g}$$

The second one is the global analog of the first null hypothesis to test if the direct treatment effect $\beta_j(\mathbf{g})$ is zero for a collection of vectors of number of treated neighbors, i.e.

$$\mathcal{H}_0 : \beta_j(\mathbf{g}) = 0 \text{ for } \mathbf{g} \in \mathcal{G}$$

where $\mathcal{G}$ could be for example the set of $\mathbf{g}$ with $\|\mathbf{g}\|_1 = g_0$, i.e. we are interested in whether the average direct treatment effect is zero whenever there are $g_0$ treated neighbors. Note that we have the multiple hypothesis issue for this type of global test, and we can use Bonferroni-corrected $p$-values from the first local hypothesis test to address this issue. Beyond testing zero direct treatment effects, our results also allow us to test if the direct treatment effect equals a particular value.

The third type of null hypothesis is to test if the direct treatment effect is the same for two specific vectors of number of treated neighbors $\mathbf{g}$ and $\mathbf{g}'$, i.e.

$$\mathcal{H}_0 : \beta_j(\mathbf{g}) - \beta_j(\mathbf{g}') = 0 \text{ for fixed } \mathbf{g} \text{ and } \mathbf{g}'$$

We need to generalize Theorem 2 to estimate the variance of $\beta_j(\mathbf{g}) - \beta_j(\mathbf{g}')$ and construct the test statistic, but it is a straightforward extension. Given the influence function $\phi_{j,\mathbf{g}}(\mathbf{Y}_c, \mathbf{Z}_c, \mathbf{X}_c)$ of $\beta_j(\mathbf{g})$ (provided in the proof of Theorem 2), the influence function for $\beta_j(\mathbf{g}) - \beta_j(\mathbf{g}')$ is $\phi_{j,\mathbf{g}}(\mathbf{Y}_c, \mathbf{Z}_c, \mathbf{X}_c) - \phi_{j,\mathbf{g}'}(\mathbf{Y}_c, \mathbf{Z}_c, \mathbf{X}_c)$. The variance of this influence function is the asymptotic variance of $\beta_j(\mathbf{g}) - \beta_j(\mathbf{g}')$. Another related null hypothesis is

$$\mathcal{H}_0 : \beta_j(\mathbf{g}) - \beta_j(\mathbf{g}') = 0 \text{ for any } \mathbf{g} \text{ and } \mathbf{g}'$$

Rejecting this null hypothesis allows us to conclude that there is interference in the data.

Finally, for average direct effects $\beta_j(\mathbf{g})$, we may also be interested in the average effects across all units, i.e. the average of $\beta_j(\mathbf{g})$ weighted by the fraction of units in subset $j$:

$$\mathcal{H}_0 : \sum_j \frac{|\mathcal{I}_j|}{n} \beta_j(\mathbf{g}) = 0 \text{ for fixed } \mathbf{g}$$

or in comparisons between $\beta_j(\mathbf{g})$ and $\beta_{j'}(\mathbf{g})$:

$$\mathcal{H}_0 : \beta_j(\mathbf{g}) - \beta_{j'}(\mathbf{g}) = 0 \text{ for fixed } \mathbf{g}$$

which allows us to test whether effects are heterogeneous across different subsets of units. Moreover, it

provides evidence on the validity of a particular partial exchangeability specification. Such null hypotheses involving multiple $j$'s can be tested easily using feasible variance estimators, since units in different subsets $\mathcal{I}_j$ and $\mathcal{I}_{j'}$ are conditionally independent, so that the asymptotic variance is simply the (weighted) sum of individual asymptotic variances $V_{\beta_j}(\mathbf{g})$ over $j$.

For the average spillover effects we have analogous null hypotheses. First, we can test if $\tau_j(z, \mathbf{g}, \mathbf{g}')$ is zero given specific treatment status $z$ and vectors of number of treated neighbors $\mathbf{g}$ and $\mathbf{g}'$, i.e.

$$\mathcal{H}_0 : \tau_j(z, \mathbf{g}, \mathbf{g}') = 0 \text{ for fixed } z, \mathbf{g}, \text{ and } \mathbf{g}'$$

A particularly interesting spillover effect is $\tau_j(0, \{|\mathcal{I}_j|\}_{j=1}^m, \mathbf{0}_m)$, where $\mathbf{0}_m$ is a vector of zeros with length $m$. $\tau_j(0, \{|\mathcal{I}_j|\}_{j=1}^m, \mathbf{0}_m)$ measures the spillovers of switching all neighbors from control to treatment, with ego not exposed to treatment directly.

The second null hypothesis is the (partial) global analog of the previous null hypothesis to test if the spillover effect is zero for a set of neighbors' treatments, i.e.

$$\mathcal{H}_0 : \tau_j(z, \mathbf{g}, \mathbf{g}') = 0 \text{ for } (\mathbf{g}, \mathbf{g}') \in \mathcal{G}$$

Here $\mathcal{G}$ could be the set of $(\mathbf{g}, \mathbf{g}')$ that satisfies $\|\mathbf{g}\|_1 = g_0, \|\mathbf{g}'\|_1 = g_1$, i.e. we are interested in whether the average spillover effect of having $g_0$ treated neighbors vs. having $g_1$ treated neighbors is always equal to 0 regardless of the exact treatment configuration. Alternatively, we may be interested in testing whether the average spillover effect is zero beyond a certain number $g_0$ of treated neighbors. In this case, $\mathcal{G}$ is the set of $(\mathbf{g}, \mathbf{g}')$ that satisfies $\|\mathbf{g}\|_1 \geq g_0, \|\mathbf{g}'\|_1 = g_0$, and the corresponding null hypothesis is written more concretely as

$$\mathcal{H}_0 : \tau_j(z, \mathbf{g}, \mathbf{g}_0) - \tau_j(z, \mathbf{g}', \mathbf{g}_0) = \tau_j(z, \mathbf{g}, \mathbf{g}') = 0 \text{ for } \|\mathbf{g}\|_1 \geq g_0, \|\mathbf{g}'\|_1 = g_0$$

Similar as before, we can use Bonferroni correction to address the multiple hypothesis testing issues for hypotheses involving multiple $\mathbf{g}, \mathbf{g}'$.

Finally, null hypotheses involving linear combinations of multiple $j$ can be tested using (weighted) sums of feasible variance estimators for each $\hat{\tau}_j$, since conditional independence implies additivity of asymptotic variances.

# Appendix D    Proofs

**Notations**. $|\cdot|$ denotes the Frobenius matrix norm $|A| = \sqrt{\text{trace}(A^\top A)}$ and $\|\cdot\|_\infty$ denotes the sup-norm in all arguments for functions.

## D.1    Proof of Lemma 1

*Proof of Lemma 1.* By definition and from Assumption 5, for any permutation $\pi$,

$$\begin{aligned}
&\mathbb{E}[Y_{c,i}(z_{c,i}, \mathbf{z}_{c,(i)}) \mid X_{c,i}, \mathbf{X}_{c,(i),1}, \cdots, \mathbf{X}_{c,(i),m}] \\
=&\mathbb{E}[Y_{c,i}(z_{c,i}, \mathbf{z}_{c,(i)}) \mid X_{c,i}, \pi_1(\mathbf{X}_{c,(i),1}), \cdots, \pi_m(\mathbf{X}_{c,(i),m})] \\
=&\mathbb{E}[Y_{c,i}(z_{c,i}, \pi_1(\mathbf{z}_{c,(i),1}), \cdots, \pi_m(\mathbf{z}_{c,(i),m})) \mid X_{c,i}, \pi_1(\mathbf{X}_{c,(i),1}), \cdots, \pi_m(\mathbf{X}_{c,(i),m})] \\
=&\mathbb{E}[Y_{c,i}(z_{c,i}, \pi_1(\mathbf{z}_{c,(i),1}), \cdots, \pi_m(\mathbf{z}_{c,(i),m})) \mid X_{c,i}, \mathbf{X}_{c,(i)}].
\end{aligned}$$

48

The last inequality implies that only the number of treated neighbors in each subset matters instead of which neighborhood units in each subset are treated. Therefore,

$$\mathbb{E}[Y_{c,i}(z_{c,i}, \mathbf{z}_{c,(i)}) \mid X_{c,i}, \mathbf{X}_{c,(i)}] = \mathbb{E}[Y_{c,i}(z_{c,i}, \|\mathbf{z}_{c,(i),1}\|_1, \cdots, \|\mathbf{z}_{c,(i),m}\|_1) \mid X_{c,i}, \mathbf{X}_{c,(i)}],$$

and the desired result follows.

□

## D.2  Proof of Proposition 1 for General $\psi_j(z, \mathbf{g}) - \psi_j(z', \mathbf{g}')$

*Proof of Proposition 1.* In this proof, we show the unbiasedness and consistency of $\psi_j^{\text{aipw}}(z, \mathbf{g}) - \psi_j^{\text{aipw}}(z', \mathbf{g}')$ for every subset $j \in \{1, \cdots, m\}$. Then as special cases, the unbiasedness and consistency of $\beta_j^{\text{aipw}}(\mathbf{g})$ and $\tau_j^{\text{aipw}}(z, \mathbf{g}, \mathbf{g}')$ directly follow. Recall the definition

$$\psi_j^{\text{aipw}}(z, \mathbf{g}) - \psi_j^{\text{aipw}}(z', \mathbf{g}') := \frac{1}{M|\mathcal{I}_j|} \sum_{c=1}^M \sum_{i \in \mathcal{I}_j} \left( \frac{\mathbf{1}\{Z_{c,i} = z, \mathbf{G}_{c,i} = \mathbf{g}\}(Y_{c,i} - \mu_{i,(z,\mathbf{g})}(\mathbf{X}_c))}{p_{i,(z,\mathbf{g})}(\mathbf{X}_c)} \right.$$
$$\left. - \frac{\mathbf{1}\{Z_{c,i} = z', \mathbf{G}_{c,i} = \mathbf{g}'\}(Y_{c,i} - \mu_{i,(z',\mathbf{g}')}(\mathbf{X}_c))}{p_{i,(z',\mathbf{g}')}(\mathbf{X}_c)} + \mu_{i,(z,\mathbf{g})}(\mathbf{X}_c) - \mu_{i,(z',\mathbf{g}')}(\mathbf{X}_c) \right)$$

Note that for all $z$, $\mathbf{g}$, $c$, and $i$,

$$\mathbb{E}\left[ \frac{\mathbf{1}\{Z_{c,i} = z, \mathbf{G}_{c,i} = \mathbf{g}\}(Y_{c,i} - \mu_{i,(z,\mathbf{g})}(\mathbf{X}_c))}{p_{i,(z,\mathbf{g})}(\mathbf{X}_c)} + \mu_{i,(z,\mathbf{g})}(\mathbf{X}_c) \right]$$
$$= \mathbb{E}\left[ \frac{\mathbf{1}\{Z_{c,i} = z, \mathbf{G}_{c,i} = \mathbf{g}\} Y_{c,i}}{p_{i,(z,\mathbf{g})}(\mathbf{X}_c)} \right] + \mathbb{E}\left[ \left(1 - \frac{\mathbf{1}\{Z_{c,i} = z, \mathbf{G}_{c,i} = \mathbf{g}\}}{p_{i,(z,\mathbf{g})}(\mathbf{X}_c)}\right) \mu_{i,(z,\mathbf{g})}(\mathbf{X}_c) \right]$$
$$= \mathbb{E}\left[ Y_{c,i}(z, \mathbf{g}) \right] + 0 = \mathbb{E}\left[ Y_{c,i}(z, \mathbf{g}) \right]$$

Then $\psi_j^{\text{aipw}}(z, \mathbf{g})$ is unbiased and same for $\psi_j^{\text{aipw}}(z', \mathbf{g}')$, and then their difference is unbiased, i.e.

$$\mathbb{E}[\psi_j^{\text{aipw}}(z, \mathbf{g}) - \psi_j^{\text{aipw}}(z', \mathbf{g}')] = \mathbb{E}[Y_{c,i}(z, \mathbf{g}) - Y_{c,i}(z', \mathbf{g}')] = \psi_j(z, \mathbf{g}) - \psi_j(z', \mathbf{g}').$$

Since clusters are i.i.d., we have

$$\text{Var}\left(\psi_j^{\text{aipw}}(z, \mathbf{g}) - \psi_j^{\text{aipw}}(z', \mathbf{g}')\right) = \frac{1}{M^2 |\mathcal{I}_j|^2} \sum_{c=1}^M \sum_{i,i' \in \mathcal{I}_j}$$
$$\text{Cov}\left( \frac{\mathbf{1}\{Z_{c,i} = z, \mathbf{G}_{c,i} = \mathbf{g}\}(Y_{c,i} - \mu_{i,(z,\mathbf{g})}(\mathbf{X}_c))}{p_{i,(z,\mathbf{g})}(\mathbf{X}_c)} - \frac{\mathbf{1}\{Z_{c,i} = z', \mathbf{G}_{c,i} = \mathbf{g}'\}(Y_{c,i} - \mu_{i,(z',\mathbf{g}')}(\mathbf{X}_c))}{p_{i,(z',\mathbf{g}')}(\mathbf{X}_c)} \right.$$
$$+ \mu_{i,(z,\mathbf{g})}(\mathbf{X}_c) - \mu_{i,(z',\mathbf{g}')}(\mathbf{X}_c), \frac{\mathbf{1}\{Z_{c,i'} = z, \mathbf{G}_{c,i'} = \mathbf{g}\}(Y_{c,i'} - \mu_{i',(z,\mathbf{g})}(\mathbf{X}_c))}{p_{i',(z,\mathbf{g})}(\mathbf{X}_c)}$$
$$\left. - \frac{\mathbf{1}\{Z_{c,i'} = z', \mathbf{G}_{c,i'} = \mathbf{g}'\}(Y_{c,i'} - \mu_{i',(z',\mathbf{g}')}(\mathbf{X}_c))}{p_{i',(z',\mathbf{g}')}(\mathbf{X}_c)} + \mu_{i',(z,\mathbf{g})}(\mathbf{X}_c) - \mu_{i',(z',\mathbf{g}')}(\mathbf{X}_c) \right)$$
$$= O\left( \frac{M|\mathcal{I}_j|^2}{M^2|\mathcal{I}_j|^2} \right) = O\left( \frac{1}{M} \right).$$

By Chebyshev's inequality, for any $t > 0$,

$$P(|\psi_j^{\text{aipw}}(z, \mathbf{g}) - \psi_j^{\text{aipw}}(z', \mathbf{g}')| > t) \leq \frac{\text{Var}(\psi_j^{\text{aipw}}(z, \mathbf{g}) - \psi_j^{\text{aipw}}(z', \mathbf{g}'))}{t^2} = O\left(\frac{1}{M}\right).$$

Therefore,

$$\psi_j^{\text{aipw}}(z, \mathbf{g}) - \psi_j^{\text{aipw}}(z', \mathbf{g}') \xrightarrow{P} \psi_j(z, \mathbf{g}) - \psi_j(z', \mathbf{g}').$$

$\square$

## D.3 Proof of Theorem 1 for General $\psi_j(z, \mathbf{g}) - \psi_j(z', \mathbf{g}')$

**Lemma 2** (Uniform Rate of Convergence of MLSE, Adapted from Theorem B-1 in Cattaneo (2010) and Newey (1997)). *Suppose Assumptions 1-7 hold. Let $\boldsymbol{\gamma}_{i,K}$ be the population parameters in the sieve estimator, $\hat{\boldsymbol{\gamma}}_{i,K}$ be the sieve estimators, $p_{i,K,(z,\mathbf{g})}^0$ be the probability with the population parameters in the sieve estimator and $\hat{p}_{i,(z,\mathbf{g})}$ be the estimated probability. Then for $i = 1, \cdots, n$ and for all $(z, \mathbf{g})$*

1. $\left\|p_{i,K,(z,\mathbf{g})}^0 - p_{i,(z,\mathbf{g})}\right\|_\infty = O(K^{-s/k_x})$

2. $|\hat{\boldsymbol{\gamma}}_{i,K} - \boldsymbol{\gamma}_{i,K}| = O_P(K^{1/2}M^{-1/2} + K^{1/2}K^{-s/k_x})$

3. $\left\|\hat{p}_{i,(z,\mathbf{g})} - p_{i,(z,\mathbf{g})}\right\|_\infty = O_P(\zeta(K)K^{1/2}M^{-1/2} + \zeta(K)K^{1/2}K^{-s/k_x})$

*Proof of Theorem 1.* In this proof, we show the unbiasedness and consistency of $\hat{\psi}_j^{\text{aipw}}(z, \mathbf{g}) - \hat{\psi}_j^{\text{aipw}}(z', \mathbf{g}')$ for every subset $j \in \{1, \cdots, m\}$. Then as special cases, the unbiasedness and consistency of $\hat{\beta}_j^{\text{aipw}}(\mathbf{g})$ and $\hat{\tau}_j^{\text{aipw}}(z, \mathbf{g}, \mathbf{g}')$ directly follow.

We show the consistency of the AIPW estimator $\hat{\psi}_j^{\text{aipw}}(z, \mathbf{g}) - \hat{\psi}_j^{\text{aipw}}(z', \mathbf{g}')$ if either the propensity or outcome is estimated from the nonparametric series estimator, i.e. if either the estimated propensity or outcome converges uniformly in probability.

If we estimate the conditional outcome $\hat{\mu}_{i,(z,\mathbf{g})}(\mathbf{x})$ by the nonparametric series estimator (26), from Newey (1997), we have

$$\sup_{\mathbf{x} \in \mathcal{X}} |\hat{\mu}_{i,(z,\mathbf{g})}(\mathbf{x}) - \mu_{i,(z,\mathbf{g})}(\mathbf{x})| = O_P(K^\eta K^{1/2}M^{-1/2} + K^\eta K^{1/2}K^{-s/k_x}) = o_p(1).$$

Similarly, if we use ordinary least squares where covariates are $\mathbf{X}_c$ (without any transformation), we also have $\sup_{\mathbf{x} \in \mathcal{X}} |\hat{\mu}_{i,(z,\mathbf{g})}(\mathbf{x}) - \mu_{i,(z,\mathbf{g})}(\mathbf{x})| = o_p(1)$.

If the estimated propensity $\hat{p}_{i,(z,\mathbf{g})}(\mathbf{X}_c)$ converges uniformly in probability, then the difference between

$\hat{\psi}_j^{\text{aipw}}(z, \mathbf{g}) - \hat{\psi}_j^{\text{aipw}}(z', \mathbf{g}')$ and $\psi_j^{\text{aipw}}(z, \mathbf{g}) - \psi_j^{\text{aipw}}(z', \mathbf{g}')$ can be bounded as

$$\left| (\hat{\psi}_j^{\text{aipw}}(z, \mathbf{g}) - \hat{\psi}_j^{\text{aipw}}(z', \mathbf{g}')) - (\psi_j^{\text{aipw}}(z, \mathbf{g}) - \psi_j^{\text{aipw}}(z', \mathbf{g}')) \right|$$

$$\leq \frac{2}{|\mathcal{I}_j|} \sum_{i \in \mathcal{I}_j} \max_{z, \mathbf{g}} \left[ \underbrace{\left| \frac{1}{M} \sum_{c=1}^{M} \left( \frac{\mathbf{1}\{Z_{c,i} = z, \mathbf{G}_{c,i} = \mathbf{g}\}}{\hat{p}_{i,(z,\mathbf{g})}(\mathbf{X}_c)} - \frac{\mathbf{1}\{Z_{c,i} = z, \mathbf{G}_{c,i} = \mathbf{g}\}}{p_{i,(z,\mathbf{g})}(\mathbf{X}_c)} \right) Y_{c,i} \right|}_{(1a)} \right.$$

$$+ \underbrace{\left| \frac{1}{M} \sum_{c=1}^{M} \left( \frac{\mathbf{1}\{Z_{c,i} = z, \mathbf{G}_{c,i} = \mathbf{g}\}}{\hat{p}_{i,(z,\mathbf{g})}(\mathbf{X}_c)} - \frac{\mathbf{1}\{Z_{c,i} = z, \mathbf{G}_{c,i} = \mathbf{g}\}}{p_{i,(z,\mathbf{g})}(\mathbf{X}_c)} \right) \hat{\mu}_{i,(z,\mathbf{g})}(\mathbf{X}_c) \right|}_{(1b)}$$

$$+ \underbrace{\left| \frac{1}{M} \sum_{c=1}^{M} \left( 1 - \frac{\mathbf{1}\{Z_{c,i} = z, \mathbf{G}_{c,i} = \mathbf{g}\}}{p_{i,(z,\mathbf{g})}(\mathbf{X}_c)} \right) \hat{\mu}_{i,(z,\mathbf{g})}(\mathbf{X}_c) \right|}_{(1c)}$$

$$+ \underbrace{\left| \frac{1}{M} \sum_{c=1}^{M} \left( 1 - \frac{\mathbf{1}\{Z_{c,i} = z, \mathbf{G}_{c,i} = \mathbf{g}\}}{p_{i,(z,\mathbf{g})}(\mathbf{X}_c)} \right) \mu_{i,(z,\mathbf{g})}(\mathbf{X}_c) \right|}_{(1d)} \right]$$

For (1a), since $|Y_{c,i}(z, \mathbf{g})| < B$, we have

$$(1a) \leq \max_{\mathbf{X}_c} \left| \frac{\mathbf{1}\{Z_{c,i} = z, \mathbf{G}_{c,i} = \mathbf{g}\} Y_{c,i} \big( p_{i,(z,\mathbf{g})}(\mathbf{X}_c) - \hat{p}_{i,(z,\mathbf{g})}(\mathbf{X}_c) \big)}{\hat{p}_{i,(z,\mathbf{g})}(\mathbf{X}_c) p_{i,(z,\mathbf{g})}(\mathbf{X}_c)} \right|$$

$$\leq B \cdot \max_{\mathbf{X}_c} \left| \frac{p_{i,(z,\mathbf{g})}(\mathbf{X}_c) - \hat{p}_{i,(z,\mathbf{g})}(\mathbf{X}_c)}{\underline{p}^2} \right| + o_P(1) = o_P(1)$$

following the uniform convergence of $\hat{p}_{i,(z,\mathbf{g})}(\mathbf{X}_c)$ in probability. Similarly, by the same argument, we can show that (1b) $= o_P(1)$ following the uniform convergence of $\hat{p}_{i,(z,\mathbf{g})}(\mathbf{X}_c)$ in probability. For (1c), we have $\mathbb{E}[(1c)] = \frac{1}{M} \sum_{c=1}^{M} \mathbb{E}\left[ \mathbb{E}\left[ 1 - \frac{\mathbf{1}\{Z_{c,i}=z, \mathbf{G}_{c,i}=\mathbf{g}\}}{p_{i,(z,\mathbf{g})}(\mathbf{X}_c)} \Big| \mathbf{X}_c \right] \hat{\mu}_{i,(z,\mathbf{g})}(\mathbf{X}_c) \right] = 0$. Furthermore, by the same argument used to show $\text{Var}(\psi_j^{\text{aipw}}(z, \mathbf{g}) - \psi_j^{\text{aipw}}(z', \mathbf{g}')) = O\big(\frac{1}{M}\big)$, it is verified that $\text{Var}((1c)) = O\big(\frac{1}{M}\big)$ and therefore, (1c) $= o_P(1)$. With the same argument as the one to show (1c) $= o_P(1)$, we have (1d) $= o_P(1)$. Since $n$ is finite, we have

$$\left| (\hat{\psi}_j^{\text{aipw}}(z, \mathbf{g}) - \hat{\psi}_j^{\text{aipw}}(z', \mathbf{g}')) - (\psi_j^{\text{aipw}}(z, \mathbf{g}) - \psi_j^{\text{aipw}}(z', \mathbf{g}')) \right| = o_P(1).$$

If the estimated outcome $\hat{\mu}_{i,(z,\mathbf{g})}(\mathbf{X}_c)$ converges uniformly in probability, then the difference between

$\hat{\psi}_j^{\text{aipw}}(z, \mathbf{g}) - \hat{\psi}_j^{\text{aipw}}(z', \mathbf{g}')$ and $\psi_j^{\text{aipw}}(z, \mathbf{g}) - \psi_j^{\text{aipw}}(z', \mathbf{g}')$ can be bounded as

$$\left| (\hat{\psi}_j^{\text{aipw}}(z, \mathbf{g}) - \hat{\psi}_j^{\text{aipw}}(z', \mathbf{g}')) - (\psi_j^{\text{aipw}}(z, \mathbf{g}) - \psi_j^{\text{aipw}}(z', \mathbf{g}')) \right|$$

$$\leq \frac{2}{n} \sum_{i=1}^n \max_{z,g} \Bigg[ \Bigg| \underbrace{\frac{1}{M} \sum_{c=1}^M \frac{\mathbf{1}\{Z_{c,i} = z, \mathbf{G}_{c,i} = \mathbf{g}\}(\hat{\mu}_{i,(z,\mathbf{g})}(\mathbf{X}_c) - \mu_{i,(z,\mathbf{g})}(\mathbf{X}_c))}{\hat{p}_{i,(z,\mathbf{g})}(\mathbf{X}_c)}}_{(2a)} \Bigg|$$

$$+ \Bigg| \underbrace{\frac{1}{M} \sum_{c=1}^M \frac{\mathbf{1}\{Z_{c,i} = z, \mathbf{G}_{c,i} = \mathbf{g}\}(Y_{c,i} - \mu_{i,(z,\mathbf{g})}(\mathbf{X}_c))}{\hat{p}_{i,(z,\mathbf{g})}(\mathbf{X}_c)}}_{(2b)} \Bigg|$$

$$+ \Bigg| \underbrace{\frac{1}{M} \sum_{c=1}^M (\hat{\mu}_{i,(z,\mathbf{g})}(\mathbf{X}_c) - \mu_{i,(z,\mathbf{g})}(\mathbf{X}_c))}_{(2c)} \Bigg|$$

$$+ \Bigg| \underbrace{\frac{1}{M} \sum_{c=1}^M \frac{\mathbf{1}\{Z_{c,i} = z, \mathbf{G}_{c,i} = \mathbf{g}\}(Y_{c,i} - \mu_{i,(z,\mathbf{g})}(\mathbf{X}_c))}{p_{i,(z,\mathbf{g})}(\mathbf{X}_c)}}_{(2d)} \Bigg| \Bigg]$$

By the same argument used to show (1a) = $o_P(1)$, it is verified that (2a) = $o_P(1)$ and (2c) = $o_P(1)$ following the uniform convergence of $\hat{\mu}_{i,(z,\mathbf{g})}(\mathbf{X}_c)$. For (2b), we have $\mathbb{E}[(1b)] = \frac{1}{M} \sum_{c=1}^M \mathbb{E}\left[\frac{p_{i,(z,\mathbf{g})}(\mathbf{X}_c)}{\hat{p}_{i,(z,\mathbf{g})}(\mathbf{X}_c)} \mathbb{E}\left[Y_{c,i} - \mu_{i,(z,\mathbf{g})}(\mathbf{X}_c) \Big| \mathbf{X}_c\right]\right] = 0$. Furthermore, by the same argument used to show $\text{Var}(\psi_j^{\text{aipw}}(z, \mathbf{g}) - \psi_j^{\text{aipw}}(z', \mathbf{g}')) = O(\frac{1}{M})$, it is verified that $\text{Var}((2b)) = O(\frac{1}{M})$ and therefore, (2b) = $o_P(1)$. With the same argument as the one to show (2b) = $o_P(1)$, we have (2d) = $o_P(1)$. Since $n$ is finite, we have

$$\left| (\hat{\psi}_j^{\text{aipw}}(z, \mathbf{g}) - \hat{\psi}_j^{\text{aipw}}(z', \mathbf{g}')) - (\psi_j^{\text{aipw}}(z, \mathbf{g}) - \psi_j^{\text{aipw}}(z', \mathbf{g}')) \right| = o_P(1).$$

Together with the consistency of $\psi_j^{\text{aipw}}(z, \mathbf{g}) - \psi_j^{\text{aipw}}(z', \mathbf{g}')$ from Proposition 1, we have

$$\hat{\psi}_j^{\text{aipw}}(z, \mathbf{g}) - \hat{\psi}_j^{\text{aipw}}(z', \mathbf{g}') \xrightarrow{P} \psi_j(z, \mathbf{g}) - \psi_j(z', \mathbf{g}').$$

□

### D.4 Proof of Theorem 6 (Semiparametric Efficiency Bound)

*Proof of Theorem 6 (Semiparametric Efficiency Bound).* The derivation of the semiparametric efficiency bound in Theorem 6 has two steps. The first step is to provide the influence function for $\hat{\psi}_j^{\text{aipw}}(z, \mathbf{g}) - \hat{\psi}_j^{\text{aipw}}(z', \mathbf{g}')$ that satisfies the assumptions in Newey (1994). The second step is to compute the asymptotic variance of the influence function, which is the semiparametric bound. For the first step, since units in a cluster are heterogeneous and dependent, it is not feasible to find an influence function of a unit, as is commonly considered in the literature (e.g. Hahn (1998); Hirano et al. (2003)), in order to satisfy the assumptions in Newey (1994). Instead, given clusters are i.i.d., we consider the influence function of a *cluster* that takes the

form

$$\phi_{j,(z,z',\mathbf{g},\mathbf{g}')}(\mathbf{Y}_c, \mathbf{Z}_c, \mathbf{X}_c)$$
$$= \frac{1}{|\mathcal{I}_j|} \sum_{i \in \mathcal{I}_j} \left( \frac{\mathbf{1}\{Z_{c,i} = z, \mathbf{G}_{c,i} = \mathbf{g}\}(Y_{c,i} - \mu_{i,(z,\mathbf{g})}(\mathbf{X}_c))}{p_{i,(z,\mathbf{g})}(\mathbf{X}_c)} - \frac{\mathbf{1}\{Z_{c,i} = z', \mathbf{G}_{c,i} = \mathbf{g}'\}(Y_{c,i} - \mu_{i,(z',\mathbf{g}')}(\mathbf{X}_c))}{p_{i,(z',\mathbf{g}')}(\mathbf{X}_c)} \right) \tag{29}$$

$$+ \frac{1}{|\mathcal{I}_j|} \sum_{i \in \mathcal{I}_j} \left( \mu_{i,(z,\mathbf{g})}(\mathbf{X}_c) - \mu_{i,(z',\mathbf{g}')}(\mathbf{X}_c) \right) - \left( \mu_{i,(z,\mathbf{g})} - \mu_{i,(z',\mathbf{g}')} \right). \tag{30}$$

We will verify $\phi_{j,(z,z',\mathbf{g},\mathbf{g}')}(\mathbf{Y}_c, \mathbf{Z}_c, \mathbf{X}_c)$ is a valid influence function for $\psi_j(z, \mathbf{g}) - \psi_j(z', \mathbf{g}')$ and show that $\frac{1}{\sqrt{M}} \sum_{c=1}^{M} \phi_{j,(z,z',\mathbf{g},\mathbf{g}')}(\mathbf{Y}_c, \mathbf{Z}_c, \mathbf{X}_c)$ is asymptotically normal with variance $V_{j,z,z',\mathbf{g},\mathbf{g}'}$.

**Step 1** We first write down the probability density of a cluster and form the corresponding score function, then find and verify a valid influence function. The density of $(\mathbf{Y}_c, \mathbf{Z}_c, \mathbf{X}_c)$ is equal to (the factorization is a result of condition (11))

$$\mathcal{L}(\mathbf{Y}_c, \mathbf{Z}_c, \mathbf{X}_c) = \prod_{\mathbf{z}} [\prod_i f_{i,(z_i,\mathbf{g})}(Y_i|X_i, \mathbf{X}_{(i)}) p_{\mathbf{z}}(\mathbf{X}_c)]^{\mathbf{1}\{\mathbf{Z}_c = \mathbf{z}\}} f(\mathbf{X}_c),$$

where $p_{\mathbf{z}}(\mathbf{x}) = P(\mathbf{Z}_c = \mathbf{z} \mid \mathbf{X}_c = \mathbf{x})$, $f(\mathbf{x}) = P(\mathbf{X}_c = \mathbf{x})$, and $f_{i,(z,\mathbf{g})}(y|x_i, \mathbf{x}_{(i)}) = P(Y_i(z, \mathbf{g}) = y \mid x_i, \mathbf{x}_{(i)})$ are understood to be probability densities. Consider a regular parametric submodel specified by the following density, with $\theta \in \Theta \subset \mathbb{R}^K$ for some $K < \infty$:

$$\mathcal{L}_s(\mathbf{Y}_c, \mathbf{Z}_c, \mathbf{X}_c; \theta) = \prod_{\mathbf{z}} [\prod_i f_{i,(z_i,\mathbf{g})}(Y_i|X_i, \mathbf{X}_{(i)}; \theta) p_{\mathbf{z}}(\mathbf{X}_c; \theta)]^{\mathbf{1}\{\mathbf{Z}_c = \mathbf{z}\}} f(\mathbf{X}_c; \theta),$$

and assume that it is equal to $\mathcal{L}$ when $\theta = \theta_0$. The corresponding score function $s(\mathbf{Y}_c, \mathbf{Z}_c, \mathbf{X}_c; \theta) = \partial_\theta \log \mathcal{L}_s(\mathbf{Y}_c, \mathbf{Z}_c, \mathbf{X}_c; \theta)$ is given by

$$s(\mathbf{Y}_c, \mathbf{Z}_c, \mathbf{X}_c; \theta) = \sum_{\mathbf{z}} \mathbf{1}\{\mathbf{Z}_c = \mathbf{z}\} s_{\mathbf{z}}(\mathbf{Y}_c|\mathbf{X}_c; \theta) + \sum_{\mathbf{z}} \frac{\mathbf{1}\{\mathbf{Z}_c = \mathbf{z}\}}{p_{\mathbf{z}}(\mathbf{X}_c; \theta)} \dot{p}_{\mathbf{z}}(\mathbf{X}_c; \theta) + t(\mathbf{X}_c; \theta)$$
$$= \sum_{\mathbf{z}} \mathbf{1}\{\mathbf{Z}_c = \mathbf{z}\} \left( \sum_i s_{i,(z_i,\mathbf{g})}(Y_i|\mathbf{X}_c; \theta) \right) + \sum_{\mathbf{z}} \frac{\mathbf{1}\{\mathbf{Z}_c = \mathbf{z}\}}{p_{\mathbf{z}}(\mathbf{X}_c; \theta)} \dot{p}_{\mathbf{z}}(\mathbf{X}_c; \theta) + t(\mathbf{X}_c; \theta)$$

where

$$s_{i,(z,\mathbf{g})}(y|\mathbf{X}_c; \theta) = \frac{\partial}{\partial \theta} \log f_{i,(z,\mathbf{g})}(y|\mathbf{X}_c; \theta), \quad \dot{p}_{\mathbf{z}}(\mathbf{X}_c; \theta) = \frac{\partial}{\partial \theta} p_{\mathbf{z}}(\mathbf{X}_c; \theta), \quad t(\mathbf{X}_c; \theta) = \frac{\partial}{\partial \theta} \log f(\mathbf{X}_c; \theta).$$

Now recall our estimand $\psi_j(z, \mathbf{g}) - \psi_j(z', \mathbf{g}') = \frac{1}{|\mathcal{I}_j|} \sum_{i \in \mathcal{I}_j} \mathbb{E}[Y_i(z, \mathbf{g}) - Y_i(z', \mathbf{g}')]$, and under the parametric submodel, define the corresponding parameterized estimand $\psi_j(z, z', \mathbf{g}, \mathbf{g}'; \theta) = \frac{1}{|\mathcal{I}_j|} \sum_{i \in \mathcal{I}_j} \psi_{ji}(z, z', \mathbf{g}, \mathbf{g}'; \theta)$ where

$$\psi_{j,i}(z, z', \mathbf{g}, \mathbf{g}'; \theta) := \int \int y_i f_{i,(z,\mathbf{g})}(y_i|\mathbf{x}_c; \theta) f(\mathbf{x}_c; \theta) dy_i d\mathbf{x}_c - \int \int y_i f_{i,(z',\mathbf{g}')}(y_i|\mathbf{x}_c; \theta) f(\mathbf{x}_c; \theta) dy_i d\mathbf{x}_c$$

Following Newey (1994), our task is to find an influence $\phi_{j,(z,z',\mathbf{g},\mathbf{g}')}(\mathbf{Y}_c, \mathbf{Z}_c, \mathbf{X}_c)$ such that

$$\frac{\partial \psi_j(z, z', \mathbf{g}, \mathbf{g}'; \theta_0)}{\partial \theta} = \mathbb{E}[\phi_{j,(z,z',\mathbf{g},\mathbf{g}')}(\mathbf{Y}_c, \mathbf{Z}_c, \mathbf{X}_c) \cdot s(\mathbf{Y}_c, \mathbf{Z}_c, \mathbf{X}_c; \theta_0)]. \tag{31}$$

The influence function that we propose is $\phi_{j,(z,z',\mathbf{g},\mathbf{g}')}(\mathbf{Y}_c, \mathbf{Z}_c, \mathbf{X}_c) = \frac{1}{|\mathcal{I}_j|}\sum_{i\in\mathcal{I}_j} \phi_{j,i,(z,z',\mathbf{g},\mathbf{g}')}(Y_i, Z_i, \mathbf{G}_{c,i}, \mathbf{X}_c)$ with

$$\begin{aligned}
&\phi_{j,i,(z,z',\mathbf{g},\mathbf{g}')}(Y_i, Z_i, \mathbf{G}_{c,i}, \mathbf{X}_c) \\
&:= \frac{\mathbf{1}\{Z_{c,i}=z, \mathbf{G}_{c,i}=\mathbf{g}\}(Y_{c,i} - \mu_{i,(z,\mathbf{g})}(\mathbf{X}_c))}{p_{i,(z,\mathbf{g})}(\mathbf{X}_c)} - \frac{\mathbf{1}\{Z_{c,i}=z', \mathbf{G}_{c,i}=\mathbf{g}\}(Y_{c,i} - \mu_{i,(z',\mathbf{g}')}(\mathbf{X}_c))}{p_{i,(z',\mathbf{g}')}(\mathbf{X}_c)} \\
&\quad + (\mu_{i,(z,\mathbf{g})}(\mathbf{X}_c) - \mu_{i,(z',\mathbf{g}')}(\mathbf{X}_c)) - (\mu_{i,(z,\mathbf{g})} - \mu_{i,(z',\mathbf{g}')}),
\end{aligned}$$

where $\mu_{i,(z,\mathbf{g})}(\mathbf{X}_c) = \mathbb{E}[Y_i(z,\mathbf{g}) \mid \mathbf{X}_c]$ and $\mu_{i,(z,\mathbf{g})} = \mathbb{E}[\mu_{i,(z,\mathbf{g})}(\mathbf{X}_c)]$. With this influence function, we now verify that

$$\frac{\partial \psi_{j,i}(z, z', \mathbf{g}, \mathbf{g}'; \theta_0)}{\partial \theta} = \mathbb{E}[\phi_{j,i,(z,z',\mathbf{g},\mathbf{g}')}(Y_i, Z_i, \mathbf{G}_{c,i}, \mathbf{X}_c) \cdot s(\mathbf{Y}_c, \mathbf{Z}_c, \mathbf{X}_c; \theta_0)],$$

from which we can conclude that (31) holds for $\phi_{j,(z,z',\mathbf{g},\mathbf{g}')}(\mathbf{Y}_c, \mathbf{Z}_c, \mathbf{X}_c)$ by linearity.

First, from the definition of $\psi_{j,i}(z, z', \mathbf{g}, \mathbf{g}'; \theta)$, we have

$$\begin{aligned}
\frac{\partial \psi_{j,i}(z, z', \mathbf{g}, \mathbf{g}'; \theta)}{\partial \theta} &= \int\int y_i s_{i,(z,\mathbf{g})}(y_i|\mathbf{x}_c;\theta) \cdot f_{i,(z,\mathbf{g})}(y_i|\mathbf{x}_c;\theta) \cdot f(\mathbf{x}_c;\theta) dy_i d\mathbf{x}_c \\
&\quad + \int\int y_i f_{i,(z,\mathbf{g})}(y_i|\mathbf{x}_c;\theta) \cdot t(\mathbf{x}_c;\theta) \cdot f(\mathbf{x}_c;\theta) dy_i d\mathbf{x}_c \\
&\quad - \int\int y_i s_{i,(z',\mathbf{g}')}(y_i|\mathbf{x}_c;\theta) \cdot f_{i,(z',\mathbf{g}')}(y_i|\mathbf{x}_c;\theta) \cdot f(\mathbf{x}_c;\theta) dy_i d\mathbf{x}_c \\
&\quad - \int\int y_i f_{i,(z',\mathbf{g}')}(y_i|\mathbf{x}_c;\theta) \cdot t(\mathbf{x}_c;\theta) \cdot f(\mathbf{x}_c;\theta) dy_i d\mathbf{x}_c
\end{aligned}$$

Decompose $\phi_{j,i,(z,z',\mathbf{g},\mathbf{g}')}(Y_i, Z_i, \mathbf{G}_{c,i}, \mathbf{X}_c) = \phi_{j,i,(z,z',\mathbf{g},\mathbf{g}'),1} + \phi_{j,i,(z,z',\mathbf{g},\mathbf{g}'),2} + \phi_{j,i,(z,z',\mathbf{g},\mathbf{g}'),3}$ and $s(\mathbf{Y}_c, \mathbf{Z}_c, \mathbf{X}_c; \theta_0) = S_1 + S_2 + S_3$ in their definitions. We will compute the expectations of all cross terms, starting with $\mathbb{E}[S_1 \cdot \phi_{j,i,(z,z',\mathbf{g},\mathbf{g}'),1}]$: note that

$$\begin{aligned}
\mathbb{E}[S_1 \cdot \phi_{j,i,(z,z',\mathbf{g},\mathbf{g}'),1}|\mathbf{X}_c] &= \mathbb{E}\left[\sum_{\mathbf{z}} \mathbf{1}\{\mathbf{Z}_c=\mathbf{z}\} s_{\mathbf{z}}(\mathbf{Y}_c|\mathbf{X}_c;\theta_0) \cdot \frac{\mathbf{1}\{Z_{c,i}=z, \mathbf{G}_{c,i}=\mathbf{g}\}(Y_{c,i} - \mu_{i,(z,\mathbf{g})}(\mathbf{X}_c))}{p_{i,(z,\mathbf{g})}(\mathbf{X}_c)}\bigg|\mathbf{X}_c\right] \\
&= \frac{\mathbb{E}[\mathbf{1}\{Z_{c,i}=z, \mathbf{G}_{c,i}=\mathbf{g}\}|\mathbf{X}_c]}{p_{i,(z,\mathbf{g})}(\mathbf{X}_c)} \mathbb{E}\left[\sum_{i'} s_{i',(z,\mathbf{g})}(Y_{i'}|\mathbf{X}_c;\theta_0) \cdot (Y_i(z,\mathbf{g}) - \mu_{i,(z,\mathbf{g})}(\mathbf{X}_c))\bigg|\mathbf{X}_c\right] \\
&= \mathbb{E}\left[s_{i,(z,\mathbf{g})}(Y_i|\mathbf{X}_c;\theta_0) Y_i(z,\mathbf{g})|\mathbf{X}_c\right],
\end{aligned}$$

using three properties:

1. For all $i$, $\mathbb{E}[s_{i,(z,\mathbf{g})}(Y_i|\mathbf{X}_c;\theta)|\mathbf{X}_c] = \int \frac{\partial}{\partial \theta}\log f_{i,(z,\mathbf{g})}(y|\mathbf{X}_c;\theta) \cdot f_{i,(z,\mathbf{g})}(y|\mathbf{X}_c;\theta) dy = \int \frac{\partial}{\partial \theta}f_{i,(z,\mathbf{g})}(y|\mathbf{X}_c;\theta) dy = \frac{\partial}{\partial \theta}\int f_{i,(z,\mathbf{g})}(y|\mathbf{X}_c;\theta) dy = \frac{\partial}{\partial \theta}1 = 0$

2. For $i \neq i'$, $\mathbb{E}[s_{i',(z,\mathbf{g})}(Y_{i'}|\mathbf{X}_c;\theta) \cdot (Y_i(z,\mathbf{g}) - \mu_{i,(z,\mathbf{g})}(\mathbf{X}_c))|\mathbf{X}_c] = \mathbb{E}[s_{i',(z,\mathbf{g})}(Y_{i'}|\mathbf{X}_c;\theta)|\mathbf{X}_c] \cdot \mathbb{E}[Y_i(z,\mathbf{g}) - \mu_{i,(z,\mathbf{g})}(\mathbf{X}_c)|\mathbf{X}_c] = 0$ from the assumption (11)

3. $\mathbb{E}[s_{i,(z,\mathbf{g})}(Y_i|\mathbf{X}_c;\theta)\mu_{i,(z,\mathbf{g})}(\mathbf{X}_c)|\mathbf{X}_c] = \mathbb{E}[s_{i,(z,\mathbf{g})}(Y_i|\mathbf{X}_c;\theta)|\mathbf{X}_c] \cdot \mu_{i,(z,\mathbf{g})}(\mathbf{X}_c) = 0$.

Then for the term $\mathbb{E}[S_1 \cdot \phi_{j,i,(z,z',\mathbf{g},\mathbf{g}'),1}]$, we have

$$\mathbb{E}[S_1 \cdot \phi_{j,i,(z,z',\mathbf{g},\mathbf{g}'),1}] = \int\int y_i s_{i,(z,\mathbf{g})}(y_i|\mathbf{x}_c;\theta_0) \cdot f_{i,(z,\mathbf{g})}(y_i|\mathbf{x}_c;\theta_0) \cdot f(\mathbf{x}_c;\theta_0) dy_i d\mathbf{x}_c.$$

In addition, for $\mathbb{E}[S_2 \cdot \phi_{j,i,(z,z',\mathbf{g},\mathbf{g}'),1}]$ and $\mathbb{E}[S_3 \cdot \phi_{j,i,(z,z',\mathbf{g},\mathbf{g}'),1}]$, we have conditional on $\mathbf{X}_c$,

$$\mathbb{E}[S_2 \cdot \phi_{j,i,(z,z',\mathbf{g},\mathbf{g}'),1}|\mathbf{X}_c] = \mathbb{E}\left[\sum_{\mathbf{z}}\frac{\mathbf{1}\{\mathbf{Z}_c = \mathbf{z}\}}{p_{\mathbf{z}}(\mathbf{X}_c;\theta_0)}\dot{p}_{\mathbf{z}}(\mathbf{X}_c;\theta_0) \cdot \frac{\mathbf{1}\{Z_{c,i} = z, \mathbf{G}_{c,i} = \mathbf{g}\}(Y_{c,i} - \mu_{i,(z,\mathbf{g})}(\mathbf{X}_c))}{p_{i,(z,\mathbf{g})}(\mathbf{X}_c)}\bigg|\mathbf{X}_c\right]$$
$$= \mathbb{E}\left[\frac{\dot{p}_{\mathbf{z}}(\mathbf{X}_c;\theta_0)\mathbf{1}\{Z_{c,i} = z, \mathbf{G}_{c,i} = \mathbf{g}\}}{p_{\mathbf{z}}(\mathbf{X}_c;\theta_0) \cdot p_{i,(z,\mathbf{g})}(\mathbf{X}_c)}\bigg|\mathbf{X}_c\right] \cdot \mathbb{E}[Y_i(z,\mathbf{g}) - \mu_{i,(z,\mathbf{g})}(\mathbf{X}_c)|\mathbf{X}_c] = 0$$

and

$$\mathbb{E}[S_3 \cdot \phi_{j,i,(z,z',\mathbf{g},\mathbf{g}'),1}|\mathbf{X}_c] = \mathbb{E}\left[t(\mathbf{X}_c;\theta_0) \cdot \frac{\mathbf{1}\{Z_{c,i} = z, \mathbf{G}_{c,i} = \mathbf{g}\}(Y_{c,i} - \mu_{i,(z,\mathbf{g})}(\mathbf{X}_c))}{p_{i,(z,\mathbf{g})}(\mathbf{X}_c)}\bigg|\mathbf{X}_c\right]$$
$$= \mathbb{E}\left[\frac{t(\mathbf{X}_c;\theta_0) \cdot \mathbf{1}\{Z_{c,i} = z, \mathbf{G}_{c,i} = \mathbf{g}\}}{p_{i,(z,\mathbf{g})}(\mathbf{X}_c)}\bigg|\mathbf{X}_c\right] \cdot \mathbb{E}\left[Y_i(z,\mathbf{g}) - \mu_{i,(z,\mathbf{g})}(\mathbf{X}_c)|\mathbf{X}_c\right] = 0$$

Then unconditional on $\mathbf{X}_c$, we have $\mathbb{E}[S_2 \cdot \phi_{j,i,(z,z',\mathbf{g},\mathbf{g}'),1}] = 0$ and $\mathbb{E}[S_3 \cdot \phi_{j,i,(z,z',\mathbf{g},\mathbf{g}'),1}] = 0$.

Similarly, we can show

$$\mathbb{E}[(S_1 + S_2 + S_3) \cdot \phi_{j,i,(z,z',\mathbf{g},\mathbf{g}'),2}] = \mathbb{E}[S_1 \cdot \phi_{j,i,(z,z',\mathbf{g},\mathbf{g}'),2}]$$
$$= -\int\int y_i s_{i,(z',\mathbf{g}')}(y_i|\mathbf{x}_c;\theta) \cdot f_{i,(z',\mathbf{g}')}(y_i|\mathbf{x}_c;\theta) \cdot f(\mathbf{x}_c;\theta) dy_i d\mathbf{x}_c$$

Finally, we consider terms involving $\phi_{j,i,(z,z',\mathbf{g},\mathbf{g}'),3}$. For the term $\mathbb{E}[S_1 \cdot \phi_{j,i,(z,z',\mathbf{g},\mathbf{g}'),3}]$, conditional on $\mathbf{X}_c$, we have

$$\mathbb{E}[S_1 \cdot \phi_{j,i,(z,z',\mathbf{g},\mathbf{g}'),3}|\mathbf{X}_c]$$
$$= \mathbb{E}\left[\sum_{\mathbf{z}}\mathbf{1}\{\mathbf{Z}_c = \mathbf{z}\}s_{\mathbf{z}}(\mathbf{Y}_c|\mathbf{X}_c;\theta) \cdot \left((\mu_{i,(z,\mathbf{g})}(\mathbf{X}_c) - \mu_{i,(z',\mathbf{g}')}(\mathbf{X}_c)) - (\mu_{i,(z,\mathbf{g})} - \mu_{i,(z',\mathbf{g}')})\right)\bigg|\mathbf{X}_c\right]$$
$$= \left((\mu_{i,(z,\mathbf{g})}(\mathbf{X}_c) - \mu_{i,(z',\mathbf{g}')}(\mathbf{X}_c)) - (\mu_{i,(z,\mathbf{g})} - \mu_{i,(z',\mathbf{g}')})\right) \cdot \sum_{\mathbf{z}}\mathbb{E}\left[\mathbf{1}\{\mathbf{Z}_c = \mathbf{z}\}s_{\mathbf{z}}(\mathbf{Y}_c|\mathbf{X}_c;\theta)|\mathbf{X}_c\right] = 0$$

where the last equality follows the unconfoundedness assumption (Assumption 3) and $\mathbb{E}\left[\mathbf{1}\{\mathbf{Z}_c = \mathbf{z}\}s_{\mathbf{z}}(\mathbf{Y}_c|\mathbf{X}_c;\theta)|\mathbf{X}_c\right] = \mathbb{E}[\mathbf{1}\{\mathbf{Z}_c = \mathbf{z}\}|\mathbf{X}_c] \cdot \mathbb{E}[s_{\mathbf{z}}(\mathbf{Y}_c|\mathbf{X}_c;\theta)|\mathbf{X}_c] = \mathbb{E}[\mathbf{1}\{\mathbf{Z}_c = \mathbf{z}\}|\mathbf{X}_c] \cdot 0 = 0$. Then unconditional on $\mathbf{X}_c$, we have $\mathbb{E}[S_1 \cdot \phi_{j,i,(z,z',\mathbf{g},\mathbf{g}'),3}] = 0$.

Next for the term $\mathbb{E}[S_2 \cdot \phi_{j,i,(z,z',\mathbf{g},\mathbf{g}'),3}]$, conditional on $\mathbf{X}_c$, we have

$$\mathbb{E}[S_2 \cdot \phi_{j,i,(z,z',\mathbf{g},\mathbf{g}'),3}|\mathbf{X}_c]$$
$$= \mathbb{E}\left[\sum_{\mathbf{z}}\frac{\mathbf{1}\{\mathbf{Z}_c = \mathbf{z}\}}{p_{\mathbf{z}}(\mathbf{X}_c;\theta)}\dot{p}_{\mathbf{z}}(\mathbf{X}_c;\theta) \cdot \left((\mu_{i,(z,\mathbf{g})}(\mathbf{X}_c) - \mu_{i,(z',\mathbf{g}')}(\mathbf{X}_c)) - (\mu_{i,(z,\mathbf{g})} - \mu_{i,(z',\mathbf{g}')})\right)\bigg|\mathbf{X}_c\right]$$
$$= \left((\mu_{i,(z,\mathbf{g})}(\mathbf{X}_c) - \mu_{i,(z',\mathbf{g}')}(\mathbf{X}_c)) - (\mu_{i,(z,\mathbf{g})} - \mu_{i,(z',\mathbf{g}')})\right) \cdot \sum_{\mathbf{z}}\mathbb{E}\left[\frac{\mathbf{1}\{\mathbf{Z}_c = \mathbf{z}\}}{p_{\mathbf{z}}(\mathbf{X}_c;\theta)}\dot{p}_{\mathbf{z}}(\mathbf{X}_c;\theta)\bigg|\mathbf{X}_c\right]$$
$$= \left((\mu_{i,(z,\mathbf{g})}(\mathbf{X}_c) - \mu_{i,(z',\mathbf{g}')}(\mathbf{X}_c)) - (\mu_{i,(z,\mathbf{g})} - \mu_{i,(z',\mathbf{g}')})\right) \cdot \sum_{\mathbf{z}}\dot{p}_{\mathbf{z}}(\mathbf{X}_c;\theta_0) = 0$$

where the last equality uses the property that $\sum_{\mathbf{z}} \dot{p}_{\mathbf{z}}(\mathbf{X}_c; \theta_0) = \sum_{\mathbf{z}} \frac{\partial}{\partial \theta} p_{\mathbf{z}}(\mathbf{X}_c; \theta_0) = \frac{\partial}{\partial \theta} \sum_{\mathbf{z}} p_{\mathbf{z}}(\mathbf{X}_c; \theta_0) = \frac{\partial}{\partial \theta} 1 = 0$. Then unconditional on $\mathbf{X}_c$, we have $\mathbb{E}[S_2 \cdot \phi_{j,i,(z,z',\mathbf{g},\mathbf{g}'),3}] = 0$.

Lastly,

$$\mathbb{E}[S_3 \cdot \phi_{j,i,(z,z',\mathbf{g},\mathbf{g}'),3}]$$
$$= \mathbb{E}[t(\mathbf{X}_c; \theta_0) \cdot (\mu_{i,(z,\mathbf{g})}(\mathbf{X}_c) - \mu_{i,(z',\mathbf{g}')}(\mathbf{X}_c))] - \mathbb{E}[t(\mathbf{X}_c; \theta_0)] \cdot (\mu_{i,(z,\mathbf{g})} - \mu_{i,(z',\mathbf{g}')})$$
$$= \mathbb{E}[t(\mathbf{X}_c; \theta_0) \cdot (\mu_{i,(z,\mathbf{g})}(\mathbf{X}_c) - \mu_{i,(z',\mathbf{g}')}(\mathbf{X}_c))] \qquad (32)$$
$$= \int t(\mathbf{x}_c, \theta) \left( \mathbb{E}[Y_i(z, \mathbf{g}) - Y_i(z', \mathbf{g}') | \mathbf{x}_c] \right) f(\mathbf{x}_c; \theta_0) d\mathbf{x}_c$$
$$= \int \int y_i f_{i,(z,\mathbf{g})}(y_i | \mathbf{x}_c; \theta_0) \cdot t(\mathbf{x}_c; \theta) \cdot f(\mathbf{x}_c; \theta_0) dy_i d\mathbf{x}_c - \int \int y_i f_{i,(z',\mathbf{g}')}(y_i | \mathbf{x}_c; \theta_0) \cdot t(\mathbf{x}_c; \theta_0) \cdot f(\mathbf{x}_c; \theta_0) dy_i d\mathbf{x}_c$$

where (32) follows from $\mathbb{E}[t(\mathbf{X}_c; \theta_0)] = \mathbb{E}[\frac{\partial}{\partial \theta} \log f(\mathbf{X}_c; \theta_0)] = \int \frac{\partial}{\partial \theta} \log f(\mathbf{X}_c; \theta_0) f(\mathbf{X}_c; \theta_0) d\mathbf{X}_c = \int \frac{\partial}{\partial \theta} f(\mathbf{X}_c; \theta_0) d\mathbf{X}_c = \frac{\partial}{\partial \theta} \int f(\mathbf{X}_c; \theta_0) d\mathbf{X}_c = \frac{\partial}{\partial \theta} 1 = 0$.

Summing up the expectations of the cross terms, we have

$$\frac{\partial \psi_{j,i}(z, z', \mathbf{g}, \mathbf{g}'; \theta_0)}{\partial \theta} = \mathbb{E}[\phi_{j,i,(z,z',\mathbf{g},\mathbf{g}')}(Y_i, Z_i, \mathbf{G}_{c,i}, \mathbf{X}_c) \cdot s(\mathbf{Y}_c, \mathbf{Z}_c, \mathbf{X}_c; \theta_0)],$$

as desired, and so our proposed influence function $\phi_{j,(z,z',\mathbf{g},\mathbf{g}')}(Y_i, Z_i, \mathbf{G}_{c,i}, \mathbf{X}_c)$ is a valid influence function for $\psi_j(z, \mathbf{g}) - \psi_j(z', \mathbf{g}')$, and its asymptotic variance is the semiparametric bound for $\psi_j(z, \mathbf{g}) - \psi_j(z', \mathbf{g}')$.

**Step 2** In this step, we would like to calculate the variance of the influence function $\phi_{j,(z,z',\mathbf{g},\mathbf{g}')}(\mathbf{Y}_c, \mathbf{Z}_c, \mathbf{X}_c)$. Using the unconfoundedness assumption (Assumption 3), and the definition of $\mu_{i,(z',\mathbf{g}')}(\mathbf{X}_c)$ and $\mu_{i,(z',\mathbf{g}')}$, we have $\mathbb{E}\left[\frac{\mathbf{1}\{Z_{c,i}=z, \mathbf{G}_{c,i}=\mathbf{g}\}(Y_{c,i} - \mu_{i,(z,\mathbf{g})}(\mathbf{X}_c))}{p_{i,(z,\mathbf{g})}(\mathbf{X}_c)}\right] = 0$ and $\mathbb{E}[\mu_{i,(z,\mathbf{g})}(\mathbf{X}_c)] = \mu_{i,(z,\mathbf{g})}$. Therefore, the influence function has mean 0, i.e. $\mathbb{E}[\phi_{j,(z,z',\mathbf{g},\mathbf{g}')}(\mathbf{Y}_c, \mathbf{Z}_c, \mathbf{X}_c)] = 0$.

Note that the covariance between the following terms are zero for any $i$ and $i'$ ($i$ and $i'$ can be the same) because

$$\mathbb{E}\left[\frac{\mathbf{1}\{Z_{c,i}=z, \mathbf{G}_{c,i}=\mathbf{g}\}(Y_{c,i} - \mu_{i,(z,\mathbf{g})}(\mathbf{X}_c))\left((\mu_{i',(z,\mathbf{g})}(\mathbf{X}_c) - \mu_{i',(z',\mathbf{g}')}(\mathbf{X}_c)) - (\mu_{i',(z,\mathbf{g})} - \mu_{i',(z',\mathbf{g}')})\right)}{p_{i,(z,\mathbf{g})}(\mathbf{X}_c)}\right]$$
$$= \mathbb{E}\left[\mathbb{E}\left[\frac{\mathbf{1}\{Z_{c,i}=z, \mathbf{G}_{c,i}=\mathbf{g}\}(Y_{c,i} - \mu_{i,(z,\mathbf{g})}(\mathbf{X}_c))((\mu_{i,(z,\mathbf{g})}(\mathbf{X}_c) - \mu_{i,(z',\mathbf{g}')}(\mathbf{X}_c)) - (\mu_{i,(z,\mathbf{g})} - \mu_{i,(z',\mathbf{g}')}))}{p_{i,(z,\mathbf{g})}(\mathbf{X}_c)}\bigg| \mathbf{X}_c\right]\right]$$
$$= \mathbb{E}\left[((\mu_{i,(z,\mathbf{g})}(\mathbf{X}_c) - \mu_{i,(z',\mathbf{g}')}(\mathbf{X}_c)) - (\mu_{i,(z,\mathbf{g})} - \mu_{i,(z',\mathbf{g}')}))\mathbb{E}\left[Y_{c,i}(z, \mathbf{g}) - \mu_{i,(z,\mathbf{g})}(\mathbf{X}_c) | \mathbf{X}_c\right]\right] = 0$$

Then in the influence function $\phi_{j,(z,z',\mathbf{g},\mathbf{g}')}(\mathbf{Y}_c, \mathbf{Z}_c, \mathbf{X}_c)$, the covariance between (29) and (30) is 0. Then the variance of $\phi_{j,(z,z',\mathbf{g},\mathbf{g}')}(\mathbf{Y}_c, \mathbf{Z}_c, \mathbf{X}_c)$ equals the sum of the variance of (29) and the variance of (30).

Let us first calculate the variance of (29). Note that we have $Y_{c,i}(Z_{c,i}, \mathbf{Z}_{c,(i)}) \perp Y_{c,i'}(Z_{c,i'}, \mathbf{Z}_{c,(i')}) | \mathbf{Z}_c, \mathbf{X}_c, \forall i \neq i'$, then conditional on $\mathbf{Z}_c, \mathbf{X}_c$, the covariance between $\frac{\mathbf{1}\{Z_{c,i}=z, \mathbf{G}_{c,i}=\mathbf{g}\}(Y_{c,i} - \mu_{i,(z,\mathbf{g})}(\mathbf{X}_c))}{p_{i,(z,\mathbf{g})}(\mathbf{X}_c)}$ and $\frac{\mathbf{1}\{Z_{c,i'}=z^\dagger, \mathbf{G}_{c,i'}=\mathbf{g}^\dagger\}(Y_{c,i'} - \mu_{i',(z^\dagger,\mathbf{g}^\dagger)}(\mathbf{X}_c))}{p_{i,(z^\dagger,\mathbf{g}^\dagger)}(\mathbf{X}_c)}$ is 0 for any $z, z^\dagger, \mathbf{g}, \mathbf{g}^\dagger$ if $i \neq i'$.

Then the variance of (29) equals the sum of the variance of $\frac{\mathbf{1}\{Z_{c,i}=z, \mathbf{G}_{c,i}=\mathbf{g}\}(Y_{c,i} - \mu_{i,(z,\mathbf{g})}(\mathbf{X}_c))}{p_{i,(z,\mathbf{g})}(\mathbf{X}_c)} - \frac{\mathbf{1}\{Z_{c,i}=z', \mathbf{G}_{c,i}=\mathbf{g}'\}(Y_{c,i} - \mu_{i,(z',\mathbf{g}')}(\mathbf{X}_c))}{p_{i,(z',\mathbf{g}')}(\mathbf{X}_c)}$

across all $i \in \mathcal{I}_j$.

$$\text{Var}\left(\frac{\mathbf{1}\{Z_{c,i} = z, \mathbf{G}_{c,i} = \mathbf{g}\}(Y_{c,i} - \mu_{i,(z,\mathbf{g})}(\mathbf{X}_c))}{p_{i,(z,\mathbf{g})}(\mathbf{X}_c)} - \frac{\mathbf{1}\{Z_{c,i} = z', \mathbf{G}_{c,i} = \mathbf{g}'\}(Y_{c,i} - \mu_{i,(z',\mathbf{g}')}(\mathbf{X}_c))}{p_{i,(z',\mathbf{g}')}(\mathbf{X}_c)}\right)$$

$$=\text{Var}\left(\frac{\mathbf{1}\{Z_{c,i} = z, \mathbf{G}_{c,i} = \mathbf{g}\}(Y_{c,i} - \mu_{i,(z,\mathbf{g})}(\mathbf{X}_c))}{p_{i,(z,\mathbf{g})}(\mathbf{X}_c)}\right) + \text{Var}\left(\frac{\mathbf{1}\{Z_{c,i} = z', \mathbf{G}_{c,i} = \mathbf{g}'\}(Y_{c,i} - \mu_{i,(z',\mathbf{g}')}(\mathbf{X}_c))}{p_{i,(z',\mathbf{g}')}(\mathbf{X}_c)}\right)$$

because each term is mean-zero and their covariance is 0 following $\mathbb{E}[\mathbf{1}\{Z_{c,i} = z, \mathbf{G}_{c,i} = \mathbf{g}\}\mathbf{1}\{Z_{c,i} = z', \mathbf{G}_{c,i} = \mathbf{g}'\}] = 0$ always holds for $(z, \mathbf{g}) \neq (z', \mathbf{g}')$. The variance of $\frac{\mathbf{1}\{Z_{c,i}=z,\mathbf{G}_{c,i}=\mathbf{g}\}(Y_{c,i}-\mu_{i,(z,\mathbf{g})}(\mathbf{X}_c))}{p_{i,(z,\mathbf{g})}(\mathbf{X}_c)}$ equals

$$\text{Var}\left(\frac{\mathbf{1}\{Z_{c,i} = z, \mathbf{G}_{c,i} = \mathbf{g}\}(Y_{c,i} - \mu_{i,(z,\mathbf{g})}(\mathbf{X}_c))}{p_{i,(z,\mathbf{g})}(\mathbf{X}_c)}\right)$$

$$=\mathbb{E}\left[\mathbb{E}\left[\frac{\mathbf{1}\{Z_{c,i} = z, \mathbf{G}_{c,i} = \mathbf{g}\}(Y_{c,i} - \mu_{i,(z,\mathbf{g})}(\mathbf{X}_c))^2}{p_{i,(z,\mathbf{g})}(\mathbf{X}_c)^2}\bigg|\mathbf{X}_c\right]\right]$$

$$=\mathbb{E}\left[\frac{1}{p_{i,(z,\mathbf{g})}(\mathbf{X}_c)}\mathbb{E}\left[(Y_{c,i} - \mu_{i,(z,\mathbf{g})}(\mathbf{X}_c))^2|\mathbf{X}_c\right]\right] = \mathbb{E}\left[\frac{\sigma^2_{i,(z,\mathbf{g})}(\mathbf{X}_c)}{p_{i,(z,\mathbf{g})}(\mathbf{X}_c)}\right].$$

Similarly, the variance of $\frac{\mathbf{1}\{Z_{c,i}=z',\mathbf{G}_{c,i}=\mathbf{g}'\}(Y_{c,i}-\mu_{i,(z',\mathbf{g}')}(\mathbf{X}_c))}{p_{i,(z',\mathbf{g}')}(\mathbf{X}_c)}$ equals $\mathbb{E}\left[\frac{\sigma^2_{i,(z',\mathbf{g}')}(\mathbf{X}_c)}{p_{i,(z',\mathbf{g}')}(\mathbf{X}_c)}\right]$.

The variance of (30) equals

$$\text{Var}\left(\frac{1}{|\mathcal{I}_j|}\sum_{i\in\mathcal{I}_j}\left(\mu_{i,(z,\mathbf{g})}(\mathbf{X}_c) - \mu_{i,(z',\mathbf{g}')}(\mathbf{X}_c)\right) - \left(\mu_{i,(z,\mathbf{g})} - \mu_{i,(z',\mathbf{g}')}\right)\right)$$

$$=\mathbb{E}\left[\left(\frac{1}{|\mathcal{I}_j|}\sum_{i\in\mathcal{I}_j}\left(\mu_{i,(z,\mathbf{g})}(\mathbf{X}_c) - \mu_{i,(z',\mathbf{g}')}(\mathbf{X}_c)\right) - \left(\mu_{i,(z,\mathbf{g})} - \mu_{i,(z',\mathbf{g}')}\right)\right)^2\right]$$

$$=\frac{1}{|\mathcal{I}_j|^2}\sum_{i,i'\in\mathcal{I}_j}\mathbb{E}\big[\big(\mu_{i,(z,\mathbf{g})}(\mathbf{X}_c) - \mu_{i,(z',\mathbf{g}')}(\mathbf{X}_c)\big) - \big(\mu_{i,(z,\mathbf{g})} - \mu_{i,(z',\mathbf{g}')}\big)\big)$$

$$\cdot \big(\mu_{i',(z,\mathbf{g})}(\mathbf{X}_c) - \mu_{i',(z',\mathbf{g}')}(\mathbf{X}_c)\big) - \big(\mu_{i',(z,\mathbf{g})} - \mu_{i',(z',\mathbf{g}')}\big)\big)\big]$$

Summing the variance of (29) and (30), we finish showing the variance of $\phi_{j,(z,z',\mathbf{g},\mathbf{g}')}(\mathbf{Y}_c, \mathbf{Z}_c, \mathbf{X}_c)$ equals (12).

□

### D.5 Proof of Theorem 6 (AIPW Asymptotic Normality)

*Proof of Theorem 6 (AIPW Asymptotic Normality).* The results follow from Theorem 8 in Cattaneo (2010) that the Condition (5.3) in Theorem 5 in Cattaneo (2010) holds when the propensity score is estimated from the multinomial logistic series estimator. As a preparation, let us first show the asymptotic normality of the IPW estimator which is defined as

$$\hat{\psi}_j^{\text{ipw}}(z, \mathbf{g}) - \hat{\psi}_j^{\text{ipw}}(z', \mathbf{g}') = \frac{1}{M|\mathcal{I}_j|}\sum_{c=1}^{M}\sum_{i\in\mathcal{I}_j}\left\{\frac{\mathbf{1}\{Z_{c,i} = z, \mathbf{G}_{c,i} = \mathbf{g}\}Y_{c,i}}{\hat{p}_{i,(z,\mathbf{g})}(\mathbf{X}_c)} - \frac{\mathbf{1}\{Z_{c,i} = z', \mathbf{G}_{c,i'} = \mathbf{g}\}Y_{c,i}}{\hat{p}_{i,(z',\mathbf{g}')}(\mathbf{X}_c)}\right\} \quad (33)$$

**Lemma 3.** *Theorem 6 continues to hold with $\hat{\psi}_j^{\text{aipw}}(z, \mathbf{g}) - \hat{\psi}_j^{\text{aipw}}(z', \mathbf{g}')$ replaced by $\hat{\psi}_j^{\text{ipw}}(z, \mathbf{g}) - \hat{\psi}_j^{\text{ipw}}(z', \mathbf{g}')$.*

*Proof of Lemma 3 (IPW Asymptotic Normality).* In this proof, we consider both using the sieve and simple multinomial logistic regression to estimate the propensity $p_{i,(z,\mathbf{g})}(\mathbf{X}_c)$ for all $i = 1, \cdots, n$ and $(z, \mathbf{g})$.

57

The results follow from Theorem 8 in Cattaneo (2010) that the Condition (4.2) in Theorem 4 in Cattaneo (2010) holds when the propensity score is estimated from the multinomial logistic series estimator. Let $\mu_{i,(z,\mathbf{g})}(\mathbf{X}_c) = \mathbb{E}[Y_{c,i}(z,\mathbf{g})|\mathbf{X}_c]$. Our objective is to show

$$\varepsilon = \frac{1}{\sqrt{M}} \sum_{c=1}^{M} \frac{1}{n} \sum_{i} \left[ \left( \frac{\mathbf{1}\{Z_{c,i}=z, \mathbf{G}_{c,i}=\mathbf{g}\} Y_{c,i}}{\hat{p}_{i,(z,\mathbf{g})}(\mathbf{X}_c)} - \mu_{i,(z,\mathbf{g})} \right) - \left( \frac{\mathbf{1}\{Z_{c,i}=z, \mathbf{G}_{c,i}=\mathbf{g}\} Y_{c,i}}{p_{i,(z,\mathbf{g})}(\mathbf{X}_c)} - \mu_{i,(z,\mathbf{g})} \right) \right.$$
$$\left. + \frac{\mu_{i,(z,\mathbf{g})}(\mathbf{X}_c)}{p_{i,(z,\mathbf{g})}(\mathbf{X}_c)} \left( \mathbf{1}\{Z_{c,i}=z, \mathbf{G}_{c,i}=\mathbf{g}\} - p_{i,(z,\mathbf{g})}(\mathbf{X}_c) \right) \right] = \frac{1}{n} \sum_{i=1}^{n} \varepsilon_i = o_P(1),$$

where $\varepsilon_i = \frac{1}{\sqrt{M}} \sum_{c=1}^{M} \left[ \left( \frac{\mathbf{1}\{Z_{c,i}=z, \mathbf{G}_{c,i}=\mathbf{g}\} Y_{c,i}}{\hat{p}_{i,(z,\mathbf{g})}(\mathbf{X}_c)} - \mu_{i,(z,\mathbf{g})} \right) - \left( \frac{\mathbf{1}\{Z_{c,i}=z, \mathbf{G}_{c,i}=\mathbf{g}\} Y_{c,i}}{p_{i,(z,\mathbf{g})}(\mathbf{X}_c)} - \mu_{i,(z,\mathbf{g})} \right) \right.$
$\left. + \frac{\mu_{i,(z,\mathbf{g})}(\mathbf{X}_c)}{p_{i,(z,\mathbf{g})}(\mathbf{X}_c)} \left( \mathbf{1}\{Z_{c,i}=z, \mathbf{G}_{c,i}=\mathbf{g}\} - p_{i,(z,\mathbf{g})}(\mathbf{X}_c) \right) \right]$. If we can show $|\varepsilon_i| = O_P(1)$, then $\varepsilon = O_P(1)$ holds for finite $n$.

We decompose $\varepsilon_i$ as $\varepsilon_i = R_{i,1M} + R_{i,2M} + R_{i,3M}$ where

$$R_{i,1M} = \left| \frac{1}{\sqrt{M}} \sum_{c=1}^{M} \left[ \frac{\mathbf{1}\{Z_{c,i}=z, \mathbf{G}_{c,i}=\mathbf{g}\} Y_{c,i}}{\hat{p}_{i,(z,\mathbf{g})}(\mathbf{X}_c)} - \frac{\mathbf{1}\{Z_{c,i}=z, \mathbf{G}_{c,i}=\mathbf{g}\} Y_{c,i}}{p_{i,(z,\mathbf{g})}(\mathbf{X}_c)} \right. \right.$$
$$\left. \left. + \frac{\mathbf{1}\{Z_{c,i}=z, \mathbf{G}_{c,i}=\mathbf{g}\} Y_{c,i}}{p_{i,(z,\mathbf{g})}(\mathbf{X}_c)^2} \left( \hat{p}_{i,(z,\mathbf{g})}(\mathbf{X}_c) - p_{i,(z,\mathbf{g})}(\mathbf{X}_c) \right) \right] \right|$$

$$R_{i,2M} = \left| \frac{1}{\sqrt{M}} \sum_{c=1}^{M} \left[ \left( -\frac{\mathbf{1}\{Z_{c,i}=z, \mathbf{G}_{c,i}=\mathbf{g}\} Y_{c,i}}{p_{i,(z,\mathbf{g})}(\mathbf{X}_c)^2} + \frac{\mu_{i,(z,\mathbf{g})}(\mathbf{X}_c)}{p_{i,(z,\mathbf{g})}(\mathbf{X}_c)} \right) \left( \hat{p}_{i,(z,\mathbf{g})}(\mathbf{X}_c) - p_{i,(z,\mathbf{g})}(\mathbf{X}_c) \right) \right] \right|$$

$$R_{i,3M} = \left| \frac{1}{\sqrt{M}} \sum_{c=1}^{M} \left[ -\frac{\mu_{i,(z,\mathbf{g})}(\mathbf{X}_c)}{p_{i,(z,\mathbf{g})}(\mathbf{X}_c)} \left( \hat{p}_{i,(z,\mathbf{g})}(\mathbf{X}_c) - p_{i,(z,\mathbf{g})}(\mathbf{X}_c) \right) + \frac{\mu_{i,(z,\mathbf{g})}(\mathbf{X}_c)}{p_{i,(z,\mathbf{g})}(\mathbf{X}_c)} \left( \mathbf{1}\{Z_{c,i}=z, \mathbf{G}_{c,i}=\mathbf{g}\} - p_{i,(z,\mathbf{g})}(\mathbf{X}_c) \right) \right] \right|$$

We use a similar proof as Theorem 8 in Cattaneo (2010) to bound $R_{i,1M}$, $R_{i,2M}$ and $R_{i,3M}$. For the first term $R_{i,1M}$

$$R_{i,1M} \leq C\sqrt{M} \left\| \hat{p}_{i,(z,\mathbf{g})} - p_{i,(z,\mathbf{g})} \right\|_\infty^2 \frac{1}{M} \sum_{c=1}^{M} \frac{\mathbf{1}\{Z_{c,i}=z, \mathbf{G}_{c,i}=\mathbf{g}\} |Y_{c,i}|}{p_{i,(z,\mathbf{g})}(\mathbf{X}_c)}$$
$$= O_P(\sqrt{M}(K^\eta K^{1/2} M^{-1/2} + K^\eta K^{1/2} K^{-s/k_x})^2) = o_P(1)$$

following the boundedness of $Y_{c,i}$, overlap assumption of $p_{i,(z,\mathbf{g})}(\mathbf{X}_c)$, and uniform convergence of $\hat{p}_{i,(z,\mathbf{g})}$. In the case of simple multinomial logit, $R_{i,1M} = O_P(M^{-1/2}) = o_P(1)$.

For the second term $R_{i,2M}$, we have the expansion

$$R_{i,2M} \leq \left| \frac{1}{\sqrt{M}} \sum_{c=1}^{M} \left[ \left( -\frac{\mathbf{1}\{Z_{c,i}=z, \mathbf{G}_{c,i}=\mathbf{g}\} Y_{c,i}}{p_{i,(z,\mathbf{g})}(\mathbf{X}_c)^2} + \frac{\mu_{i,(z,\mathbf{g})}(\mathbf{X}_c)}{p_{i,(z,\mathbf{g})}(\mathbf{X}_c)} \right) \left( \hat{p}_{i,(z,\mathbf{g})}(\mathbf{X}_c) - p^0_{i,K,(z,\mathbf{g})}(\mathbf{X}_c) \right) \right] \right| \quad (34)$$

$$+ \left| \frac{1}{\sqrt{M}} \sum_{c=1}^{M} \left[ \left( -\frac{\mathbf{1}\{Z_{c,i}=z, \mathbf{G}_{c,i}=\mathbf{g}\} Y_{c,i}}{p_{i,(z,\mathbf{g})}(\mathbf{X}_c)^2} + \frac{\mu_{i,(z,\mathbf{g})}(\mathbf{X}_c)}{p_{i,(z,\mathbf{g})}(\mathbf{X}_c)} \right) \left( p^0_{i,K,(z,\mathbf{g})}(\mathbf{X}_c) - p_{i,(z,\mathbf{g})}(\mathbf{X}_c) \right) \right] \right|, \quad (35)$$

where $p^0_{i,K,(z,\mathbf{g})}$ is the probability with the population parameters in the sieve estimator. The term (34) characterizes the estimation error of the sieve estimator and we can use a second-order Taylor expansion to bound (34). Suppose the estimated and true sieve parameters are $\hat{\boldsymbol{\gamma}}_{i,K}$ and $\boldsymbol{\gamma}^0_{i,K}$. Then there exists some $\tilde{\boldsymbol{\gamma}}_{i,K}$ such that $|\tilde{\boldsymbol{\gamma}}_{i,K} - \boldsymbol{\gamma}^0_{i,K}| \leq |\hat{\boldsymbol{\gamma}}_{i,K} - \boldsymbol{\gamma}^0_{i,K}|$ and, together with $|\hat{\boldsymbol{\gamma}}_{i,K} - \boldsymbol{\gamma}^0_{i,K}| = O_P(K^{1/2}M^{-1/2} + K^{1/2}K^{-s/k_x})$

from Lemma 2,

$$(34) \leq |\hat{\boldsymbol{\gamma}}_{i,K} - \boldsymbol{\gamma}_{i,K}^0| \cdot \left| \frac{1}{\sqrt{M}} \sum_{c=1}^{M} \left[ \left( -\frac{\mathbf{1}\{Z_{c,i} = z, \mathbf{G}_{c,i} = \mathbf{g}\} Y_{c,i}}{p_{i,(z,\mathbf{g})}(\mathbf{X}_c)^2} + \frac{\mu_{i,(z,\mathbf{g})}(\mathbf{X}_c)}{p_{i,(z,\mathbf{g})}(\mathbf{X}_c)} \right) [\dot{\mathbf{L}}_{i,(z,\mathbf{g})}(h_{-(0,0)}(\mathbf{X}_c, \boldsymbol{\gamma}_{i,K}^0)) \otimes R_K(\mathbf{X}_c)^\top] \right] \right| \quad (36)$$

$$+ \sqrt{M} |\hat{\boldsymbol{\gamma}}_{i,K} - \boldsymbol{\gamma}_{i,K}^0|^2 \cdot \frac{1}{M} \sum_{c=1}^{M} \left| -\frac{\mathbf{1}\{Z_{c,i} = z, \mathbf{G}_{c,i} = \mathbf{g}\} Y_{c,i}}{p_{i,(z,\mathbf{g})}(\mathbf{X}_c)^2} + \frac{\mu_{i,(z,\mathbf{g})}(\mathbf{X}_c)}{p_{i,(z,\mathbf{g})}(\mathbf{X}_c)} \right| |\mathbf{I} \otimes R_K(\mathbf{X}_c) R_K(\mathbf{X}_c)^\top| \quad (37)$$

$$= O_P(K^{1/2} M^{-1/2} + K^{1/2} K^{-s/k_x}) O_P(K^{1/2}) + O_P(\sqrt{M}(K^{1/2} M^{-1/2} + K^{1/2} K^{-s/k_x})^2) O_P(K) = o_P(1) \quad (38)$$

where $h_{-(0,0)}(\mathbf{X}_c, \boldsymbol{\gamma}_{i,K}^0) = [R_K(\mathbf{X}_c)^\top \boldsymbol{\gamma}_{i,K,(0,1)}, \cdots, R_K(\mathbf{X}_c)^\top \boldsymbol{\gamma}_{i,K,(1,G)}]^\top$ (we set $(z, \mathbf{g}) = (0, \mathbf{0})$ as the base level), $\dot{\mathbf{L}}_{i,(z,\mathbf{g})}(\cdot)$ is the gradient of the log-likelihood function, and $\mathbf{I}$ is the identity matrix. The second term in the right-hand side of (36) is $O_P(K^{1/2})$ following

$$\mathbb{E}\left[ \left( -\frac{\mathbf{1}\{Z_{c,i} = z, \mathbf{G}_{c,i} = \mathbf{g}\} Y_{c,i}}{p_{i,(z,\mathbf{g})}(\mathbf{X}_c)^2} + \frac{\mu_{i,(z,\mathbf{g})}(\mathbf{X}_c)}{p_{i,(z,\mathbf{g})}(\mathbf{X}_c)} \right) [\dot{\mathbf{L}}_{i,(z,\mathbf{g})}(h_{-(0,0)}(\mathbf{X}_c, \boldsymbol{\gamma}_{i,K}^0)) \otimes R_K(\mathbf{X}_c)^\top] \right]$$

$$= \mathbb{E}\left[ \mathbb{E}\left[ \left( -\frac{\mathbf{1}\{Z_{c,i} = z, \mathbf{G}_{c,i} = \mathbf{g}\} Y_{c,i}}{p_{i,(z,\mathbf{g})}(\mathbf{X}_c)^2} + \frac{\mu_{i,(z,\mathbf{g})}(\mathbf{X}_c)}{p_{i,(z,\mathbf{g})}(\mathbf{X}_c)} \right) \middle| \mathbf{X}_c \right] [\dot{\mathbf{L}}_{i,(z,\mathbf{g})}(h_{-(0,0)}(\mathbf{X}_c, \boldsymbol{\gamma}_{i,K}^0)) \otimes R_K(\mathbf{X}_c)^\top] \right] = 0$$

and

$$\mathbb{E}\left| \frac{1}{\sqrt{M}} \sum_{c=1}^{M} \left[ \left( -\frac{\mathbf{1}\{Z_{c,i} = z, \mathbf{G}_{c,i} = \mathbf{g}\} Y_{c,i}}{p_{i,(z,\mathbf{g})}(\mathbf{X}_c)^2} + \frac{\mu_{i,(z,\mathbf{g})}(\mathbf{X}_c)}{p_{i,(z,\mathbf{g})}(\mathbf{X}_c)} \right) [\dot{\mathbf{L}}_{i,(z,\mathbf{g})}(h_{-(0,0)}(\mathbf{X}_c, \boldsymbol{\gamma}_{i,K}^0)) \otimes R_K(\mathbf{X}_c)^\top] \right] \right|^2$$

$$\leq \max_{\mathbf{X}_c} \left| \dot{\mathbf{L}}_{i,(z,\mathbf{g})}(h_{-(0,0)}(\mathbf{X}_c, \boldsymbol{\gamma}_{i,K}^0)) \otimes R_K(\mathbf{X}_c)^\top \right|^2 \cdot \frac{1}{M} \sum_{c=1}^{M} \mathbb{E}\left| -\frac{\mathbf{1}\{Z_{c,i} = z, \mathbf{G}_{c,i} = \mathbf{g}\} Y_{c,i}}{p_{i,(z,\mathbf{g})}(\mathbf{X}_c)^2} + \frac{\mu_{i,(z,\mathbf{g})}(\mathbf{X}_c)}{p_{i,(z,\mathbf{g})}(\mathbf{X}_c)} \right|^2 = O(K).$$

The second term in the right-hand side of (37) is $O_P(K)$ following the boundedness of $Y_{c,i}$ and $|\mathbf{I} \otimes R_K(\mathbf{X}_c) R_K(\mathbf{X}_c)^\top| = O(K)$ (the dimension of $\mathbf{I} \otimes R_K(\mathbf{X}_c) R_K(\mathbf{X}_c)^\top$ is $S_j K \times S_j K$, where $S_j$ is the number of possible realizations of $(Z_{c,i}, \mathbf{G}_{c,i})$ for unit $i \in \mathcal{I}_j$ and $S_j$ is finite).

For the term (35), recall $\left\| p_{i,K,(z,\mathbf{g})}^0(\mathbf{X}_c) - p_{i,(z,\mathbf{g})}(\mathbf{X}_c) \right\|_\infty = K^{-s/k_x}$

$$\mathbb{E}[(35)]^2 \leq \left\| p_{i,K,(z,\mathbf{g})}^0(\mathbf{X}_c) - p_{i,(z,\mathbf{g})}(\mathbf{X}_c) \right\|_\infty^2 \cdot \frac{1}{M} \sum_{c=1}^{M} \mathbb{E}\left| -\frac{\mathbf{1}\{Z_{c,i} = z, \mathbf{G}_{c,i} = \mathbf{g}\} Y_{c,i}}{p_{i,(z,\mathbf{g})}(\mathbf{X}_c)^2} + \frac{\mu_{i,(z,\mathbf{g})}(\mathbf{X}_c)}{p_{i,(z,\mathbf{g})}(\mathbf{X}_c)} \right|^2$$

$$= O_P(K^{-2s/k_x}) = o_P(1)$$

In the case of simple multinomial logit, we do not have the term (35) and the only term in $R_{i,2M}$ has $(34) = O_P(M^{-1/2})$.

For the last term $R_{i,3M}$, we use the property that the first order condition of MLSE has $\sum_{c=1}^{M} \left( \mathbf{1}\{Z_{c,i} = z, \mathbf{G}_{c,i} = \mathbf{g}\} - p_{i,(z,\mathbf{g})} \right) = 0$. Then with a properly chosen $\theta$ similar as Cattaneo (2010) ($\tilde{\boldsymbol{\gamma}}$ is the projection of $\frac{\mu_{i,(z,\mathbf{g})}(\mathbf{X}_c)}{p_{i,(z,\mathbf{g})}(\mathbf{X}_c)}$ on $R_K(\mathbf{X}_c)$), we have

$$R_{i,3M} = \left| \frac{1}{\sqrt{M}} \sum_{c=1}^{M} \left( \frac{\mu_{i,(z,\mathbf{g})}(\mathbf{X}_c)}{p_{i,(z,\mathbf{g})}(\mathbf{X}_c)} - R_K(\mathbf{X}_c)^\top \tilde{\boldsymbol{\gamma}} \right) \left( \mathbf{1}\{Z_{c,i} = z, \mathbf{G}_{c,i} = \mathbf{g}\} - p_{i,(z,\mathbf{g})}(\mathbf{X}_c) \right) \right| \quad (39)$$

$$+ \left| \frac{1}{\sqrt{M}} \sum_{c=1}^{M} \left( \frac{\mu_{i,(z,\mathbf{g})}(\mathbf{X}_c)}{p_{i,(z,\mathbf{g})}(\mathbf{X}_c)} - R_K(\mathbf{X}_c)^\top \tilde{\boldsymbol{\gamma}} \right) \left( p_{i,(z,\mathbf{g})}(\mathbf{X}_c) - \hat{p}_{i,(z,\mathbf{g})}(\mathbf{X}_c) \right) \right| \quad (40)$$

59

For the term (39),

$$\mathbb{E}\left[\left(\frac{\mu_{i,(z,\mathbf{g})}(\mathbf{X}_c)}{p_{i,(z,\mathbf{g})}(\mathbf{X}_c)} - R_K(\mathbf{X}_c)^\top \tilde{\gamma}\right)\left(\mathbf{1}\{Z_{c,i}=z, \mathbf{G}_{c,i}=\mathbf{g}\} - p_{i,(z,\mathbf{g})}(\mathbf{X}_c)\right)\right]$$
$$=\mathbb{E}\left[\left(\frac{\mu_{i,(z,\mathbf{g})}(\mathbf{X}_c)}{p_{i,(z,\mathbf{g})}(\mathbf{X}_c)} - R_K(\mathbf{X}_c)^\top \tilde{\gamma}\right)\mathbb{E}\left[\mathbf{1}\{Z_{c,i}=z, \mathbf{G}_{c,i}=\mathbf{g}\} - p_{i,(z,\mathbf{g})}(\mathbf{X}_c)\Big|\mathbf{X}_c\right]\right]=0$$

and

$$\mathbb{E}[(39)]^2 \leq \max_{\mathbf{X}_c}\left(\frac{\mu_{i,(z,\mathbf{g})}(\mathbf{X}_c)}{p_{i,(z,\mathbf{g})}(\mathbf{X}_c)} - R_K(\mathbf{X}_c)^\top \theta\right)^2 \cdot \frac{1}{M}\sum_{c=1}^M \mathbb{E}\left|\left(\mathbf{1}\{Z_{c,i}=z, \mathbf{G}_{c,i}=\mathbf{g}\} - p_{i,(z,\mathbf{g})}(\mathbf{X}_c)\right|^2 = O_P(K^{-2s/k_x}),$$

where $\max_{\mathbf{X}_c}\left|\frac{\mu_{i,(z,\mathbf{g})}(\mathbf{X}_c)}{p_{i,(z,\mathbf{g})}(\mathbf{X}_c)} - R_K(\mathbf{X}_c)^\top \tilde{\gamma}\right| = O_P(K^{-s/k_x})$ follows from Assumption 7 and Newey (1997) (similar as the argument in Cattaneo (2010)).

For the term (40),

$$(40) = M^{1/2}\max_{\mathbf{X}_c}\left|\frac{\mu_{i,(z,\mathbf{g})}(\mathbf{X}_c)}{p_{i,(z,\mathbf{g})}(\mathbf{X}_c)} - R_K(\mathbf{X}_c)^\top \tilde{\gamma}\right| \cdot \max_{\mathbf{X}_c}\left|p_{i,(z,\mathbf{g})}(\mathbf{X}_c) - \hat{p}_{i,(z,\mathbf{g})}(\mathbf{X}_c)\right|$$
$$= M^{1/2}O_P(K^{-s/k_x})O_P(K^\eta K^{1/2}n^{-1/2} + K^\eta K^{1/2}K^{-s/k_x})$$

Therefore we have

$$R_{i,3M} = O_P(K^{-s/k_x}) + M^{1/2}O_P(K^{-s/k_x})O_P(K^\eta K^{1/2}n^{-1/2} + K^\eta K^{1/2}K^{-s/k_x}) = o_P(1)$$

$\square$

Given Lemma 3, we are ready to show the asymptotic distribution of the AIPW estimator. It is equivalent to showing

$$\varepsilon = \frac{1}{\sqrt{M}}\sum_{c=1}^M \frac{1}{n}\sum_i \left[\left(\frac{\mathbf{1}\{Z_{c,i}=z, \mathbf{G}_{c,i}=\mathbf{g}\}(Y_{c,i}-\hat{\mu}_{i,(z,\mathbf{g})})}{\hat{p}_{i,(z,\mathbf{g})}(\mathbf{X}_c)} + \hat{\mu}_{i,(z,\mathbf{g})}\right)\right.$$
$$\left. - \left(\frac{\mathbf{1}\{Z_{c,i}=z, \mathbf{G}_{c,i}=\mathbf{g}\}(Y_{c,i}-\mu_{i,(z,\mathbf{g})})}{p_{i,(z,\mathbf{g})}(\mathbf{X}_c)} + \mu_{i,(z,\mathbf{g})}\right)\right] = \frac{1}{n}\sum_{i=1}^n \varepsilon_i = o_P(1).$$

where $\varepsilon_i = \frac{1}{\sqrt{M}}\sum_{c=1}^M \left[\left(\frac{\mathbf{1}\{Z_{c,i}=z,\mathbf{G}_{c,i}=\mathbf{g}\}Y_{c,i}}{\hat{p}_{i,(z,\mathbf{g})}(\mathbf{X}_c)} - \mu_{i,(z,\mathbf{g})}\right) - \left(\frac{\mathbf{1}\{Z_{c,i}=z,\mathbf{G}_{c,i}=\mathbf{g}\}Y_{c,i}}{p_{i,(z,\mathbf{g})}(\mathbf{X}_c)} - \mu_{i,(z,\mathbf{g})}\right)\right.$
$\left. + \frac{\mu_{i,(z,\mathbf{g})}(\mathbf{X}_c)}{p_{i,(z,\mathbf{g})}(\mathbf{X}_c)}\left(\mathbf{1}\{Z_{c,i}=z, \mathbf{G}_{c,i}=\mathbf{g}\} - p_{i,(z,\mathbf{g})}(\mathbf{X}_c)\right)\right]$. If we can show $|\varepsilon_i| = O_P(1)$, then $\varepsilon = O_P(1)$ holds for finite $n$.

We decompose $\varepsilon_i$ as $\varepsilon_i = R_{i,4M} + R_{i,5M} + 2\cdot R_{i,6M} + o_P(1)$ using the identity $\hat{a}/\hat{b} = a/b + (\hat{a}-a)/b -$

$a(\hat{b} - b)/b^2 + a(\hat{b} - b)/(b^2\hat{b}) - (\hat{a} - a)(\hat{b} - b)/(b\hat{b})$ where

$$R_{i,4M} = \left| \frac{1}{\sqrt{M}} \sum_{c=1}^{M} \left[ \frac{\mathbf{1}\{Z_{c,i} = z, \mathbf{G}_{c,i} = \mathbf{g}\}(Y_{c,i} - \mu_{i,(z,\mathbf{g})}(\mathbf{X}_c))}{p_{i,(z,\mathbf{g})}(\mathbf{X}_c)^2} \left( \hat{p}_{i,(z,\mathbf{g})}(\mathbf{X}_c) - p_{i,(z,\mathbf{g})}(\mathbf{X}_c) \right) \right] \right|$$

$$R_{i,5M} = \left| \frac{1}{\sqrt{M}} \sum_{c=1}^{M} \left[ \frac{\mathbf{1}\{Z_{c,i} = z, \mathbf{G}_{c,i} = \mathbf{g}\} - p_{i,(z,\mathbf{g})}(\mathbf{X}_c)}{p_{i,(z,\mathbf{g})}(\mathbf{X}_c)} \left( \hat{\mu}_{i,(z,\mathbf{g})}(\mathbf{X}_c) - \mu_{i,(z,\mathbf{g})}(\mathbf{X}_c) \right) \right] \right|$$

$$R_{i,6M} = \left| \frac{1}{\sqrt{M}} \sum_{c=1}^{M} \left( \hat{\mu}_{i,(z,\mathbf{g})}(\mathbf{X}_c) - \mu_{i,(z,\mathbf{g})}(\mathbf{X}_c) \right) \right|$$

For the first term $R_{i,4M}$,

$$R_{i,4M} \leq \left| \frac{1}{\sqrt{M}} \sum_{c=1}^{M} \left[ \frac{\mathbf{1}\{Z_{c,i} = z, \mathbf{G}_{c,i} = \mathbf{g}\}(Y_{c,i} - \mu_{i,(z,\mathbf{g})}(\mathbf{X}_c))}{p_{i,(z,\mathbf{g})}(\mathbf{X}_c)^2} \left( \hat{p}_{i,(z,\mathbf{g})}(\mathbf{X}_c) - p^0_{i,K,(z,\mathbf{g})}(\mathbf{X}_c) \right) \right] \right| \quad (41)$$

$$+ \left| \frac{1}{\sqrt{M}} \sum_{c=1}^{M} \left[ \frac{\mathbf{1}\{Z_{c,i} = z, \mathbf{G}_{c,i} = \mathbf{g}\}(Y_{c,i} - \mu_{i,(z,\mathbf{g})}(\mathbf{X}_c))}{p_{i,(z,\mathbf{g})}(\mathbf{X}_c)^2} \left( p^0_{i,K,(z,\mathbf{g})}(\mathbf{X}_c) - p_{i,(z,\mathbf{g})}(\mathbf{X}_c) \right) \right] \right| \quad (42)$$

We can use the same proof as the term $R_{i,2M}$ in the proof of asymptotic normality of IPW to show that $(41) = O_P(K^{1/2}M^{-1/2} + K^{1/2}K^{-s/k_x})O_P(K^{1/2})$ and $(42) = O_P(\sqrt{M}(K^{1/2}M^{-1/2} + K^{1/2}K^{-s/k_x})^2)O_P(K)$, and therefore

$$R_{i,4M} = O_P(K^{1/2}M^{-1/2} + K^{1/2}K^{-s/k_x})O_P(K^{1/2}) + O_P(\sqrt{M}(K^{1/2}M^{-1/2} + K^{1/2}K^{-s/k_x})^2)O_P(K) = o_P(1)$$

For the second term $R_{i,5M}$,

$$R_{i,5M} \leq \left| \frac{1}{\sqrt{M}} \sum_{c=1}^{M} \left[ \frac{\mathbf{1}\{Z_{c,i} = z, \mathbf{G}_{c,i} = \mathbf{g}\} - p_{i,(z,\mathbf{g})}(\mathbf{X}_c)}{p_{i,(z,\mathbf{g})}(\mathbf{X}_c)} \left( \hat{\mu}_{i,(z,\mathbf{g})}(\mathbf{X}_c) - \mu^0_{i,K,(z,\mathbf{g})}(\mathbf{X}_c) \right) \right] \right| \quad (43)$$

$$+ \left| \frac{1}{\sqrt{M}} \sum_{c=1}^{M} \left[ \frac{\mathbf{1}\{Z_{c,i} = z, \mathbf{G}_{c,i} = \mathbf{g}\} - p_{i,(z,\mathbf{g})}(\mathbf{X}_c)}{p_{i,(z,\mathbf{g})}(\mathbf{X}_c)} \left( \mu^0_{i,K,(z,\mathbf{g})}(\mathbf{X}_c) - \mu_{i,(z,\mathbf{g})}(\mathbf{X}_c) \right) \right] \right| \quad (44)$$

where $\mu^0_{i,K,(z,\mathbf{g})}(\mathbf{X}_c)$ is the conditional expected outcome with the population parameters in the sieve estimator. Then similar as $R_{i,5M}$, we have

$$R_{i,5M} = O_P(K^{1/2}M^{-1/2} + K^{1/2}K^{-s/k_x})O_P(K^{1/2}) + O_P(\sqrt{M}(K^{1/2}M^{-1/2} + K^{1/2}K^{-s/k_x})^2)O_P(K) = o_P(1)$$

For the last term $R_{i,6M}$, following the same argument as the proof of $R_{6n} = o_P(1)$ in Theorem 8 in Cattaneo (2010) and $R_{i,3M} = o_P(1)$ in the proof of asymptotic normality of IPW, we have $R_{i,6M} = o_P(1)$. □

### D.6 Proof of Theorem 3

*Proof of Theorem 3.* Recall that $\hat{\beta}^{\text{soph}}_j(\mathbf{h}) = \sum_{\mathbf{g} \in \mathcal{G}_A(\mathbf{h})} \hat{\omega}_A(\mathbf{g}) \cdot \hat{\beta}_j(\mathbf{g})$. It is consistent as long as the fine partition satisfies Assumption 5, since each $\hat{\beta}_j(\mathbf{g})$ is consistent for $\beta_j(\mathbf{g})$ by Theorem 1 and empirical distributions $\hat{\omega}_A(\mathbf{g})$ are consistent for $\omega_A(\mathbf{g})$. To check whether $\hat{\beta}^{\text{simp}}_j(\mathbf{h})$ is consistent when the coarse partition also satisfies Assumption 5, we can follow the same reasoning in the proof of Theorem 1, and reduce to checking whether

$$\mathbb{E}\left[ \frac{\mathbf{1}\{Z_{c,i} = z, \mathbf{G}_{c,i} \in \mathcal{G}_A(\mathbf{h})\}Y_{c,i}}{\sum_{\mathbf{g} \in \mathcal{G}_A(\mathbf{h})} p_{i,(z,\mathbf{g})}(\mathbf{X}_c)} \right] = \sum_{\mathbf{g} \in \mathcal{G}_A(\mathbf{h})} \omega_A(\mathbf{g}) \mathbb{E}\left[ Y_{c,i}(z,\mathbf{g}) \right]$$

When the coarse partition satisfies Assumption 5, for all $\mathbf{g}, \mathbf{g}' \in \mathcal{G}_A(\mathbf{h})$, we have $\mathbb{E}[Y_{c,i}(z,\mathbf{g})|\mathbf{X}_c] = \mathbb{E}[Y_{c,i}(z,\mathbf{g}')|\mathbf{X}_c]$ (denote this quantity by $\mu_{z,\mathbf{h}}(\mathbf{X}_c)$), so that

$$\mathbb{E}\left[\frac{\mathbf{1}\{Z_{c,i}=z, \mathbf{G}_{c,i} \in \mathcal{G}_A(\mathbf{h})\} Y_{c,i}}{\sum_{\mathbf{g} \in \mathcal{G}_A(\mathbf{h})} p_{i,(z,\mathbf{g})}(\mathbf{X}_c)}\right] = \sum_{\mathbf{g} \in \mathcal{G}_A(\mathbf{h})} \mathbb{E}\left[\frac{\mathbf{1}\{Z_{c,i}=z, \mathbf{G}_{c,i} = \mathbf{g}\} Y_{c,i}(z,\mathbf{g})}{\sum_{\mathbf{g} \in \mathcal{G}_A(\mathbf{h})} p_{i,(z,\mathbf{g})}(\mathbf{X}_c)}\right]$$

$$= \sum_{\mathbf{g} \in \mathcal{G}_A(\mathbf{h})} \mathbb{E}\left[\frac{\mathbb{E}\left[\mathbf{1}\{Z_{c,i}=z, \mathbf{G}_{c,i} = \mathbf{g}\}|\mathbf{X}_c\right]}{\sum_{\mathbf{g} \in \mathcal{G}_A(\mathbf{h})} p_{i,(z,\mathbf{g})}(\mathbf{X}_c)} \mathbb{E}\left[Y_{c,i}(z,\mathbf{g})|\mathbf{X}_c\right]\right]$$

$$= \mathbb{E}\left[\mu_{z,\mathbf{h}}(\mathbf{X}_c) \frac{\sum_{\mathbf{g} \in \mathcal{G}_A(\mathbf{h})} \mathbb{E}\left[\mathbf{1}\{Z_{c,i}=z, \mathbf{G}_{c,i} = \mathbf{g}\}|\mathbf{X}_c\right]}{\sum_{\mathbf{g} \in \mathcal{G}_A(\mathbf{h})} p_{i,(z,\mathbf{g})}(\mathbf{X}_c)}\right]$$

$$= \mathbb{E}\left[Y_{c,i}(z,\mathbf{g})\right], \forall \mathbf{g} \in \mathcal{G}_A(\mathbf{h}),$$

and so the last term can be rewritten as $\sum_{\mathbf{g} \in \mathcal{G}_A(\mathbf{h})} \omega_A(\mathbf{g}) \mathbb{E}[Y_{c,i}(z,\mathbf{g})]$. If the coarse partition does not satisfy Assumption 5, then there exist $\mathbf{g}, \mathbf{g}' \in \mathcal{G}_A(\mathbf{h})$, such that $\mathbb{E}[Y_{c,i}(z,\mathbf{g})|\mathbf{X}_c] \neq \mathbb{E}[Y_{c,i}(z,\mathbf{g}')|\mathbf{X}_c]$, so that the last two equalities above no longer hold, unless

$$\frac{p_{i,(z,\mathbf{g})}(\mathbf{X}_c)}{\sum_{\mathbf{g} \in \mathcal{G}_A(\mathbf{h})} p_{i,(z,\mathbf{g})}(\mathbf{X}_c)} \equiv \omega_A(\mathbf{g}), \forall \mathbf{X}_c, z, \mathbf{g},$$

i.e. when treatment is assigned independently across units and do not depend on $\mathbf{X}_c$.

Now we move on to prove the variance result when both partitions satisfy Assumption 5. To get the asymptotic variance of $\hat{\beta}_j^{\text{soph}}(\mathbf{h})$, note that

$$\sqrt{M}\Big(\sum_{\mathbf{g} \in \mathcal{G}_A(\mathbf{h})} \hat{\omega}_A(\mathbf{g}) \cdot \hat{\beta}_j(\mathbf{g}) - \sum_{\mathbf{g} \in \mathcal{G}_A(\mathbf{h})} \omega_A(\mathbf{g}) \cdot \beta_j(\mathbf{g})\Big)$$
$$= \sqrt{M} \sum_{\mathbf{g} \in \mathcal{G}_A(\mathbf{h})} \hat{\omega}_A(\mathbf{g}) \cdot (\hat{\beta}_j(\mathbf{g}) - \beta_j(\mathbf{g})) + \sqrt{M} \sum_{\mathbf{g} \in \mathcal{G}_A(\mathbf{h})} (\hat{\omega}_A(\mathbf{g}) - \omega_A(\mathbf{g})) \cdot \beta_j(\mathbf{g})$$

and the two terms are asymptotically independent, so that the asymptotic variance of $\hat{\beta}_j^{\text{soph}}(\mathbf{h})$ is the sum of the asymptotic variances of the two terms. For the first term, Theorem 6 implies

$$\sqrt{M} \sum_{\mathbf{g} \in \mathcal{G}_A(\mathbf{h})} \hat{\omega}_A(\mathbf{g}) \cdot (\hat{\beta}_j(\mathbf{g}) - \beta_j(\mathbf{g})) \xrightarrow{d} \mathcal{N}\big(0, V_{j,\mathbf{h}}^{\text{soph}}\big) \quad (45)$$

where $V_{j,\mathbf{h}}^{\text{soph}}$ equals

$$V_{j,\mathbf{h}} = \frac{1}{|\mathcal{I}_j|^2} \sum_{i \in \mathcal{I}_j} \sum_{\mathbf{g} \in \mathcal{G}_A(\mathbf{h})} \omega_A^2(\mathbf{g}) \mathbb{E}\left[\frac{\sigma_{i,(1,\mathbf{g})}^2(\mathbf{X}_c)}{p_{i,(1,\mathbf{g})}(\mathbf{X}_c)} + \frac{\sigma_{i,(0,\mathbf{g})}^2(\mathbf{X}_c)}{p_{i,(0,\mathbf{g})}(\mathbf{X}_c)}\right]$$
$$+ \frac{1}{|\mathcal{I}_j|^2} \sum_{i,i' \in \mathcal{I}_j} \sum_{\mathbf{g}, \mathbf{g}' \in \mathcal{G}_A(\mathbf{h})} \omega_A(\mathbf{g}) \omega_A(\mathbf{g}') \mathbb{E}\big[\big(\beta_{i,\mathbf{g}}(\mathbf{X}_c) - \beta_{i,\mathbf{g}}\big)\big(\beta_{i',\mathbf{g}'}(\mathbf{X}_c) - \beta_{i',\mathbf{g}'}\big)\big] \quad (46)$$

with $\beta_{i,\mathbf{g}}(\mathbf{X}_c) = \mathbb{E}[Y_{c,i}(1,\mathbf{g}) - Y_{c,i}(0,\mathbf{g}) \mid \mathbf{X}_c]$ and $\beta_{i,\mathbf{g}} = \mathbb{E}[Y_{c,i}(1,\mathbf{g}) - Y_{c,i}(0,\mathbf{g})]$. For the second term, we only need that it is bounded below by 0.

For $\hat{\beta}_j^{\text{simp}}(\mathbf{h})$, we can apply Theorem 2 and get

$$\sqrt{M}\big(\hat{\beta}_j^{\text{simp}}(\mathbf{h}) - \beta_j(\mathbf{h})\big) \xrightarrow{d} \mathcal{N}\big(0, V_{j,\mathbf{h}}^{\text{simp}}\big) \quad (47)$$

where $V_{j,\mathbf{h}}^{\text{simp}}$ equals

$$V_{j,\mathbf{h}}^{\text{simp}} = \frac{1}{|\mathcal{I}_j|^2} \sum_{i \in \mathcal{I}_j} \mathbb{E}\left[\frac{\sigma_{i,(1,\mathbf{h})}^2(\mathbf{X}_c)}{p_{i,(1,\mathbf{h})}(\mathbf{X}_c)} + \frac{\sigma_{i,(0,\mathbf{h})}^2(\mathbf{X}_c)}{p_{i,(0,\mathbf{h})}(\mathbf{X}_c)}\right] \\ + \frac{1}{|\mathcal{I}_j|^2} \sum_{i,i' \in \mathcal{I}_j} \mathbb{E}\left[\big(\beta_{i,\mathbf{h}}(\mathbf{X}_c) - \beta_{i,\mathbf{h}}\big)\big(\beta_{i',\mathbf{h}}(\mathbf{X}_c) - \beta_{i',\mathbf{h}}\big)\right] \quad (48)$$

where we again note that $\beta_{i,\mathbf{h}}(\mathbf{X}_c) = \beta_{i,\mathbf{g}}(\mathbf{X}_c)$ for all $\mathbf{g} \in \mathcal{G}_A(\mathbf{h})$ by exchangeability, and $\beta_{i,\mathbf{h}} = \beta_{i,\mathbf{g}}$ for all $\mathbf{g} \in \mathcal{G}_A(\mathbf{h})$. In particular, we can write

$$\beta_{i,\mathbf{h}}(\mathbf{X}_c) - \beta_{i,\mathbf{h}} = \sum_{\mathbf{g} \in \mathcal{G}_A(\mathbf{h})} \omega_A(\mathbf{g}) \cdot \big(\beta_{i,\mathbf{g}}(\mathbf{X}_c) - \beta_{i,\mathbf{g}}\big)$$

and using this we can show the second term in the RHS of (46) equals the second term of the RHS of (48) as follows. For any $i, i' \in \mathcal{I}_j$,

$$\sum_{\mathbf{g},\mathbf{g}' \in \mathcal{G}_A(\mathbf{h})} \omega_A(\mathbf{g})\omega_A(\mathbf{g}')\mathbb{E}\left[\big(\beta_{i,\mathbf{g}}(\mathbf{X}_c) - \beta_{i,\mathbf{g}}\big)\big(\beta_{i',\mathbf{g}'}(\mathbf{X}_c) - \beta_{i',\mathbf{g}'}\big)\right] \\ = \mathbb{E}\left[\Big(\sum_{\mathbf{g} \in \mathcal{G}_A(\mathbf{h})} \omega_A(\mathbf{g}) \cdot \big(\beta_{i,\mathbf{g}}(\mathbf{X}_c) - \beta_{i,\mathbf{g}}\big)\Big)\Big(\sum_{\mathbf{g}' \in \mathcal{G}_A(\mathbf{h})} \omega_A(\mathbf{g}') \cdot \big(\beta_{i',\mathbf{g}'}(\mathbf{X}_c) - \beta_{i',\mathbf{g}'}\big)\Big)\right] \\ = \mathbb{E}\left[\big(\beta_{i,\mathbf{h}}(\mathbf{X}_c) - \beta_{i,\mathbf{h}}\big)\big(\beta_{i',\mathbf{h}}(\mathbf{X}_c) - \beta_{i',\mathbf{h}}\big)\right].$$

Now show the key part of the result, that the first term in the RHS of (46) is lower bounded by the first term in the RHS of (48). For any $i \in \mathcal{I}_j$ and $z$,

$$\sum_{\mathbf{g} \in \mathcal{G}_A(\mathbf{h})} \omega_A^2(\mathbf{g})\mathbb{E}\left[\frac{\sigma_{i,(z,\mathbf{g})}^2(\mathbf{X}_c)}{p_{i,(z,\mathbf{g})}(\mathbf{X}_c)}\right] = \mathbb{E}\left[\sum_{\mathbf{g} \in \mathcal{G}_A(\mathbf{h})} \omega_A(\mathbf{g}) \cdot \Big(\frac{\sigma_{i,(z,\mathbf{g})}^2(\mathbf{X}_c)}{p_{i,(z,\mathbf{g})}(\mathbf{X}_c)/\omega_A(\mathbf{g})}\Big)\right] \\ = \mathbb{E}\left[\sum_{\mathbf{g} \in \mathcal{G}_A(\mathbf{h})} \omega_A(\mathbf{g}) \cdot \Big(\frac{\sigma_{i,(z,\mathbf{h})}^2(\mathbf{X}_c)}{p_{i,(z,\mathbf{g})}(\mathbf{X}_c)/\omega_A(\mathbf{g})}\Big)\right] \\ \geq \mathbb{E}\left[\frac{\sigma_{i,(z,\mathbf{h})}^2(\mathbf{X}_c)}{\sum_{\mathbf{g} \in \mathcal{G}_A(\mathbf{h})} \omega_A(\mathbf{g}) \cdot p_{i,(z,\mathbf{g})}(\mathbf{X}_c)/\omega_A(\mathbf{g})}\right] \\ = \mathbb{E}\left[\frac{\sigma_{i,(z,\mathbf{h})}^2(\mathbf{X}_c)}{p_{i,(z,\mathbf{h})}(\mathbf{X}_c)}\right]$$

where we have used the convexity of the function $x :\to 1/x$ for $x > 0$ and

$$p_{i,(z,\mathbf{h})}(\mathbf{X}_c) = \sum_{\mathbf{g} \in \mathcal{G}_A(\mathbf{h})} p_{i,(z,\mathbf{g})}(\mathbf{X}_c)$$

$$\sigma_{i,(z,\mathbf{h})}^2(\mathbf{X}_c) = \sigma_{i,(z,\mathbf{g})}^2(\mathbf{X}_c), \forall \mathbf{g} \in \mathcal{G}_A(\mathbf{h})$$

Moreover, equality holds if and only if for any $\mathbf{X}_c$ and $z$, $\frac{\omega_A(\mathbf{g})}{p_{i,(z,\mathbf{g})}(\mathbf{X}_c)}$ is the same across all $\mathbf{g}$.

□

63

## D.7 Proof of Theorem 4

*Proof of Theorem 4.* We first prove the case of $l = 2$ for matching-based variance estimators. Letting $Y_i^{(1)}(z, \mathbf{g})$ and $Y_i^{(2)}(z, \mathbf{g})$ denote the outcomes of the closest and second-closest matched unit in $\mathcal{J}_{2,(z,\mathbf{g})}(c, i)$, we can write

$$\hat{\beta}_{i,\mathbf{g}}(\mathbf{X}_c) := \overline{Y}_i(1, \mathbf{g}) - \overline{Y}_i(0, \mathbf{g})$$
$$\hat{\sigma}^2_{i,(z,\mathbf{g})}(\mathbf{X}_c) := (Y_i^{(1)}(z, \mathbf{g}) - \overline{Y}_i(z, \mathbf{g}))^2 + (Y_i^{(2)}(z, \mathbf{g}) - \overline{Y}_i(z, \mathbf{g}))^2$$

where $\overline{Y}_i(z, \mathbf{g}) := \frac{Y_i^{(1)}(z,\mathbf{g}) + Y_i^{(2)}(z,\mathbf{g})}{2}$ is the average of the observed outcomes of the two matched units. Note that for any unit with covariates $\mathbf{X}_c$ we can write $Y_i(z, \mathbf{g}) = \mathbb{E}[Y_i(z, \mathbf{g}) \mid \mathbf{X}_c] + \varepsilon_i$ with $\mathbb{E}[\varepsilon_i \mid \mathbf{X}_c] = 0$ (which actually holds regardless of whether $i$ is in $c$), so that

$$\overline{Y}_i(z, \mathbf{g}) = \frac{Y_i^{(1)}(z, \mathbf{g}) + Y_i^{(2)}(z, \mathbf{g})}{2}$$
$$= \frac{\mathbb{E}[Y_i^{(1)}(z, \mathbf{g}) \mid \mathbf{X}^{(1)}] + \mathbb{E}[Y_i^{(2)}(z, \mathbf{g}) \mid \mathbf{X}^{(2)}] + \varepsilon_i^{(1)} + \varepsilon_i^{(2)}}{2}$$
$$= \frac{\mu_{z,\mathbf{g}}(\mathbf{X}^{(1)}) + \mu_{z,\mathbf{g}}(\mathbf{X}^{(2)})}{2} + \frac{\varepsilon_i^{(1)} + \varepsilon_i^{(2)}}{2}$$

where the superscript $(j)$ denotes quantities of the $j$-th matched unit for $i$ with treatment $(z, \mathbf{g})$.

The key convergence property we use is $\sup_{\mathbf{X}_c} \mathbb{E}[d(\mathbf{X}_c, \mathbf{X}^{(j)})^p \mid \mathbf{X}_c] = o(1)$ as $N \to \infty$ for all $j$ and $p = 1, 2$. That is, as the number of samples goes to infinity, the expected distance and distance squared (conditional on $\mathbf{X}_c$) between $\mathbf{X}_c$ and the covariates $\mathbf{X}^{(j)}$ of any matched sample converge to 0 uniformly in $\mathbf{X}_c$. This property can be proved using the same argument as Lemma 2 in Abadie and Imbens (2006) and is therefore omitted here. Using the Lipschitz continuity of $\mu_{z,\mathbf{g}}$, and with $\mathbb{E}_{\mathbf{X}_c}$ denoting the conditional expectation given $\mathbf{X}_c$,

$$\sup_{\mathbf{X}_c} \left| \mathbb{E}[\overline{Y}_i(z, \mathbf{g}) \mid \mathbf{X}_c] - \mu_{z,\mathbf{g}}(\mathbf{X}_c) \right| = \sup_{\mathbf{X}_c} \left| \mathbb{E}[\frac{\mu_{z,\mathbf{g}}(\mathbf{X}^{(1)}) + \mu_{z,\mathbf{g}}(\mathbf{X}^{(2)})}{2} - \mu_{z,\mathbf{g}}(\mathbf{X}_c) \mid \mathbf{X}_c] \right|$$
$$\leq \frac{1}{2} \sup_{\mathbf{X}_c} \mathbb{E}_{\mathbf{X}_c} \left( |\mu_{z,\mathbf{g}}(\mathbf{X}^{(1)}) - \mu_{z,\mathbf{g}}(\mathbf{X}_c)| + |\mu_{z,\mathbf{g}}(\mathbf{X}^{(2)}) - \mu_{z,\mathbf{g}}(\mathbf{X}_c)| \right)$$
$$\leq \frac{L}{2} \sup_{\mathbf{X}_c} \mathbb{E}_{\mathbf{X}_c} \left( d(\mathbf{X}^{(1)}, \mathbf{X}_c) + d(\mathbf{X}^{(2)}, \mathbf{X}_c) \right) = o(1)$$

In other words, the matching estimator $\overline{Y}_i(z, \mathbf{g})$ for $\mu_{z,\mathbf{g}}(\mathbf{X}_c)$ is uniformly (in $\mathbf{X}_c$) asymptotically unbiased, and so is $\hat{\beta}_{i,\mathbf{g}}(\mathbf{X}_c)$ for $\beta_{i,\mathbf{g}}(\mathbf{X}_c)$.

Next, for $\hat{\sigma}^2_{i,z,\mathbf{g}}(\mathbf{X}_c)$, we have

$$\mathbb{E}_{\mathbf{X}_c}(Y_i^{(1)}(z, \mathbf{g}) - \overline{Y}_i(z, \mathbf{g}))^2 = \mathbb{E}_{\mathbf{X}_c} \left( \frac{\mu_{z,\mathbf{g}}(\mathbf{X}^{(1)}) - \mu_{z,\mathbf{g}}(\mathbf{X}^{(2)})}{2} + \frac{\varepsilon_i^{(1)} - \varepsilon_i^{(2)}}{2} \right)^2$$
$$= \mathbb{E}_{\mathbf{X}_c} \left( \frac{\mu_{z,\mathbf{g}}(\mathbf{X}^{(1)}) - \mu_{z,\mathbf{g}}(\mathbf{X}^{(2)})}{2} \right)^2 + \mathbb{E}_{\mathbf{X}_c} \left( \frac{\varepsilon_i^{(1)} - \varepsilon_i^{(2)}}{2} \right)^2$$

where the cross term vanishes regardless of whether the matched units are in the same cluster as $(c, i)$. The

first term is uniformly $o(1)$, since

$$\mathbb{E}_{\mathbf{X}_c}\left(\frac{\mu_{z,\mathbf{g}}(\mathbf{X}^{(1)}) - \mu_{z,\mathbf{g}}(\mathbf{X}^{(2)})}{2}\right)^2 \leq \frac{1}{2}\mathbb{E}_{\mathbf{X}_c}\left(\mu_{z,\mathbf{g}}(\mathbf{X}^{(1)}) - \mu_{z,\mathbf{g}}(\mathbf{X}_c)\right)^2 + \frac{1}{2}\mathbb{E}_{\mathbf{X}_c}\left(\mu_{z,\mathbf{g}}(\mathbf{X}^{(2)}) - \mu_{z,\mathbf{g}}(\mathbf{X}_c)\right)^2$$

$$\leq \frac{L^2}{2}\mathbb{E}_{\mathbf{X}_c}\left(\mathrm{d}(\mathbf{X}_c, \mathbf{X}^{(1)})^2 + \mathrm{d}(\mathbf{X}_c, \mathbf{X}^{(2)})^2\right)$$

and taking the supremum over $\mathbf{X}_c$ and using $\sup_{\mathbf{X}_c} \mathbb{E}[\mathrm{d}(\mathbf{X}_c, \mathbf{X}^{(j)})^2 \mid \mathbf{X}_c] = o(1)$ as $N \to \infty$ proves the claim. Combining this with a similar calculation for $(Y_i^{(2)}(z, \mathbf{g}) - \overline{Y}_i(z, \mathbf{g}))^2$,

$$\mathbb{E}_{\mathbf{X}_c} \hat{\sigma}_{i,(z,\mathbf{g})}^2(\mathbf{X}_c) = \mathbb{E}_{\mathbf{X}_c}\left(\left(\frac{\varepsilon_i^{(1)} - \varepsilon_i^{(2)}}{2}\right)^2 + \left(\frac{\varepsilon_i^{(2)} - \varepsilon_i^{(1)}}{2}\right)^2\right) + o(1)$$

$$= \frac{1}{2}\left(\mathbb{E}_{\mathbf{X}_c}(\varepsilon_i^{(1)})^2 + \mathbb{E}_{\mathbf{X}_c}(\varepsilon_i^{(2)})^2\right) + o(1) = \sigma_{i,(z,\mathbf{g})}^2(\mathbf{X}_c) + o(1)$$

where the cross terms vanish due to the "independent errors" assumption in (11), and taking the supremum over $\mathbf{X}_c$ proves that $\hat{\sigma}_{i,(z,\mathbf{g})}^2(\mathbf{X}_c)$ is uniformly asymptotically unbiased.

Next, we show that the uniform asymptotic unbiasedness of $\hat{\sigma}_{i,(z,\mathbf{g})}^2(\mathbf{X}_c)$ and $\hat{\beta}_{i,\mathbf{g}}(\mathbf{X}_c)$, combined with the uniform consistency of $\hat{p}_{i,(z,\mathbf{g})}(\mathbf{X}_c)$, implies that the bias-corrected plug-in variance estimator $\tilde{V}_{j,\mathbf{g}}$ is consistent.

First, uniform asymptotic unbiasedness implies the unconditional expectations satisfy

$$\mathbb{E}[\hat{\beta}_{i,\mathbf{g}}(\mathbf{X}_c)] \to \mathbb{E}\beta_{i,\mathbf{g}}(\mathbf{X}_c)$$
$$\mathbb{E}[\hat{\sigma}_{i,(z,\mathbf{g})}^2(\mathbf{X}_c)] \to \mathbb{E}\sigma_{i,(z,\mathbf{g})}^2(\mathbf{X}_c)$$

as $N \to \infty$, and similarly exchanging the order of limit and expectation,

$$\mathbb{E}\left[\frac{\hat{\sigma}_{i,(1,\mathbf{g})}^2(\mathbf{X}_c)}{p_{i,(1,\mathbf{g})}(\mathbf{X}_c)} + \frac{\hat{\sigma}_{i,(0,\mathbf{g})}^2(\mathbf{X}_c)}{p_{i,(0,\mathbf{g})}(\mathbf{X}_c)}\right] \to \mathbb{E}\left[\frac{\sigma_{i,(1,\mathbf{g})}^2(\mathbf{X}_c)}{p_{i,(1,\mathbf{g})}(\mathbf{X}_c)} + \frac{\sigma_{i,(0,\mathbf{g})}^2(\mathbf{X}_c)}{p_{i,(0,\mathbf{g})}(\mathbf{X}_c)}\right]$$

The strong law of large numbers then implies (with population propensity $p$ instead of feasible $\hat{p}$)

$$\lim_{N \to \infty} \frac{1}{|\mathcal{I}_j|^2} \sum_{i \in \mathcal{I}_j} \frac{1}{M} \sum_{c=1}^M \left[\frac{\hat{\sigma}_{i,(1,\mathbf{g})}^2(\mathbf{X}_c)}{p_{i,(1,\mathbf{g})}(\mathbf{X}_c)} + \frac{\hat{\sigma}_{i,(0,\mathbf{g})}^2(\mathbf{X}_c)}{p_{i,(0,\mathbf{g})}(\mathbf{X}_c)}\right] = \frac{1}{|\mathcal{I}_j|^2} \sum_{i \in \mathcal{I}_j} \mathbb{E}\left[\frac{\sigma_{i,(1,\mathbf{g})}^2(\mathbf{X}_c)}{p_{i,(1,\mathbf{g})}(\mathbf{X}_c)} + \frac{\sigma_{i,(0,\mathbf{g})}^2(\mathbf{X}_c)}{p_{i,(0,\mathbf{g})}(\mathbf{X}_c)}\right]$$

Using the uniform asymptotic unbiasedness of $\hat{p}_{i,(z,\mathbf{g})}(\mathbf{X}_c)$, the following estimator of $1/p_{i,(z,\mathbf{g})}(\mathbf{X}_c)$

$$\widehat{p_{i,(z,\mathbf{g})}^{-1}}(\mathbf{X}_c) = \frac{1}{q_{\tilde{K}}} \prod_{\ell=1}^{\tilde{K}} \left(1 - \hat{p}_{i,(z,\mathbf{g})}^{\ell}(\mathbf{X}_c)\right)$$

is uniformly asymptotically unbiased, using the same argument as Blanchet et al. (2015); Moka et al. (2019). Using the boundedness of $\hat{\sigma}^2$ and overlap of $p$, we then have

$$\mathbb{E}\left[\frac{\hat{\sigma}_{i,(1,\mathbf{g})}^2(\mathbf{X}_c)}{p_{i,(1,\mathbf{g})}(\mathbf{X}_c)} + \frac{\hat{\sigma}_{i,(0,\mathbf{g})}^2(\mathbf{X}_c)}{p_{i,(0,\mathbf{g})}(\mathbf{X}_c)}\right] - \mathbb{E}\left[\frac{\hat{\sigma}_{i,(1,\mathbf{g})}^2(\mathbf{X}_c)}{\hat{p}_{i,(1,\mathbf{g})}(\mathbf{X}_c)} + \frac{\hat{\sigma}_{i,(0,\mathbf{g})}^2(\mathbf{X}_c)}{\hat{p}_{i,(0,\mathbf{g})}(\mathbf{X}_c)}\right] =$$
$$O(\mathbb{E}(|\frac{1}{\hat{p}_{i,(1,\mathbf{g})}(\mathbf{X}_c)} - \frac{1}{p_{i,(1,\mathbf{g})}(\mathbf{X}_c)}|) + \mathbb{E}(|\frac{1}{\hat{p}_{i,(0,\mathbf{g})}(\mathbf{X}_c)} - \frac{1}{p_{i,(0,\mathbf{g})}(\mathbf{X}_c)}|)) = o(1)$$

so that we also have the consistency property

$$\lim_{N \to \infty} \frac{1}{|\mathcal{I}_j|^2} \sum_{i \in \mathcal{I}_j} \frac{1}{M} \sum_{c=1}^{M} \left[ \frac{\hat{\sigma}^2_{i,(1,\mathbf{g})}(\mathbf{X}_c)}{\hat{p}_{i,(1,\mathbf{g})}(\mathbf{X}_c)} + \frac{\hat{\sigma}^2_{i,(0,\mathbf{g})}(\mathbf{X}_c)}{\hat{p}_{i,(0,\mathbf{g})}(\mathbf{X}_c)} \right] = \frac{1}{|\mathcal{I}_j|^2} \sum_{i \in \mathcal{I}_j} \mathbb{E} \left[ \frac{\sigma^2_{i,(1,\mathbf{g})}(\mathbf{X}_c)}{p_{i,(1,\mathbf{g})}(\mathbf{X}_c)} + \frac{\sigma^2_{i,(0,\mathbf{g})}(\mathbf{X}_c)}{p_{i,(0,\mathbf{g})}(\mathbf{X}_c)} \right]$$

Next, we deal with terms involving $\hat{\beta}_{i,\mathbf{g}}(\mathbf{X}_c) - \hat{\beta}_{i,\mathbf{g}}$ and bias correction terms in $\tilde{V}_{j,\mathbf{g}}$. Recall

$$\hat{\beta}_{i,\mathbf{g}}(\mathbf{X}_c) = \overline{Y}_i(1,\mathbf{g}) - \overline{Y}_i(0,\mathbf{g})$$

$$\overline{Y}_i(z,\mathbf{g}) = \frac{\mu_{z,\mathbf{g}}(\mathbf{X}^{(1)}) + \mu_{z,\mathbf{g}}(\mathbf{X}^{(2)})}{2} + \frac{\varepsilon_i^{(1)} + \varepsilon_i^{(2)}}{2}$$

so that expanding the squares,

$$\mathbb{E}\big[(\hat{\beta}_{i,\mathbf{g}}(\mathbf{X}_c) - \beta_{i,\mathbf{g}})^2\big] = \mathbb{E} \left( \frac{\mu_{1,\mathbf{g}}(\mathbf{X}^{(1)}_{1,\mathbf{g}}) + \mu_{1,\mathbf{g}}(\mathbf{X}^{(2)}_{1,\mathbf{g}})}{2} - \frac{\mu_{0,\mathbf{g}}(\mathbf{X}^{(1)}_{0,\mathbf{g}}) + \mu_{0,\mathbf{g}}(\mathbf{X}^{(2)}_{0,\mathbf{g}})}{2} \right)^2$$

$$+ \mathbb{E} \left( \frac{\varepsilon^{(1)}_{i,(1,\mathbf{g})} + \varepsilon^{(2)}_{i,(1,\mathbf{g})}}{2} - \frac{\varepsilon^{(1)}_{i,(0,\mathbf{g})} + \varepsilon^{(2)}_{i,(0,\mathbf{g})}}{2} \right)^2 - 2\mathbb{E}\hat{\beta}_{i,\mathbf{g}}(\mathbf{X}_c)\beta_{i,\mathbf{g}} + \beta^2_{i,\mathbf{g}}$$

where the cross term in $(\hat{\beta}_{i,\mathbf{g}}(\mathbf{X}_c))^2$ vanishes again because $\mathbb{E}[\varepsilon_i \mid \mathbf{X}_c] = 0$. For the first term above, using the boundedness and Lipschitzness of $\mu$, we have

$$\sup_{\mathbf{X}_c} \left| \mathbb{E}_{\mathbf{X}_c} \left( \frac{\mu_{1,\mathbf{g}}(\mathbf{X}^{(1)}_{1,\mathbf{g}}) + \mu_{1,\mathbf{g}}(\mathbf{X}^{(2)}_{1,\mathbf{g}})}{2} - \frac{\mu_{0,\mathbf{g}}(\mathbf{X}^{(1)}_{0,\mathbf{g}}) + \mu_{0,\mathbf{g}}(\mathbf{X}^{(2)}_{0,\mathbf{g}})}{2} \right)^2 - \mathbb{E}_{\mathbf{X}_c} (\mu_{1,\mathbf{g}}(\mathbf{X}_c) - \mu_{0,\mathbf{g}}(\mathbf{X}_c))^2 \right|$$

$$= O \left( \sup_{\mathbf{X}_c} \mathbb{E}_{\mathbf{X}_c} d(\mathbf{X}^{(1)}_{1,\mathbf{g}}, \mathbf{X}_c) + d(\mathbf{X}^{(2)}_{1,\mathbf{g}}, \mathbf{X}_c) + d(\mathbf{X}^{(1)}_{0,\mathbf{g}}, \mathbf{X}_c) + d(\mathbf{X}^{(2)}_{0,\mathbf{g}}, \mathbf{X}_c) \right) = o(1)$$

which implies that

$$\mathbb{E} \left( \frac{\mu_{1,\mathbf{g}}(\mathbf{X}^{(1)}_{1,\mathbf{g}}) + \mu_{1,\mathbf{g}}(\mathbf{X}^{(2)}_{1,\mathbf{g}})}{2} - \frac{\mu_{0,\mathbf{g}}(\mathbf{X}^{(1)}_{0,\mathbf{g}}) + \mu_{0,\mathbf{g}}(\mathbf{X}^{(2)}_{0,\mathbf{g}})}{2} \right)^2 \to \mathbb{E}(\beta_{i,\mathbf{g}}(\mathbf{X}_c))^2$$

On the other hand, by definition of $\varepsilon$'s,

$$\mathbb{E} \left( \frac{\varepsilon^{(1)}_{i,(1,\mathbf{g})} + \varepsilon^{(2)}_{i,(1,\mathbf{g})}}{2} - \frac{\varepsilon^{(1)}_{i,(0,\mathbf{g})} + \varepsilon^{(2)}_{i,(0,\mathbf{g})}}{2} \right)^2 = \frac{1}{2}\mathbb{E}(\sigma^2_{i,(1,\mathbf{g})}(\mathbf{X}_c) + \sigma^2_{i,(0,\mathbf{g})}(\mathbf{X}_c))$$

so that combining these together, we have

$$\mathbb{E}\left[(\hat{\beta}_{i,\mathbf{g}}(\mathbf{X}_c) - \beta_{i,\mathbf{g}})^2\right] \to \mathbb{E}(\beta_{i,\mathbf{g}}(\mathbf{X}_c))^2 + 2\mathbb{E}\beta_{i,\mathbf{g}}(\mathbf{X}_c)\beta_{i,\mathbf{g}} + \beta^2_{i,\mathbf{g}} + \frac{1}{2}\mathbb{E}(\sigma^2_{i,(1,\mathbf{g})}(\mathbf{X}_c) + \sigma^2_{i,(0,\mathbf{g})}(\mathbf{X}_c))$$

$$= \mathbb{E}(\beta_{i,\mathbf{g}}(\mathbf{X}_c) - \beta_{i,\mathbf{g}})^2 + \frac{1}{2}\mathbb{E}\left(\sigma^2_{i,(1,\mathbf{g})}(\mathbf{X}_c) + \sigma^2_{i,(0,\mathbf{g})}(\mathbf{X}_c)\right)$$

Similarly, using the assumption that $\mathbf{X}_c$ has non-identical rows with probability 1, so that for large enough $N$, any two units regardless of their clusters have distinct matches with probability 1, the cross terms do not

have asymptotic bias:

$$\mathbb{E}(\hat{\beta}_{i,\mathbf{g}}(\mathbf{X}_c) - \beta_{i,\mathbf{g}})(\hat{\beta}_{i',\mathbf{g}}(\mathbf{X}_c) - \beta_{i',\mathbf{g}}) \to \mathbb{E}\left[(\beta_{i,\mathbf{g}}(\mathbf{X}_c) - \beta_{i,\mathbf{g}})(\beta_{i',\mathbf{g}}(\mathbf{X}_c) - \beta_{i',\mathbf{g}})\right]$$

so that invoking the strong law of large numbers again,

$$\hat{V}_{j,\mathbf{g}} \to_p V_{j,\mathbf{g}} + \frac{1}{2}\frac{1}{|\mathcal{I}_j|^2}\sum_{i\in\mathcal{I}_j}\mathbb{E}(\sigma^2_{i,(1,\mathbf{g})}(\mathbf{X}_c) + \sigma^2_{i,(0,\mathbf{g})}(\mathbf{X}_c))$$

and we see that the asymptotic bias of $\hat{V}_{j,\mathbf{g}}$ arising from matching is exactly corrected by the term

$$\frac{1}{2}\frac{1}{|\mathcal{I}_j|^2}\sum_{i\in\mathcal{I}_j}\frac{1}{M}\sum_{c=1}^M \left(\hat{\sigma}^2_{i,(1,\mathbf{g})}(\mathbf{X}_c) + \hat{\sigma}^2_{i,(0,\mathbf{g})}(\mathbf{X}_c)\right) \to_p \frac{1}{2}\frac{1}{|\mathcal{I}_j|^2}\sum_{i\in\mathcal{I}_j}\mathbb{E}(\sigma^2_{i,(1,\mathbf{g})}(\mathbf{X}_c) + \sigma^2_{i,(0,\mathbf{g})}(\mathbf{X}_c))$$

Finally, using the consistency of $\hat{\beta}_{i,\mathbf{g}}$ for $\beta_{i,\mathbf{g}}$,

$$\mathbb{E}\left[(\hat{\beta}_{i,\mathbf{g}}(\mathbf{X}_c) - \hat{\beta}_{i,\mathbf{g}})^2\right] - \mathbb{E}\left[(\hat{\beta}_{i,\mathbf{g}}(\mathbf{X}_c) - \beta_{i,\mathbf{g}})^2\right] = o(1)$$
$$\mathbb{E}(\hat{\beta}_{i,\mathbf{g}}(\mathbf{X}_c) - \hat{\beta}_{i,\mathbf{g}})(\hat{\beta}_{i',\mathbf{g}}(\mathbf{X}_c) - \hat{\beta}_{i',\mathbf{g}}) - \mathbb{E}(\hat{\beta}_{i,\mathbf{g}}(\mathbf{X}_c) - \beta_{i,\mathbf{g}})(\hat{\beta}_{i',\mathbf{g}}(\mathbf{X}_c) - \beta_{i',\mathbf{g}}) = o(1)$$

and we have proved the consistency result for $l = 2$:

$$\tilde{V}_\beta(\mathbf{g}) \xrightarrow{P} V_\beta(\mathbf{g})$$

For general $l$, the proof requires minimal adaptation from the $l = 2$ case for the uniform asymptotic unbiasedness of $\hat{\sigma}^2_{i,(z,\mathbf{g})}(\mathbf{X}_c)$ and $\hat{\beta}_{i,\mathbf{g}}(\mathbf{X}_c)$, and is identical for the consistency of variance components involving $\hat{\sigma}^2_{i,(z,\mathbf{g})}(\mathbf{X}_c)$. The only difference that warrants additional calculation is the bias term resulting from estimation errors of matching estimators:

$$\mathbb{E}\left(\frac{\varepsilon^{(1)}_{i,(1,\mathbf{g})} + \cdots + \varepsilon^{(l)}_{i,(1,\mathbf{g})}}{l} - \frac{\varepsilon^{(1)}_{i,(0,\mathbf{g})} + \cdots + \varepsilon^{(l)}_{i,(0,\mathbf{g})}}{l}\right)^2 = \frac{1}{l}\mathbb{E}(\sigma^2_{i,(1,\mathbf{g})}(\mathbf{X}_c) + \sigma^2_{i,(0,\mathbf{g})}(\mathbf{X}_c))$$

□

### D.8 Proof of Theorem 7

*Proof of Theorem 7.* In this proof, we show the consistency and asymptotic normality of $\hat{\psi}^{\text{aipw}}_j(z,\mathbf{g}) - \hat{\psi}^{\text{aipw}}_j(z',\mathbf{g}')$ for every subset $j \in \{1,\cdots,m\}$. Then as special cases of $\hat{\psi}^{\text{aipw}}_j(z,\mathbf{g}) - \hat{\psi}^{\text{aipw}}_j(z',\mathbf{g}')$, the unbiasedness and consistency of $\hat{\beta}^{\text{aipw}}_j(\mathbf{g})$ and $\hat{\tau}^{\text{aipw}}_j(z,\mathbf{g},\mathbf{g}')$ directly follow.

Note that the plug-in estimator is

$$\hat{\psi}^{\text{aipw}}_j(z,\mathbf{g}) - \hat{\psi}^{\text{aipw}}_j(z',\mathbf{g}') = \frac{1}{\sum_{n'\in\mathcal{S}}\hat{p}_{n'}\mathbf{1}_{\|\mathbf{g}\|_1\leq n'}}\sum_{n'\in\mathcal{S}}\hat{p}_{n'}\mathbf{1}_{\|\mathbf{g}\|_1\leq n'}\left(\hat{\psi}^{\text{aipw}}_{n',j}(z,\mathbf{g}) - \hat{\psi}^{\text{aipw}}_{n',j}(z',\mathbf{g}')\right),$$

where $\hat{p}_n = M_n/M$ and $\hat{\psi}^{\text{aipw}}_{n',j}(z,\mathbf{g}) - \hat{\psi}^{\text{aipw}}_{n',j}(z',\mathbf{g}')$ is the same estimator as in (7) but only using the samples with cluster size $n$. Since $\hat{p}_n$ is consistent, $p_n$ is bounded away from 0 for all $n \in \mathcal{S}$, and $|\mathcal{S}| < \infty$, $\frac{\hat{p}_n\mathbf{1}_{\|\mathbf{g}\|_1\leq n}}{\sum_{n'\in\mathcal{S}}\hat{p}_{n'}\mathbf{1}_{\|\mathbf{g}\|_1\leq n'}} \xrightarrow{P} \frac{p_n\mathbf{1}_{\|\mathbf{g}\|_1\leq n}}{\sum_{n'\in\mathcal{S}}p_{n'}\mathbf{1}_{\|\mathbf{g}\|_1\leq n'}}$ for all $n$. The consistency of $(\hat{\psi}^{\text{aipw}}_{n,j}(z,\mathbf{g}) - \hat{\psi}^{\text{aipw}}_{n,j}(z',\mathbf{g}'))$ and Slutsky's

67

Theorem then imply $\hat{\beta}^{\text{aipw}}(g) \xrightarrow{P} \beta(g)$. For normality, scaling up the bias by $\sqrt{M}$ yields

$$\sqrt{M}\left((\hat{\psi}_j^{\text{aipw}}(z,\mathbf{g}) - \hat{\psi}_j^{\text{aipw}}(z',\mathbf{g}')) - (\psi_j(z,\mathbf{g}) - \psi_j(z',\mathbf{g}'))\right)$$

$$= \underbrace{\sqrt{M} \sum_{n \in \mathcal{S}} \frac{\hat{p}_n \mathbf{1}_{\|\mathbf{g}\|_1 \leq n}}{\sum_{n' \in \mathcal{S}} \hat{p}_{n'} \mathbf{1}_{\|\mathbf{g}\|_1 \leq n'}} \left((\hat{\psi}_{n,j}^{\text{aipw}}(z,\mathbf{g}) - \hat{\psi}_{n,j}^{\text{aipw}}(z',\mathbf{g}')) - (\psi_{n,j}(z,\mathbf{g}) - \psi_{n,j}(z',\mathbf{g}'))\right)}_{:=D_1}$$

$$+ \underbrace{\sqrt{M} \sum_{n \in \mathcal{S}} \left(\frac{\hat{p}_n \mathbf{1}_{\|\mathbf{g}\|_1 \leq n}}{\sum_{n' \in \mathcal{S}} \hat{p}_{n'} \mathbf{1}_{\|\mathbf{g}\|_1 \leq n'}} - \frac{p_n \mathbf{1}_{\|\mathbf{g}\|_1 \leq n}}{\sum_{n' \in \mathcal{S}} p_{n'} \mathbf{1}_{\|\mathbf{g}\|_1 \leq n'}}\right)(\psi_{n,j}(z,\mathbf{g}) - \psi_{n,j}(z',\mathbf{g}'))}_{:=D_2}.$$

From Theorem 2, we have for all $n \in \mathcal{S}$,

$$\sqrt{M_n}\left((\hat{\psi}_{n,j}^{\text{aipw}}(z,\mathbf{g}) - \hat{\psi}_{n,j}^{\text{aipw}}(z',\mathbf{g}')) - (\psi_{n,j}(z,\mathbf{g}) - \psi_{n,j}(z',\mathbf{g}'))\right) \xrightarrow{d} \mathcal{N}(0, V_{n,j,z,z',\mathbf{g},\mathbf{g}'})$$

where $V_{n,j,z,z',\mathbf{g},\mathbf{g}'}$ is the semiparametric bound for estimators of $(\psi_{n,j}(z,\mathbf{g}) - \psi_{n,j}(z',\mathbf{g}'))$, and therefore

$$\sqrt{M}\left((\hat{\psi}_{n,j}^{\text{aipw}}(z,\mathbf{g}) - \hat{\psi}_{n,j}^{\text{aipw}}(z',\mathbf{g}')) - (\psi_{n,j}(z,\mathbf{g}) - \psi_{n,j}(z',\mathbf{g}'))\right) \xrightarrow{d} \mathcal{N}\left(0, \frac{1}{p_n}V_{n,j,z,z',\mathbf{g},\mathbf{g}'}\right).$$

Since clusters are independent, $\hat{\psi}_{n,j}^{\text{aipw}}(z,\mathbf{g}) - \hat{\psi}_{n,j}^{\text{aipw}}(z',\mathbf{g}')$ and $\hat{\psi}_{n',j}^{\text{aipw}}(z,\mathbf{g}) - \hat{\psi}_{n',j}^{\text{aipw}}(z',\mathbf{g}')$ are independent for any $n \neq n'$, and $(\hat{\psi}_{s_1,j}^{\text{aipw}}(z,\mathbf{g}) - \hat{\psi}_{s_1,j}^{\text{aipw}}(z',\mathbf{g}'), \hat{\psi}_{s_2,j}^{\text{aipw}}(z,\mathbf{g}) - \hat{\psi}_{s_2,j}^{\text{aipw}}(z',\mathbf{g}'), \cdots, \hat{\psi}_{s_{\bar{n}},j}^{\text{aipw}}(z,\mathbf{g}) - \hat{\psi}_{s_{\bar{n}},j}^{\text{aipw}}(z',\mathbf{g}'))$ are jointly asymptotically normal, where $s_1, \cdots, s_{\bar{n}}$ is an ordering of all the possible values in $\mathcal{S}$ with $\bar{n} = |\mathcal{S}|$. Therefore,

$$D_1 \xrightarrow{d} \mathcal{N}\left(0, \underbrace{\sum_{n \in \mathcal{S}} \left(\frac{p_n \mathbf{1}_{\|\mathbf{g}\|_1 \leq n}}{\sum_{n' \in \mathcal{S}} p_{n'} \mathbf{1}_{\|\mathbf{g}\|_1 \leq n'}}\right)^2 \frac{1}{p_n} V_{n,j,z,z',\mathbf{g},\mathbf{g}'}}_{:=V_{n,j,z,z',\mathbf{g},\mathbf{g}'}^{(1)}}\right).$$

For $D_2$, note that $\hat{p}_n = \sum_{c=1}^M \mathbf{1}\{n_c = n\}/M$, and more generally

$$\begin{bmatrix} \hat{p}_{s_1} \\ \hat{p}_{s_2} \\ \vdots \\ \hat{p}_{s_{\bar{n}}} \end{bmatrix} = \frac{1}{M} \sum_c \mathbf{I}_c = \frac{1}{M} \sum_c \begin{bmatrix} \mathbf{1}\{n_c = s_1\} \\ \mathbf{1}\{n_c = s_2\} \\ \vdots \\ \mathbf{1}\{n_c = s_{\bar{n}}\} \end{bmatrix},$$

with $\mathbb{E}\left[(\mathbf{1}\{n_c = n\} - p_n)^2\right] = (1 - p_n)p_n$ and $\mathbb{E}[(\mathbf{1}\{n_c = n\} - p_n)(\mathbf{1}\{n_c = n'\} - p_{n'})] = -p_n p_{n'}$ for any $n \neq n'$. Since $\mathbf{I_c}$ are i.i.d. random vectors with finite covariance matrix, the standard multivariate CLT then implies

$$\sqrt{M}\left(\begin{bmatrix} \hat{p}_{s_1} \\ \hat{p}_{s_2} \\ \vdots \\ \hat{p}_{s_{\bar{n}}} \end{bmatrix} - \begin{bmatrix} p_{s_1} \\ p_{s_2} \\ \vdots \\ p_{s_{\bar{n}}} \end{bmatrix}\right) \xrightarrow{P} \mathcal{N}\left(0, \begin{bmatrix} p_{s_1}(1 - p_{s_1}) & -p_{s_1}p_{s_2} & \cdots & -p_{s_1}p_{s_{\bar{n}}} \\ -p_{s_2}p_{s_1} & p_{s_2}(1 - p_{s_2}) & \cdots & -p_{s_2}p_{s_{\bar{n}}} \\ \vdots & \vdots & \ddots & \vdots \\ -p_{s_{\bar{n}}}p_{s_1} & -p_{s_{\bar{n}}}p_{s_2} & \cdots & p_{s_{\bar{n}}}(1 - p_{s_{\bar{n}}}) \end{bmatrix}\right).$$

We can use the standard trick to add and subtract a term to further decompose $D_2$:

$$
\begin{aligned}
D_2 &= \frac{\sqrt{M}}{\left(\sum_{n' \in \mathcal{S}} \hat{p}_{n'} \mathbf{1}_{\|\mathbf{g}\|_1 \leq n'}\right)\left(\sum_{n' \in \mathcal{S}} p_{n'} \mathbf{1}_{\|\mathbf{g}\|_1 \leq n'}\right)} \left[\left(\sum_{n' \in \mathcal{S}} p_{n'} \mathbf{1}_{\|\mathbf{g}\|_1 \leq n'}\right)\left(\sum_{n \in \mathcal{S}} (\hat{p}_n - p_n) \mathbf{1}_{\|\mathbf{g}\|_1 \leq n} (\psi_{n,j}(z, \mathbf{g}) - \psi_{n,j}(z', \mathbf{g}'))\right)\right. \\
&\quad \left. + \left(\sum_{n \in \mathcal{S}} (p_n - \hat{p}_n) \mathbf{1}_{\|\mathbf{g}\|_1 \leq n}\right)\left(\sum_{n' \in \mathcal{S}} p_{n'} \mathbf{1}_{\|\mathbf{g}\|_1 \leq n'} (\psi_{n',j}(z, \mathbf{g}) - \psi_{n',j}(z', \mathbf{g}'))\right)\right] \\
&= \frac{\sqrt{M}}{\left(\sum_{n' \in \mathcal{S}} \hat{p}_{n'} \mathbf{1}_{\|\mathbf{g}\|_1 \leq n'}\right)\left(\sum_{n' \in \mathcal{S}} p_{n'} \mathbf{1}_{\|\mathbf{g}\|_1 \leq n'}\right)} \left[\sum_{n \in \mathcal{S}} (\hat{p}_n - p_n) \right. \\
&\quad \left. \times \underbrace{\mathbf{1}_{\|\mathbf{g}\|_1 \leq n} \left[(\psi_{n,j}(z, \mathbf{g}) - \psi_{n,j}(z', \mathbf{g}')) \left(\sum_{n' \in \mathcal{S}} p_{n'} \mathbf{1}_{\|\mathbf{g}\|_1 \leq n'}\right) - \left(\sum_{n' \in \mathcal{S}} p_{n'} \mathbf{1}_{\|\mathbf{g}\|_1 \leq n'} (\psi_{n',j}(z, \mathbf{g}) - \psi_{n',j}(z', \mathbf{g}'))\right)\right]}_{:= c_{n,j,z,z',\mathbf{g},\mathbf{g}'}}\right] \\
&\xrightarrow{d} \mathcal{N}\left(0, \underbrace{\frac{\sum_{n \in \mathcal{S}} c_{n,j,z,z',\mathbf{g},\mathbf{g}'}^2 (1-p_n)p_n - \sum_{n \neq n'} c_{n,j,z,z',\mathbf{g},\mathbf{g}'} c_{n',j,z,z',\mathbf{g},\mathbf{g}'} p_n p_{n'}}{\left(\sum_{n' \in \mathcal{S}} p_{n'} \mathbf{1}_{\|\mathbf{g}\|_1 \leq n'}\right)^4}}_{:= V^{(2)}_{n,j,z,z',\mathbf{g},\mathbf{g}'}}\right).
\end{aligned}
$$

Since the asymptotic distribution of $D_1$ depends on that of $(\mathbf{X}_c, \mathbf{Z}_c, \mathbf{Y}_c)$, while the asymptotic distribution of $D_2$ depends only on that of $n_c$, which is assumed to be independent of the data generating process $(\mathbf{X}_c, \mathbf{Z}_c, \mathbf{Y}_c)$, $D_1$ and $D_2$ are thus asymptotically independent. Hence, for all $j \in \{1, \cdots, m\}$,

$$\sqrt{M}\left((\hat{\psi}_j^{\text{aipw}}(z, \mathbf{g}) - \hat{\psi}_j^{\text{aipw}}(z', \mathbf{g}')) - (\psi_j(z, \mathbf{g}) - \psi_j(z', \mathbf{g}'))\right) \xrightarrow{d} \mathcal{N}\left(0, V^{(1)}_{n,j,z,z',\mathbf{g},\mathbf{g}'} + V^{(2)}_{n,j,z,z',\mathbf{g},\mathbf{g}'}\right)$$

$\square$

# Appendix E  Additional Simulation Results

## E.1  Additional Results for AIPW

**Figure 3:** Histograms of standardized direct treatment effect $\hat{\beta}_1^{\text{aipw}}(\mathbf{g})$ with Partial Interference

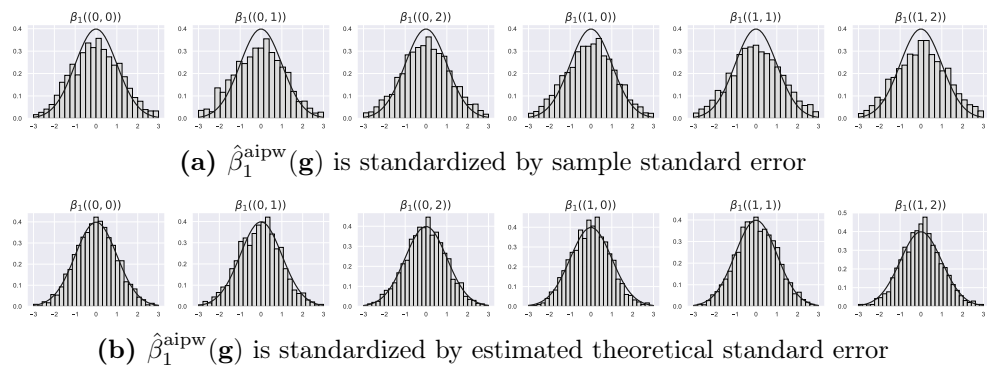

**(a)** $\hat{\beta}_1^{\text{aipw}}(\mathbf{g})$ is standardized by sample standard error

**(b)** $\hat{\beta}_1^{\text{aipw}}(\mathbf{g})$ is standardized by estimated theoretical standard error

These figures show histograms of standardized estimated direct treatment effects $\hat{\beta}_1^{\text{aipw}}(\mathbf{g})$ for the first subset under conditional exchangeability with $m = 2$ for $\mathbf{g} = (0,0), (0,1), (0,2), (1,0), (1,1), (1,2)$.

**Figure 4:** Histograms of standardized direct treatment effect $\hat{\beta}_1^{\text{aipw}}(\mathbf{g})$ with Homogeneous Interference

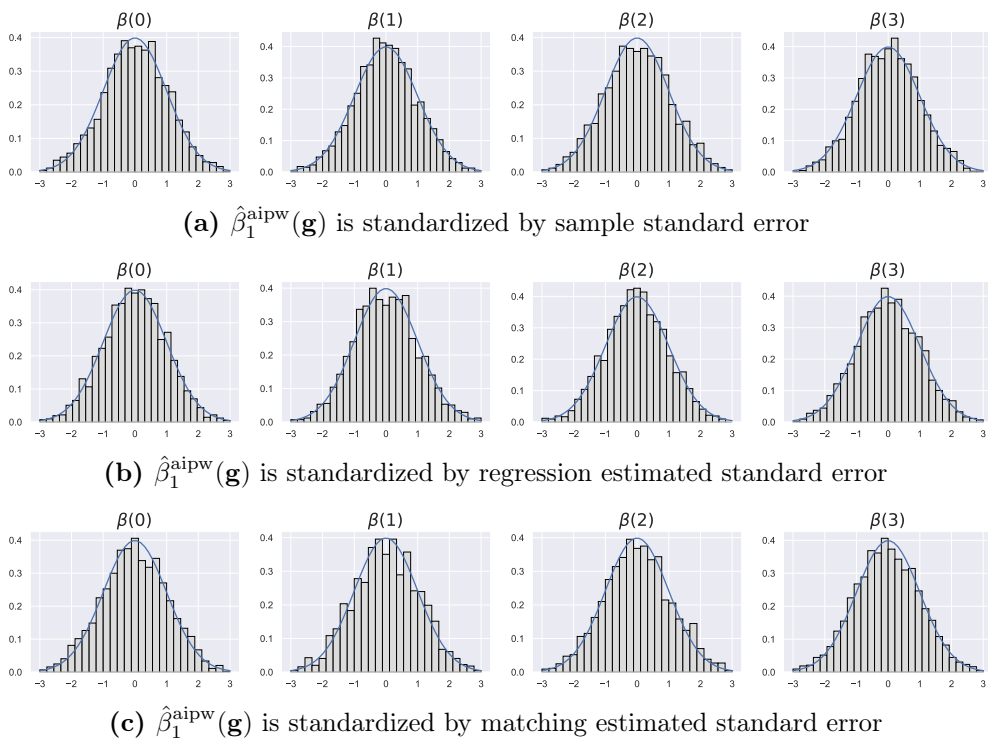

**(a)** $\hat{\beta}_1^{\text{aipw}}(\mathbf{g})$ is standardized by sample standard error

**(b)** $\hat{\beta}_1^{\text{aipw}}(\mathbf{g})$ is standardized by regression estimated standard error

**(c)** $\hat{\beta}_1^{\text{aipw}}(\mathbf{g})$ is standardized by matching estimated standard error

These figures show histograms of standardized estimated direct treatment effects $\hat{\beta}^{\text{aipw}}(g)$ for clusters of size four under full exchangeability. We run 2,000 Monte Carlo simulations.

## E.2 Varying Cluster Size

We study the finite sample properties of Theorem 5, where we allow for varying cluster sizes (see Section 7.1). For each Monte Carlo replication, we generate 5,000 clusters of size 2 and 3 and 2,000 clusters of size 5. Treatment assignments and outcomes are generated using the same models as the base case. Figure 5 shows the histogram of the standardized $\hat{\beta}^{\text{aipw}}(g)$ using the feasible variance estimator and MLR for neighborhood propensities. It similarly demonstrates the good finite sample properties of Theorem 5 and the validity of our variance estimators.

**Figure 5:** Histograms of standardized direct treatment effect $\hat{\beta}^{\text{aipw}}(g)$ with Varying Cluster Size

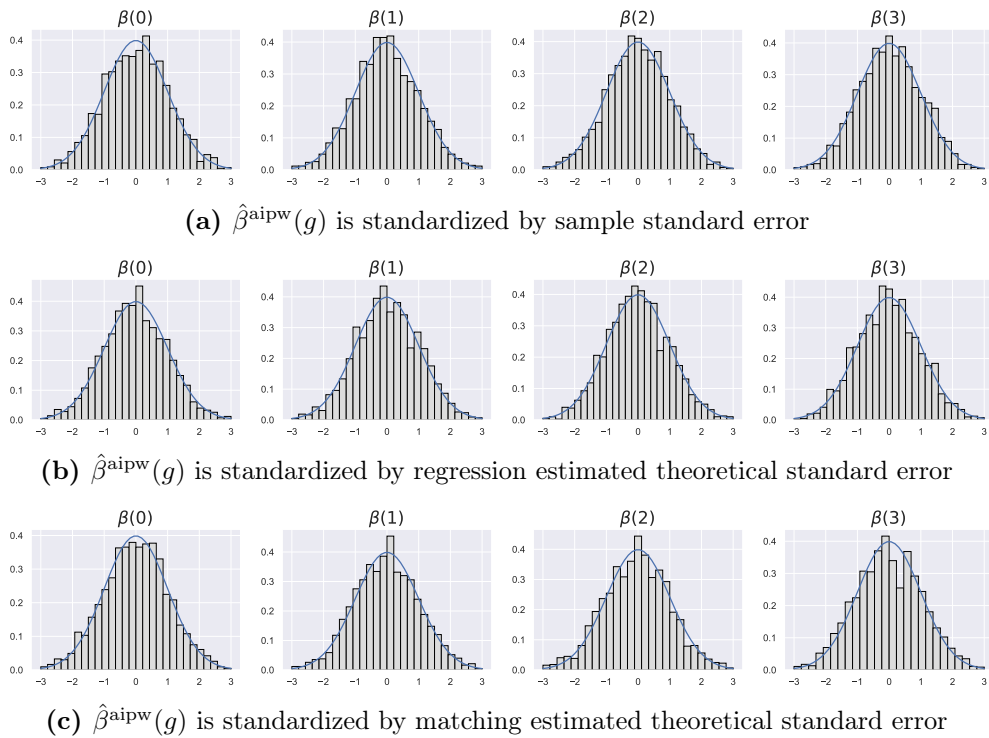

(a) $\hat{\beta}^{\text{aipw}}(g)$ is standardized by sample standard error

(b) $\hat{\beta}^{\text{aipw}}(g)$ is standardized by regression estimated theoretical standard error

(c) $\hat{\beta}^{\text{aipw}}(g)$ is standardized by matching estimated theoretical standard error

These figures show histograms of standardized estimated direct treatment effects $\hat{\beta}^{\text{aipw}}(g)$ for $g = 0, \cdots, 4$ with varying cluster size. $\hat{\beta}^{\text{aipw}}(g)$ is standardized by the feasible standard error estimator based on Theorems 4 and 5. The propensity model is the same across all units with the same cluster size. Neighborhood propensities are estimated with multinomial logistic regression. We run 2,000 Monte Carlo simulations.



\end{document}